\documentclass[11pt]{article}
\usepackage[letterpaper,margin=1in]{geometry}

\usepackage{color}
\usepackage{amsmath,amssymb,amsthm}
\usepackage{mathtools}
\usepackage{algorithm}
\usepackage{algpseudocode}
\newcommand{\es}[0]{\mathsf{J}}

\newcommand{\deltaH}[0]{\epsilon_H}

\usepackage{tikz}

\usepackage{authblk}

\DeclareMathOperator{\tr}{\operatorname{Tr}}
\newcommand{\I}{\mathbb{I}}

\newcommand{\bJ}[0]{\mathbb{J}}

\newcommand{\cO}{\mathcal{O}}
\usepackage{thmtools,thm-restate}

\newcommand{\he}[0]{\lambda}

\usepackage{hyperref}
\hypersetup{
  colorlinks=true,
  linkcolor=blue,
  linktoc=section
}

\newcommand{\norm}[1]{\left\lVert #1 \right\rVert}
\newcommand{\cL}{\mathcal{L}}
\newcommand{\cF}{\mathcal{F}}
\newcommand{\cE}{\mathcal{E}}
\DeclareMathOperator{\gap}{gap}
\newcommand{\ip}[3]{\left\langle #1, #2 \right\rangle_{#3}}
\newcommand{\wnorm}[2]{\left\lVert #1 \right\rVert_{#2}}
\newcommand{\R}{\mathbb{R}}
\newcommand{\dd}{\,\mathrm d}
\newcommand{\abs}[1]{\lvert #1\rvert}

\newcommand{\E}{\mathbb E}

\DeclareMathOperator{\Tr}{Tr}
\DeclareMathOperator{\spanop}{span}
\DeclareMathOperator{\Var}{Var}

\newcommand{\ket}[1]{| #1 \rangle}
\newcommand{\bra}[1]{\langle #1 |}

\newcommand{\C}{\mathbb{C}}
\newtheorem{theorem}{Theorem}[section]

\newtheorem{lemma}[theorem]{Lemma}
\newtheorem{claim}[theorem]{Claim}

\newtheorem{corollary}[theorem]{Corollary}

\newcommand{\ms}[0]{m_*}

\definecolor{purple}{rgb}{0.5,0,0.5}

\newcommand{\oket}[1]{| #1 \rangle\!\rangle}
\newcommand{\obra}[1]{\langle\!\langle #1 |}
\newcommand{\obraket}[2]{\langle\!\langle #1 |#2 \rangle\!\rangle}

\allowdisplaybreaks[2]

\title{Quantum simulation of low-temperature impurity models}
\author[1]{Jiaqing Jiang\thanks{\texttt{jiaqingjiang95@gmail.com}}}
\author[2]{Nathan Ju\thanks{\texttt{nju@berkeley.edu}}}

\affil[1]{Simons Institute for the Theory of Computing,
UC Berkeley, USA}

\affil[2]{Department of EECS,
UC Berkeley, USA} 
\date{}
\begin{document}
\maketitle
\begin{abstract}

Quantum impurity models, which describe a small but strongly-interacting subsystem coupled to a large noninteracting  bath, are the computational core of dynamical mean field theory and play an important role in simulating strongly correlated materials. 
We develop quantum algorithms for efficiently simulating these models at low temperatures and with high precision. For any impurity model on $n$ fermionic modes with fixed impurity size:
\begin{itemize}
    \item We give a quantum algorithm that prepares the thermal state at inverse temperature $\beta$ to precision $\delta$ in time $\mathrm{poly}(n,\beta,\log(1/\delta))$.  This algorithm is the first to achieve both polynomial scaling in the inverse temperature and polylogarithmic scaling in the precision.
    When the Hamiltonian has an inverse-polynomial spectral gap, our quantum algorithm prepares its ground state to precision $\delta$ in time $\mathrm{poly}(n,\log(1/\delta))$.
    \item We further give a quantum algorithm that prepares a purification of the thermal state in time $\mathrm{poly}(n,\beta,\log(1/\delta))$. Using this purification, we estimate static and dynamical properties with error $\delta$ in time polynomial in $n$ and $\beta$ and linear in $1/\delta$ up to polylogarithmic factors, achieving the Heisenberg scaling. 
\end{itemize}  

Our algorithms build on quantum Gibbs samplers, which are quantum analogues of Markov chain Monte Carlo methods.
By using global jump operators instead of local ones, we avoid the $\exp(\beta)$ overhead commonly associated with quantum Gibbs samplers, making them efficient for impurity models even at low temperatures.

\end{abstract}

\tableofcontents

\section{Introduction}
Simulating strongly correlated materials at finite temperature is a central problem in computational chemistry and physics, underpinning our understanding of phenomena such as metal–insulator transitions and superconductivity. Quantum impurity models, which describe a large noninteracting bath coupled to a small but strongly interacting subsystem, play an important role in these simulations. Through embedding methods such as dynamical mean-field theory (DMFT)~\cite{georges1992hubbard,georges1996dynamical,RevModPhys.78.865} 
and density matrix embedding theory (DMET)~\cite{knizia2012density,wouters2016practical,sun2020finite}, one can simulate large interacting systems approximately by iteratively solving effective impurity models.
The impurity solver is therefore the key computational bottleneck in simulating many strongly correlated materials. These embedding methods have provided a qualitative description of the Mott transition~\cite{georges1992hubbard,georges1996dynamical} and advanced our understanding of superconductivity in the Fermi--Hubbard model~\cite{maier2005systematic,capone2006competition}.

Over the past  decades, a wide range of classical algorithms have been developed for quantum impurity models, including the numerical renormalization group (NRG)~\cite{wilson1975renormalization}, continuous-time quantum Monte Carlo (CT-QMC)~\cite{werner2006continuous,gull2011continuous}, and tensor-network-based methods~\cite{PhysRevLett.93.246403,grundner2024complex}. While these methods have been successful in computing many static properties, dynamical properties remain substantially harder to compute for classical algorithms~\cite{jiang2026polynomial,arunachalam2026quantum}. Dynamical properties are central to DMFT calculations and, roughly speaking, correspond to the values of observables in thermal states and ground states after real-time evolution. Quantum computers offer a promising route to overcoming this classical difficulty and efficiently estimating dynamical properties, since they can efficiently simulate real-time evolution~\cite{bauer2016hybrid,hogan2025quantum,ayral2025dynamical}. The central remaining task is then to design fast quantum algorithms for preparing the initial thermal or ground state.

In this work, we develop quantum algorithms for preparing thermal states and ground states for impurity models.
We focus on \textit{the low-temperature regime}, where impurity-model calculations are needed to study strongly correlated phenomena such as superconductivity~\cite{maier2005systematic,capone2006competition}, and on \textit{the high-precision regime}, where classical calculations can become computationally demanding. Previous quantum algorithms  offer favorable scaling in one of these regimes, but cannot achieve both simultaneously. After mapping the impurity model to a one-dimensional spin system~\cite{jiang2026polynomial,wilson1975renormalization}, algorithms based on quantum Gibbs sampling with local updates~\cite{jiang2026polynomial,wilson1975renormalization,bergamaschi2025fast} prepare thermal states to precision $\delta$ in time polynomial in the system size $n$ and $\log\delta^{-1}$, but exponential in the inverse temperature $\beta$, limiting their applicability at low temperatures. Algorithms based on Hilbert-space compression~\cite{jiang2026polynomial} have polynomial dependence on both $n$ and $\beta$, but also polynomial dependence on $\delta^{-1}$, making high-precision simulation costly. In contrast, our algorithm prepares thermal states in time \textit{polynomial} in $n$ and $\beta$ and \textit{polylogarithmic} in $\delta^{-1}$, allowing efficient simulation at both low temperatures and high precision:

\begin{theorem}[Thermal state preparation (informal)]\label{thm:intro_thermal_state}
For any impurity model on $n$ fermionic modes with fixed impurity size, there is a quantum algorithm that prepares its thermal state at inverse temperature $\beta$ to precision $\delta$ in time $\mathrm{poly}(n,\beta,\log\delta^{-1})$ using $\widetilde{O}(n)$ qubits. If the Hamiltonian has an inverse-polynomial spectral gap above its ground space, the same algorithm can prepare the ground space to precision $\delta$ in time $\mathrm{poly}(n,\log\delta^{-1})$.
\end{theorem}

The quantum algorithm in Theorem~\ref{thm:intro_thermal_state} belongs to the family of quantum Gibbs samplers~\cite{chen2025efficient,jiang2024quantum,ding2025efficient,ding2026simple}, which are quantum analogues of classical Markov chain Monte Carlo methods. We avoid the exponential $\beta$-dependence of existing quantum Gibbs samplers~\cite{jiang2026polynomial,wilson1975renormalization,bergamaschi2025fast} by replacing local jump operators with model-specific global jump operators. We bound the mixing time by reducing the spectral-gap analysis to that of a classical Markov chain. Section~\ref{sec:algo_tech_overview} gives an overview of the algorithm and its analysis. We remark that a more refined analysis improves the spectral-gap lower bound to $\Omega(1/n)$ when the bath single-particle energy is constant and the Lindbladian is normalized to have constant norm. We will include this improved analysis in the next version.

Thermal expectation values can be estimated by using
Theorem~\ref{thm:intro_thermal_state} to prepare $\cO(\delta^{-2})$
independent copies of the thermal state and then performing the
corresponding measurements.  One way to reduce this additional \(\cO(\delta^{-2})\) cost factor is to use single-trajectory approaches~\cite{jiang2026predicting,chen2026thermal} or quadratic-speedup techniques~\cite{leng2026accelerating}.
On the other hand, by further analyzing the structure of the impurity model, we show that this $\delta^{-2}$ scaling can be  improved to  \(\delta^{-1}\) up to polylogarithmic factors,  achieving the Heisenberg scaling:

\begin{theorem}[Estimating thermal properties with Heisenberg scaling (informal)]\label{thm:intro_thermal_property}
For any impurity model on $n$ fermionic modes with fixed impurity size, there is a quantum algorithm that prepares the purified thermal state at inverse temperature $\beta$ to precision $\delta$ in time $\mathrm{poly}(n,\beta,\log\delta^{-1})$. Using this preparation, a second quantum algorithm estimates static and dynamical properties to precision $\delta$ in time polynomial in $n$ and $\beta$ and linear in $\delta^{-1}$ up to polylogarithmic factors. Both quantum algorithms use $\widetilde{O}(n)$ qubits.
\end{theorem}

Compared to previous classical and quantum algorithms for estimating thermal properties of impurity models~\cite{jiang2026polynomial,arunachalam2026quantum,bravyi2017complexity}, Theorem~\ref{thm:intro_thermal_property} significantly improves the dependence on $\delta^{-1}$ to almost linear, while maintaining polynomial dependence on $\beta$ and $n$. Computing dynamical properties of impurity models has been shown to be classically hard~\cite{jiang2026polynomial,arunachalam2026quantum}. The favorable scaling of our algorithms in $n$, $\beta$, and $\delta$ may make them promising candidates for quantum advantage  when used as impurity solvers in DMFT~\cite{bauer2016hybrid,hogan2025quantum,ayral2025dynamical}.

In addition to the favorable runtime scaling, our quantum algorithms may provide a space advantage over classical algorithms. Memory requirements are a key bottleneck for many numerical methods used to simulate large impurity systems, including exact diagonalization, tensor-network methods~\cite{PhysRevLett.93.246403,grundner2024complex}, and configuration interaction methods~\cite{zgid2012truncated,mejuto2019dynamical}. Previous rigorous classical algorithms for static properties face a similar issue~\cite{jiang2026polynomial,arunachalam2026quantum,bravyi2017complexity}, as they require storing and diagonalizing matrices whose dimensions scale as large polynomials in the system size. In contrast, the quantum algorithms in Theorems~\ref{thm:intro_thermal_state} and~\ref{thm:intro_thermal_property} use only \(\widetilde{\cO}(n)\) qubits, and their classical processing involves only multiplying and diagonalizing $2n\times 2n$ matrices.

\subsection{Algorithm  and technical overview}\label{sec:algo_tech_overview}

We consider quantum impurity models consisting of a constant-size, arbitrarily interacting impurity coupled to a large free-fermion bath. 
 Let $\gamma_1,\ldots,\gamma_{2n}$ be the Majorana operators on $n$ fermionic modes. The Hamiltonian of the quantum impurity model takes the form
\begin{equation}
H=H_0+H_{\mathrm{imp}},\qquad
H_0=\alpha \I+\frac{i}{4}\sum_{j,k=1}^{2n}h_{jk}\gamma_j\gamma_k, \label{eq:impurity_model}
\end{equation}
where $\alpha\in\mathbb{R}$ is an energy shift and $h$ is a real antisymmetric matrix.
The impurity term $H_{\mathrm{imp}}$ is supported only on the first $m$ modes,
$\gamma_1,\ldots,\gamma_{2m}$, where $m=\cO(1)$, and is a sum of even-degree products of these operators. The term $H_0$ is quadratic in the Majorana operators and therefore describes noninteracting fermions. We assume $\|H\|$ and $\|h\|$ are bounded by $\mathrm{poly(n)}.$

The Jordan--Wigner transformation maps the Majorana operators to the Pauli strings $\gamma_{2j-1}=Z_1\cdots Z_{j-1}X_j$ and $\gamma_{2j}=Z_1\cdots Z_{j-1}Y_j$.

The thermal state of $H$ at inverse temperature $\beta$ is defined as 
$\rho_\beta:=e^{-\beta H}/\tr \left(e^{-\beta H}\right).$

\paragraph{Change of basis} After a canonical transformation, the impurity model is written with respect to annihilation operators $a_1,...,a_n$, where $a_j$ can be represented as $Z_1...Z_{j-1}\ket{0}\bra{1}$. 
To describe our algorithm, we first change the basis to the \textit{band-wise Krylov basis}, which has been used in classical simulations of impurity models~\cite{jiang2026polynomial,arunachalam2026quantum,wilson1975renormalization}\footnote{The band-wise Krylov basis in this manuscript differs slightly from that of~\cite{jiang2026polynomial} in how we partition the bath spectrum. We use the bands $[0,1/\beta)$ and $[2^\ell/\beta,2^{\ell+1}/\beta)$ for $\ell=0,1,\ldots$. The construction in \cite{jiang2026polynomial} uses the bands $[2^\ell\omega,2^{\ell+1}\omega)$ for $\ell=0,1,\ldots$ where $\omega$ is a lower bound of the bath single-particle energy.}. 
This new basis is specified by an $n\times n$ unitary matrix $V$ with column vectors $v_1,\ldots,v_n$, which define a new set of annihilation operators
$a(v_j):=\sum_k (v_j)_k \,a_k$ where $(v_j)_k$  is the $k$-th entry of $v_j$.

\begin{figure}[htbp]
\centering
\def\figurescale{0.8} 

\begin{tikzpicture}[
    scale=\figurescale,
    transform shape,
    font=\sffamily\large,
    box/.style={
        draw=cyan!25!black,
        line width=0.8pt,
        rounded corners=5pt,
        minimum width=1.2cm,
        minimum height=1.0cm,
        inner sep=0pt
    },
    connection/.style={draw=black, line width=0.8pt}
]
    \node[box, minimum width=2.1cm, minimum height=1.15cm]
        (impurity) at (0,0) {Impurity};

    \node[box] (upper1) at (2.85, 1.0) {};
    \node[box] (upper2) at (5.25, 1.0) {};
    \node[box] (upper3) at (7.65, 1.0) {};

    \node[box] (lower1) at (2.85,-1.0) {};
    \node[box] (lower2) at (5.25,-1.0) {};
    \node[box] (lower3) at (7.65,-1.0) {$a(v_j)$};

    \draw[connection] (impurity.east) -- (upper1.west);
    \draw[connection] (impurity.east) -- (lower1.west);

    \draw[connection] (upper1.east) -- (upper2.west);
    \draw[connection] (upper2.east) -- (upper3.west);
    \draw[connection] (upper3.east) -- ++(1.2,0);

    \draw[connection] (lower1.east) -- (lower2.west);
    \draw[connection] (lower2.east) -- (lower3.west);
    \draw[connection] (lower3.east) -- ++(1.2,0);

    \node at (2.85,0) {$\cdots$};
    \node at (5.25,0) {$\cdots$};
    \node at (7.65,0) {$\cdots$};

    \node[anchor=north] at ([yshift=-1.5mm]lower3.south) {depth $d_j$};
\end{tikzpicture}
\caption{One-dimensional structure of the band-wise Krylov basis.}\label{fig:bandwise_krylov}
\end{figure}
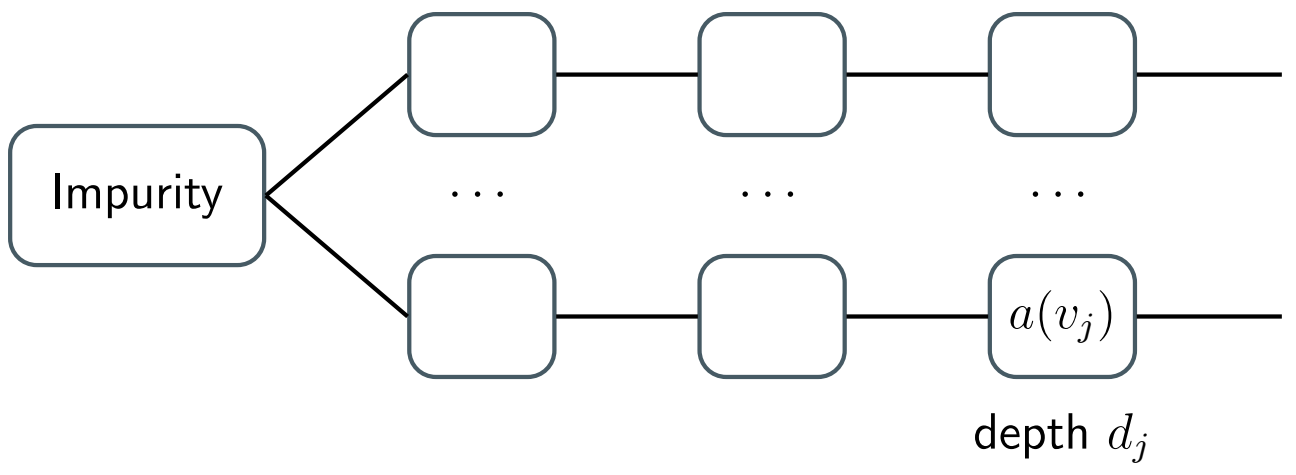

The key structural property is that, in this new basis, $H$ takes the form of a coarse-grained one-dimensional
chain, with the impurity term acting only on the first few sites.
Let $d_{j}$ be the depth of the mode $a(v_j)$ in this chain.
The band-wise modification is needed for technical reasons. The classical preprocessing to find this basis requires only diagonalizing $2n\times 2n$ matrices and block-tridiagonalizing $n\times n$ matrices.

The basis change corresponds to a unitary $\mathcal{U}(V)$ on the $2^n$-dimensional Hilbert space satisfying
$\mathcal{U}(V)a_j\mathcal{U}(V)^\dagger=a(v_j).$
Since ground-state and thermal-state properties are invariant under unitary transformations, we do not need to construct or implement $\mathcal{U}(V)$. Instead, for the algorithmic implementation, we apply the fermion-to-qubit mapping directly to the new annihilation operators, representing them as
$$a(v_j)=Z_1\cdots Z_{j-1}\lvert 0\rangle\langle 1\rvert_j.$$

\vspace{0.3em}

\paragraph{Our algorithm for thermal states}
Our algorithm belongs to the family of quantum Gibbs samplers, which are quantum analogues of classical Markov chain Monte Carlo methods. To ease the analysis, we work with the CKG Gibbs sampler~\cite{chen2023efficient,chen2025efficient} and design the model-specific global jump operators described below. We expect that the same jump operators can also be used in other quantum Gibbs samplers~\cite{ding2025efficient,jiang2024quantum,ding2026simple}.

\textit{Quantum Gibbs sampler.}
Let $\bJ$ be a set of $n$-qubit operators that is closed under taking adjoints ($\bJ=\bJ^\dagger$) and generates the full operator algebra. For example, a popular choice is to set $\bJ$ to be the set of all single-qubit Pauli operators. For a given choice of $\bJ$ (called jump operators), the CKG Gibbs sampler defines a Lindbladian whose unique stationary state is the thermal state $\rho_\beta$.  The corresponding Lindbladian dynamics can be simulated efficiently on a quantum computer; thus, one can prepare the thermal state to the desired accuracy by running the Lindbladian simulation for \textit{a sufficiently long time}. Note that we use the Metropolis transition rate instead of the Gaussian rate, because the latter can exponentially suppress the transitions needed to reach lower-energy states.

\vspace{0.3em}

\textit{Model-specific jump operators.}
The choice of jump operators $\bJ$ strongly affects the convergence rate and hence the runtime of the algorithm.  Most existing quantum Gibbs samplers use local jump operators~\cite{chen2023efficient,bergamaschi2025fast,ding2025efficient,tong2025fast}. The runtime of those algorithms is exponential in the inverse temperature $\beta$, since  crossing even a constant-height energy barrier occurs with probability $e^{-\Omega(\beta)}$. 
To circumvent this technical barrier for impurity models, we design sets of model-specific jump operators.

More specifically,
recall that $a(v_j)$ is the annihilation operator with respect to the band-wise Krylov basis, and $d_j$ is its depth in the one-dimensional chain. 
Define the Hermitian Majorana operators as $\gamma_{2j-1}=a(v_j)+a(v_j)^\dagger, \gamma_{2j}=-i(a(v_j)-a(v_j)^\dagger)$. We use a mixture of base and global jumps, with the
probabilities specifying their respective weights:

\begin{itemize}
\item \textit{Base jump:}
With probability $1/2$, we set the jump operator $J$ uniformly from $\{\gamma_1,\ldots,\gamma_{2n}\}$.
\item \textit{Global jump:} With probability $1/2$, for every $j=1,...,n$, independently  choose
\[
J_j\in\{I,\gamma_{2j-1},\gamma_{2j},-i\gamma_{2j-1}\gamma_{2j}\}
\]
with probabilities $1-3p_j/4,\,p_j/4,\,p_j/4,\,p_j/4$ respectively, where $p_j=2^{-(d_j+1)}$.  
Then  set the jump operator as $J=J_1\cdots J_n$ or
$J=(J_1\cdots J_n)^\dagger$ with equal probabilities.
\end{itemize}

\paragraph{The classical chain and its spectral gap analysis.} 
To ease the analysis, we also add  jump operators of the form $e^{-iuH}$ for real times $u$  to the CKG Gibbs sampler. These phase jumps allow us to reduce the spectral gap analysis of the quantum Gibbs sampler to that of a classical Markov chain (see Theorem \ref{thm:qtoc} for a formal statement). 
We remark that the claimed performance can also be achieved without phase jumps.

The state space of the classical chain is an eigenbasis of a rounded version of the impurity Hamiltonian, denoted as $\{\ket{x}\}_x$. For $x\ne y$, the transition probability from $\ket{x}$ to $\ket{y}$ is
\[
P(x,y):=\mathbb{E}_{J}\left(|\langle y|J|x\rangle|^2\right)
\min\{1,e^{-\beta(k_y-k_x)}\},
\]
where $J$ is sampled from the mixture of base and global jumps described above, and $k_x$ and $k_y$ are the rounded energies of $\ket{x}$ and $\ket{y}$, respectively.
By Cheeger’s inequality, one can lower bound the spectral gap of the classical chain by lower bounding its conductance, which measures the minimum (normalized) transition flow across all partitions (cuts) of the state space. Since we use a mixture of base and global jumps, the total transition flow is the average of the base-jump and global-jump transition flows.  

The key observation is that when the base-jump conductance is small, we can prove that the global-jump conductance is large. Thus, the two sets of jump operators play complementary roles.
More specifically, previous work showed that there is a  Gaussian state $\sigma$ that is close to the thermal state in the following sense: In a suitable fermionic  basis of the \textit{doubled Hilbert space}, the purification of the Gaussian state $\sigma$ is the vacuum, and the probability that a given mode is occupied  in the purified thermal state decays exponentially with its depth~\cite{arunachalam2026quantum}\footnote{For technical reasons, our reference Gaussian state $\sigma$ differs slightly from the one in \cite{arunachalam2026quantum}. For completeness, we give self-contained proofs of the properties of $\sigma$ that we use in the appendix.}. 
We prove a related occupation bound for the thermal states restricted
to either side of a cut: the thermal-mode occupations in their
purifications are controlled by the base-jump conductance, a term
that decays polynomially with depth, and a small rounding error.
Thus, when the base-jump conductance is small, the occupations of
deeper modes in both purifications are small. We then show that
these occupation bounds imply large global-jump conductance across
the cut. 

The formal statements are given in
Section~\ref{sec:spectral_gap}, with the statements  on the spectral gap and simulation cost  in Lemma~\ref{lem:uniform-classical-gap}, Corollary~\ref{cor:impurity-quantum-gap}, and Corollary~\ref{cor:impurity-gibbs-preparation}.

\paragraph{Thermal observable estimation with Heisenberg scaling}
Our quantum Gibbs sampling algorithm corresponds to a Lindbladian whose unique stationary state is the thermal state and whose KMS spectral gap on even operators is \(1/\mathrm{poly}(n,\beta)\).\footnote{Since the Lindbladian preserves the space of even operators, we choose an even initial density operator and restrict our analysis to the space of even operators.} We then map this Lindbladian to a Hermitian parent Hamiltonian on the doubled Hilbert space~\cite{chen2023efficient}. After adding a penalty to the odd-parity sector, the parent Hamiltonian has a \(1/\mathrm{poly}(n,\beta)\) spectral gap and the purified thermal state as its unique ground state. We further show that an efficiently preparable Gaussian state on the doubled Hilbert space has overlap at least $1/\mathrm{poly}(n,\beta)$ with the purified thermal state. Applying an approximate ground-state projector to this Gaussian state~\cite{lin2020near}, we can prepare the purified thermal state to precision $\delta$ in time polynomial in $n$, $\beta$, and $\log(1/\delta)$. We then use catalytic tomography~\cite{chen2025catalytic} to estimate thermal expectation values to precision $\delta$, with a cost polynomial in $n$ and $\beta$ and linear in $1/\delta$ up to polylogarithmic factors. This applies to both static and dynamical properties that can be expressed as expectations of operators with efficient block encodings.

The formal statements are given in
Section~\ref{sec:observable}.

\subsection{Related work}

Our analysis relies on bounding the mixing time of the quantum Gibbs sampler. Most existing mixing-time analyses exploit the local structure induced by local jump operators in the Gibbs sampler and apply at high temperatures~\cite{rouze2025efficient,rouze2026optimal,bakshi2026dobrushin}, in weakly interacting regimes~\cite{tong2025fast}, or to one-dimensional systems~\cite{bergamaschi2025fast}. These techniques do not directly apply to our setting, where we use global jump operators, allow strong interactions, and seek polynomial dependence on the inverse temperature. Global jump operators have  been used in special settings previously: in the classical Swendsen--Wang algorithm for the two-dimensional Ising model~\cite{swendsen1987nonuniversal}, in quantum samplers for code Hamiltonians~\cite{ding2026toric,hangleiter2025code}, and for certain commuting Hamiltonians through reductions to efficient classical sampling~\cite{hwang2026gibbs}.

Our analysis reduces the spectral-gap analysis of a modified CKG 
Gibbs sampler to that of a classical Markov chain. Previous work studied
comparisons between the spectral gap of a Davies generator and those
of the classical Markov chains induced by the same generator~\cite{temme2013lower,basso2025quantum}.
These results are not directly comparable to ours, since our reduction
modifies the Lindbladian  (hence the quantum algorithm) by adding
phase jumps to control quantum coherences. In particular, our results
do not contradict the counterexample in~\cite{temme2013lower}.

Our work focuses on rigorous quantum algorithms for preparing thermal and ground states. Several quantum impurity solvers without rigorous efficiency guarantees have also been proposed for use in DMFT and related embedding methods. Bauer et al.~\cite{bauer2016hybrid} proposed adiabatic ground-state preparation followed by phase estimation to prepare the ground state. Later approaches use variational circuits to prepare ground states~\cite{sakurai2022hybrid,mukherjee2023comparative} and excited states~\cite{sakurai2022hybrid}, or construct state-preparation circuits using classical tensor-network calculations or Gaussian subspace approximations~\cite{jamet2025anderson,hogan2025quantum}.

\subsection{AI methodology}
With assistance from Codex 5.5, we built a math agent using GPT-5.5. It found a candidate algorithm for thermally gapped impurity models that combined base jumps with Weyl cluster jumps $i^{x\cdot z}X^xZ^z$ in the occupation basis. The bits are independent, with $\Pr(x_i=1)=p_i$ and $\Pr(z_i=1)=1/2$, where $p_i$ is chosen from occupation-decay bounds. However, counterexamples showed that this algorithm fails for general impurity models.

Once we understood this algorithm, we identified occupation decay as the source of its performance. Properties of impurity models suggested that, for general models, analogous decay could occur in the doubled Hilbert space relative to a suitable reference state. This intuition guided further interactions with GPT-5.6 and GPT-6.

The second candidate replaced the Weyl cluster updates with a SWAP unitary that refreshes the system using a reference state. Its analysis relied on the fact that there is a Gaussian state dominating the thermal state. Although GPT-6 presented this as a quantum Gibbs sampling algorithm, we viewed its underlying mechanism as closer to rejection sampling than to Markov-chain convergence. This observation led us to develop a separate coherent rejection sampling algorithm based on quantum belief propagation, which involves simulating time-dependent Hamiltonians. This candidate is more complicated than the algorithm in Section \ref{sec:observable}, so we omit it.

To obtain a quantum Gibbs sampler, we used the occupation-decay intuition in the doubled Hilbert space to guide the design of the jump operators. After several unsatisfactory complicated jumps, we developed the final sampler, which combines base jumps with the global jumps presented in this work.
The authors independently checked all technical details and take full responsibility for the
correctness of the results.

\section{Quantum-to-classical reduction for spectral gap analysis}

In this section, we describe the quantum Gibbs sampler we use and reduce its spectral-gap analysis to that of a classical Markov chain. For convenience, we use $K$ to denote a general $n$-qubit Hamiltonian throughout this section; its specific form will be given when needed.
Define
\[
\rho_{\beta,K}:=\frac{e^{-\beta K}}{\Tr(e^{-\beta K})}.
\]
For the impurity Hamiltonian $H$, we write $\rho_\beta:=\rho_{\beta,H}$.

 Consider a distribution over jump operators $J\sim \bJ$
such that $J$ and $J^\dagger$ are chosen with the same probability, and
\begin{equation} \label{eq:boundedjump}
    \E_J \left[ J J^{\dagger} \right] \preceq \I ,\qquad \E_J \left[ J^{\dagger} J \right] \preceq \I .
\end{equation}
To ensure that the quantum and classical Gibbs samplers (defined later) have a unique stationary state, we also require that the jump operators $J$ with nonzero probability generate the full operator algebra. 

In this section, we treat $J\sim \bJ$ as an abstract distribution over jump operators. The specific choice of $J$ is given in the next section.

\subsection{The CKG quantum Gibbs sampler}

We use the Metropolis Gibbs sampler of Chen--Kastoryano--Gily\'en~\cite{chen2023efficient}.
Write the spectral decomposition of $K$ as $K=\sum_\he \he\Pi_\he$, where the sum runs over distinct eigenvalues. For each Bohr frequency $\nu$ (a difference of two eigenvalues), define
\begin{equation}
 J_\nu^K:=\sum_{\he-\he'=\nu}\Pi_\he J\Pi_{\he'},\qquad
\varphi_\beta(\omega):=\sqrt\beta(2\pi)^{-1/4}e^{-\beta^2\omega^2/4},\qquad
 \gamma_\beta(\omega):=e^{-\max\{\beta\omega+1/2,0\}}.
 \label{eq:filters}
\end{equation}
Here $J_\nu^K$ is called the Fourier component of $J$ with Bohr frequency $\nu$. The = $\varphi_\beta(\omega)$ is a Gaussian function used in the proof, and $ \gamma_\beta(\omega)$ is the Metropolis rate function.

Note that $[K,J_\nu^K]=\nu J_\nu^K$ and $\int\varphi_\beta(\omega)^2\dd\omega=1$. Write
\begin{equation}
 \widehat J_K(\omega):=\sum_\nu\varphi_\beta(\omega-\nu)J_\nu^K,
 \qquad
 \alpha_{\nu\nu'}:=\int_\R\gamma_\beta(\omega)\varphi_\beta(\omega-\nu)\varphi_\beta(\omega-\nu')\dd\omega.
 \label{eq:filtered-jumps}
\end{equation}
The Metropolis CKG generator in the Heisenberg-picture is written as
\begin{align}
 \cL_{\mathrm{CKG},K}(X)
 &=i[G_K,X]+\E_J\int_\R\gamma_\beta(\omega)
 \left(\widehat J_K(\omega)^\dagger X\widehat J_K(\omega)
 -\frac12\{\widehat J_K(\omega)^\dagger\widehat J_K(\omega),X\}\right)\dd\omega,
 \label{eq:ckg}\\
 G_K&:=\frac{i}{2}\E_J\sum_{\nu,\nu'}\alpha_{\nu\nu'}
 \tanh\!\left(\frac{\beta(\nu-\nu')}4\right)(J_{\nu'}^K)^\dagger J_\nu^K.
 \label{eq:coherent}
\end{align}

The CKG generator (in the Schrödinger picture) is known to have $\rho_{\beta,K}$ as its unique stationary state, and the corresponding Lindbladian dynamics can be efficiently simulated on quantum computers~\cite{chen2023efficient}.  

At fixed inverse temperature $\beta$, define the \emph{KMS inner product}
\begin{equation}
    \ip{X}{Y}{K}:=\Tr(X^\dagger\rho_{\beta,K}^{1/2}Y\rho_{\beta,K}^{1/2}),\qquad
    \wnorm{X}{K}^2:=\ip{X}{X}{K}.
    \label{eq:kms}
\end{equation}
The CKG generator $\cL_{\mathrm{CKG},K}$ satisfies the Kubo–Martin–Schwinger (KMS) detailed balance, 
\begin{align}
    \langle X,\cL_{\mathrm{CKG},K}(Y)\rangle_{K}=  \langle \cL_{\mathrm{CKG},K}(X),Y\rangle_{K}.
\end{align}

\paragraph{Spectral gaps and mixing time}

For Hermitian $X$, set
\[
\Var_K(X)=\inf_{a\in\mathbb R}\|X-aI\|_K^2
=\|X\|_K^2-|\Tr(\rho_{\beta,K}X)|^2.
\]
The minimizing scalar is $a=\Tr(\rho_{\beta,K}X)$.
For a generator $\cL$ satisfying KMS detailed balance, the \emph{Dirichlet form} is
$-\ip{X}{\cL(Y)}{K}$, and its spectral gap is
\begin{align}
    \gap(\cL)=\inf_{\substack{X=X^\dagger\\\Var_K(X)>0}}
\frac{-\langle X,\cL(X)\rangle_K} {\Var_K(X)}. \label{eq:def_gap}
\end{align}

We write $P=(-1)^{a_1^\dagger a_1 +....+a_n^\dagger a_n}$ as the fermionic parity operator; an operator $X$ is even if $[X,P]=0$.
When $[K,P]=0$ and $\cL$ preserves even operators, we define
$\gap_{\mathrm{even}}(\cL)$ by restricting the infimum above to even Hermitian observables. The corresponding mixing time is defined as
\begin{align}
    t_{mix}(\delta):=\min\{t: \|e^{t\cL^\dagger}(\rho)-\rho_{\beta,K}\|_1\leq \delta,\,\,\text{where}\,\, [\rho,P]=0\}. 
\end{align}

\subsection{The classical Metropolis chain}

For an $n$-qubit Hamiltonian $K$ and a distribution of jump operators $J\sim \bJ$, we define a classical Markov chain on its eigenbasis. More precisely, consider a spectral decomposition of $K$ as  $K=\kappa_x\ket{x}\bra{x}$. The state space of the classical chain is $\{\ket{x}\}_x$. Define
\begin{equation}
 \pi_x=\frac{e^{-\beta\kappa_x}}{\sum_z e^{-\beta\kappa_z}},\qquad
 q_{xy}=\E_J|\langle y|J|x\rangle|^2,\qquad
 P_K(x,y)=q_{xy}\min\{1,e^{-\beta(\kappa_y-\kappa_x)}\}\quad(x\ne y).
 \label{eq:classical}
\end{equation}
Set $P_K(x,x)=1-\sum_{y\ne x}P_K(x,y)$.
The normalization in~\eqref{eq:boundedjump} makes $P_K$ a transition matrix.
We refer to $P_K$ as the \emph{classical Metropolis chain}. Since $J$ and $J^\dagger$ are chosen with the same probability,
then $q_{xy}=q_{yx}$ and
\[
\pi_xP_K(x,y)=\min\{\pi_x,\pi_y\}q_{xy}=\pi_yP_K(y,x).
\]
Thus the chain is reversible with stationary distribution $\pi$.

For a reversible chain, the \emph{spectral gap} is the largest $\lambda_0$ for which
the following Poincar\'e inequality holds for every real-valued function $f$:
\begin{equation}
\frac12\sum_{x\ne y}\pi_xP_K(x,y)(f_x-f_y)^2
\geq\lambda_0\left(\sum_x\pi_xf_x^2-\left(\sum_x\pi_xf_x\right)^2\right).
\label{eq:classical-poincare}
\end{equation}

\subsection{Quantum-to-classical spectral-gap reduction via phase jumps}

We use the CKG quantum Gibbs sampler with additional phase jumps. The following theorem connects its spectral gap to the spectral gap of the classical Metropolis chain:

\begin{theorem} \label{thm:qtoc} 
    Let $0<\delta_0,\lambda_0\le1$ be two parameters.
    Assume $[H,P]=0$ and every jump operator $J$ has definite parity, $PJP=\pm J$.
    Suppose that, whenever a Hamiltonian $K$ satisfies
    \begin{equation}
        [K,H]=0,\qquad [K,P]=0,\qquad \beta \norm{K-H}\le\delta_0,
        \label{eq:robust}
    \end{equation}
    the classical Metropolis chain $P_K$ has spectral gap at least $\lambda_0$ in \emph{every} common orthonormal eigenbasis of $K$ and $P$.
    Let $\cL_{\mathrm{CKG},H}$ be the normalized Metropolis CKG generator defined in Eq.~\eqref{eq:ckg}.
    There are numerical constants $c,C>0$, independent of $H$, $\beta$, and $n$ such that the following holds. Set $\tau=\beta/\delta_0$ and
    \begin{equation}
        p_\tau(u)=\frac{\tau}{\pi(u^2+\tau^2)},\qquad
        R=\frac{C\tau}{\lambda_0}.
        \label{eq:dephasing-cutoff}
    \end{equation}
    Here $p_\tau$ is the density of the Cauchy distribution with scale $\tau$.
    The generator
    \begin{equation}
        \cL(X)=\cL_{\mathrm{CKG},H}(X)
        +\int_{-R}^R p_\tau(u)\bigl(e^{iuH}Xe^{-iuH}-X\bigr)\dd u
        \label{eq:total-physical}
    \end{equation}
    satisfies KMS detailed balance, preserves the thermal state and the even operator space, and satisfies
    \begin{equation}
        \boxed{\gap_{\mathrm{even}}(\cL)\ge c\lambda_0.}
        \label{eq:main}
    \end{equation}
\end{theorem}

Since $e^{-iuH}$ commutes with $H$, the additional term
in~\eqref{eq:total-physical} can be incorporated into the
CKG sampler by adding phase jumps $e^{-iuH}$, with a
constant rescaling of their weights.
After discretizing the integral, we can therefore apply
the CKG implementation theorem~\cite[Theorem~I.2]{chen2023efficient}
to efficiently simulate the dynamics generated by $\cL$
on a quantum computer.
The implementation cost is analyzed in
Corollary~\ref{cor:impurity-gibbs-preparation}.

Theorem~\ref{thm:qtoc} is proved by analyzing the Dirichlet form
$-\ip{X}{\cL(X)}{H}$ for $\cL$ defined in Eq.~(\ref{eq:total-physical}). It is known that the Dirichlet form of any Lindbladian satisfying KMS detailed balance can be written as a sum of nonnegative terms~\cite{jiang2026predicting,chen2025quantum,rouze2025efficient}. 
The corresponding non-negative term for the phase jumps penalizes the part of 
$X$ that does not commute with $H$.
When the total Dirichlet form is small, we can approximate $X$
by an observable that commutes with a rounded version of $H$.
Controlling the rounding and approximation errors then allows us
to reduce the spectral-gap analysis of the quantum Gibbs sampler
to that of a classical Markov chain. The detailed proof is given
in Appendix~\ref{app:gap_reduction}.

\section{Preliminaries on impurity models}

In this section, we introduce the band-wise Krylov basis and explain the details of the base and global jump operators. We also introduce the thermal modes on the doubled Hilbert space, which play an important role in proofs in later sections.

\subsection{Fermionic notations and the single-particle space}

A system of $n$ fermionic modes is described by $n$  operators $a_1,\ldots,a_n$ satisfying the \textit{canonical anticommutation relations} (CAR)
\begin{align}
    \{a_j,a_k^\dagger\}
    \coloneqq a_j a_k^\dagger+a_k^\dagger a_j
    =\delta_{jk}\I,\qquad
    \{a_j,a_k\} =0,
    \qquad j,k\in[n].
    \label{eq:annihilation-car}
\end{align}
Using the Jordan--Wigner transformation, these operators act on an $n$-qubit Hilbert space and are represented by
	$a_j = Z_1  \cdots  Z_{j-1}\ket{0}\bra{1}_j$,
where $Z_k$ represents the Pauli-$Z$ operator on qubit $k$. The operators
$a_j^\dagger$ and $a_j$ are called the \textit{creation and annihilation operators}.

It is convenient to use an $n$-dimensional vector to represent a linear
combination of annihilation operators. For any $\mu\in\C^n$, denote
\begin{align}
    a(\mu)\coloneqq\sum_{j=1}^n \mu_j a_j.
\end{align}

The space $\mathbb{C}^n$ that $\mu$ lives in is the \emph{single-particle space} of the fermions, and any orthonormal basis $\{f_k\}_{k=1}^n$ of $\mathbb{C}^n$ is called a \emph{single-particle basis}. One can check that the corresponding operators $\{a(f_k)\}_{k=1}^n$ satisfy the CAR and thus form a new set of annihilation operators.

\subsection{The band-wise Krylov basis}

We use the band-wise Krylov basis of
\cite{jiang2026polynomial}, with a different
partition of the bath spectrum adapted to the inverse temperature
$\beta$, which was used in \cite{arunachalam2026quantum}.
Below we summarize the construction; more explanations and
intuitions can be found in
\cite[Sections~2.1--3.3]{jiang2026polynomial}.

Recall that we have defined the quantum impurity model in Section \ref{sec:algo_tech_overview} Eq.~(\ref{eq:impurity_model}). After applying a canonical transformation to the bath~\cite[Sections~2.1]{jiang2026polynomial} ,  one can efficiently construct annihilation operators $a_1,\ldots,a_n$ and write $H$ in the hybridization form:

\begin{equation}
    H
    = K_0+ 
    \sum_{j,k=m+1}^n \!\! (E_0)_{jk}\,\, a_k^\dagger a_j
    +
    \sum_{j=1}^{2m}
    \left(
        a^\dagger(\mu_j)O_j
        +
        O_j^\dagger a(\mu_j)
    \right),\qquad 0\preceq E_0\preceq \es \I .\label{eq:impurity_model_hyb}
\end{equation}
Here $a_1,\ldots,a_m$ are called the impurity modes, and the remaining modes are called the bath
modes.   
The term $K_0$ only acts on impurity modes. \(E_0\) is the coefficient matrix of the quadratic terms that act only on the bath modes. 
$\mu_j$ is a vector with zero entries in the first $m$ coordinates, so $a(\mu_j):=\sum_k (\mu_j)_k \,a_k$ acts only on the bath modes.   $\{O_j\}_{j=1}^{2m}$ are the Majorana fermion operators on the impurity modes. The terms $a^\dagger(\mu_j)O_j$
and their Hermitian conjugates describe the hybridization term.
  The vectors $\{\mu_j\}_j$ need not
be orthogonal.  The scalar parameter $\es$ upper bounds the bath single-particle energies and is independent of $H_{imp}$.

\paragraph{Partition the bath spectrum.}
Let $\mathbb{C}^n$ be the single-particle space.
Let $e_j$ be the unit vector in the $j$-th coordinate and $W_{imp}:=\mathrm{span}\left\{e_1,...,e_m\right\}$ denote the impurity single-particle space. 
View $E_0$ as a matrix on the bath single-particle space $W_{bath}:=W_{imp}^\perp$.
Partition the spectrum of $E_0$ into the low band $[0,1/\beta)$ and the
bands $[2^k/\beta,2^{k+1}/\beta)$ for integers $k\geq0$.
Retain the nonempty bands and label them consecutively by $\ell=0,\ldots,L$;
if the bath is empty, use one empty band and set $L=0$. Thus $L+1$ is the
maximum number of nonempty bands, 
\begin{equation}
1\leq L+1\leq 2\log(2+\beta \es).
\label{eq:bath-band-count}
\end{equation}

For each band $\ell$, let $W_\ell\subseteq W_{\mathrm{bath}}$ be the span of the eigenvectors of $E_0$ with eigenvalues in that band, and let $\Pi_\ell$ be the orthogonal projector onto $W_\ell$.
Write $E_{(\ell)}$ for the restriction of $E_0$ to $W_\ell$ and set $W_{\ell,0}=\operatorname{span}\{\Pi_\ell\mu_1,\ldots,\Pi_\ell\mu_{2m}\}$.
Define 
\[
\mathcal K_{\ell,d}=\spanop\{E_{(\ell)}^qW_{\ell,0}:0\leq q\leq d\},
\quad \mathcal K_{\ell,-1}=\{0\},\quad
W_{\ell,d}=\mathcal K_{\ell,d}\ominus\mathcal K_{\ell,d-1}.
\]

Write 
$$s_{\ell,0}=\dim W_{\ell,0}\leq2m.$$
Since $\mathcal K_{\ell,d}$ is generated by $s_{\ell,0}$ Krylov sequences, we have
$\dim W_{\ell,d}\leq s_{\ell,0}$. Without loss of generality, we assume that
$W_\ell=\bigoplus_{d\geq0}W_{\ell,d}$, since any modes in the
orthogonal complement form a decoupled free-fermion sector
that can be treated separately; see
\cite[Section~3.1, footnote~3]{jiang2026polynomial}.
Then we have 
\begin{align}
\mathbb{C}^n=W_{\mathrm{imp}}\oplus  \bigoplus_{\ell\geq0, d\geq 0} W_{\ell,d}
\end{align}

\paragraph{Band-wise Krylov basis.}
Choose an orthonormal basis for each $W_{\ell,d}$. Together with
$e_1,\ldots,e_m$, these vectors form the band-wise Krylov basis.
For convenience, relabel the vectors in this basis as
$v_1,\ldots,v_n$. Accordingly, $a(v_1),...,a(v_n)$ define a new set of annihilation operators.

With some abuse of notation, we use the notation $\gamma_{2j-1},\gamma_{2j}$ for their associated Majorana operators, 
\begin{align}
    \gamma_{2j-1}= a(v_j)+a(v_j)^\dagger, \qquad  \gamma_{2j}= -ia(v_j)+i a(v_j)^\dagger
\end{align}
We call these $\{a(v_k)\}_k$ and $\{\gamma_{k}\}_k$ \textit{physical operators}, in contrast to the \textit{thermal  operators} defined later, which act on the doubled Hilbert space.

The most important structural property is that, when expressed in terms of
$a(v_1),\ldots,a(v_n)$, the impurity Hamiltonian $H$ has the one-dimensional
structure shown in Figure~\ref{fig:bandwise_krylov}.
We say that the mode $a(v_j)$ has depth $0$ if $v_j\in W_{\mathrm{imp}}$
or $v_j\in W_{\ell,0}$ for some $\ell$. Otherwise, we have $v_j\in W_{\ell,d}$
with $d\geq 1$, we say that the mode belongs to band $\ell$ and has depth $d$.

We refer to the set of depth-$0$ modes as the enlarged impurity, which we also call the boundary.  The size of the boundary is bounded by

\begin{equation}
\ms:=m+\sum_{\ell=0}^L s_{\ell,0}\leq m+2m(L+1)\leq3m(L+1).
\label{eq:enlarged-impurity-size}
\end{equation}

To distinguish the impurity modes and bath modes, for any $a(v)$ where $v$ is in the enlarged impurity single-particle space $W_{\mathrm{imp}}\oplus \bigoplus_{\ell} W_{\ell, 0}$, we write $a(v)$ as $b(v)$.

\paragraph{The commutator identity.}
Let $\Pi_{\ell,0}$ be the orthogonal projector in $\mathbb C^n$
onto $W_{\ell,0}$, and extend $E_0$ by zero on
$W_{\mathrm{imp}}$. For $v\in W_\ell\ominus W_{\ell,0}$,
define
\[
E_\ell v=(I-\Pi_{\ell,0})E_0v,
\qquad
T_\ell v=\Pi_{\ell,0}E_0v.
\]
Thus $E_\ell$ acts on the residual space
$W_\ell\ominus W_{\ell,0}$, while $T_\ell$ maps this space
into $W_{\ell,0}$. 

Since the original hybridization terms involve only
boundary modes, the canonical anticommutation relations give
\begin{equation}
[H,a(v)]=-a(E_0v)
       =-a(E_\ell v)-b(T_\ell v),
\qquad v\in W_\ell\ominus W_{\ell,0}.
\label{eq:boundary}
\end{equation}

For each band $\ell$, set $\omega_\ell=1/\beta$ if it is
the low band $[0,1/\beta)$, and set $\omega_\ell=2^k/\beta$
if it is the dyadic band $[2^k/\beta,2^{k+1}/\beta)$.  Thus $\omega_\ell$ is the energy scale
of band $\ell$. Block tridiagonality of the Krylov basis~\cite[Section 3.1]{jiang2026polynomial}  gives
\begin{equation}
\|T_\ell\|\leq 2\omega_\ell,\qquad
r_\ell:=\operatorname{rank}T_\ell\leq s_{\ell,0}\leq2m,\qquad
T_\ell E_\ell^qv=0\quad(0\leq q<d-1).
\label{eq:krylov}
\end{equation}
Indeed, the quadratic matrix connects only consecutive Krylov layers, and
only the first residual layer connects to the boundary.

\subsection{The mixed base and global jumps and our algorithm}
\label{sec:physical-proposal}

Recall that $\{\gamma_j\}_{j=1}^{2n}$ are the physical Majorana operators. For every $j$, set 
$$p_j=2^{-(d_j+1)}.$$

The mixed base and global jump form a distribution over jump operators $J$, defined as follows:

\begin{itemize}
	\item\textit{Base jumps:} With probability $1/2$, choose $j$ uniformly from
$\{1,\ldots,2n\}$ and set $J=\gamma_j$ to be the $j$-th physical Majorana. 
	\item\textit{Global jumps:}  With probability $1/2$, for every $j=1,...,n$, independently  choose
\[
J_j\in\{I,\gamma_{2j-1},\gamma_{2j},-i\gamma_{2j-1}\gamma_{2j}\}
\]
with probabilities $1-3p_j/4,\,p_j/4,\,p_j/4,\,p_j/4$.  
Then  set the physical global word as $J=J_1\cdots J_n$ or
$J=(J_1\cdots J_n)^\dagger$ with equal probabilities.
\end{itemize}

Our algorithm uses the CKG quantum Gibbs sampler with additional phase jumps, as in Theorem~\ref{thm:qtoc} and Eq.~\eqref{eq:total-physical}; its parameters are fixed later in Corollary~\ref{cor:impurity-quantum-gap}. By Theorem~\ref{thm:qtoc}, bounding the quantum spectral gap reduces to bounding the spectral gap of the classical Metropolis chain. The following sections establish this bound. Section~\ref{sec:thermal_operator} introduces thermal operators on the doubled Hilbert space, which play a central role in the proof. Section~\ref{sec:spectral_gap} proves the spectral-gap bound for the classical chain, and hence for the quantum sampler.

\section{Thermal operators}\label{sec:thermal_operator}

In this section, we introduce the thermal operators that act on the doubled Hilbert space.

\subsection{The doubled Hilbert space and thermal operators}

\paragraph{The doubled Hilbert space}
The Hilbert--Schmidt inner product is  $\langle X,Y\rangle=\Tr(X^\dagger Y)$.
Fix an orthonormal basis $\{|x\rangle\}_x$. Vectorization identifies
an operator with a vector in two copies of the Hilbert space:
\[
\oket{O}=\sum_{x,y}O_{xy}|x\rangle\otimes|y\rangle.
\]
It satisfies
\[
\obraket{X}{Y}=\Tr(X^\dagger Y),\qquad
\|\oket{X}\|^2=\|X\|_2^2,\qquad
\oket{AOB}=(A\otimes B^{\mathsf T})\oket{O},
\]
where the transpose is taken in the chosen basis.

For a density operator $\rho$, its canonical purification is
the normalized vector $\oket{\sqrt{\rho}}$ in the doubled
Hilbert space. Tracing out the second copy gives $\rho$.

\paragraph{Thermal annihilation operators}
Let $E_{\mathrm{ref}}$ be zero on the boundary  and equal to
$E_\ell$ on each residual band, i.e. $E_{\mathrm{ref}}=\oplus_\ell E_\ell$. Write
\[
u(t)=(1+e^{-\beta t})^{-1/2},\qquad
w(t)=(1+e^{\beta t})^{-1/2},\qquad
u=u(E_{\mathrm{ref}}),\quad w=w(E_{\mathrm{ref}}),
\]
and define the Gaussian reference state
\[
\sigma=\frac{e^{-\beta \sum_{jk} (E_{\mathrm{ref}})_{jk} \,a_k^\dagger a_j}}
{\Tr e^{-\beta \sum_{jk} (E_{\mathrm{ref}})_{jk}\, a_k^\dagger a_j}}.
\]
Recall that  $P =(-1)^{\sum_i a_i^\dagger a_i}$ is the fermion parity operator on the original physical Hilbert space.

\textit{Annihilation operators in the doubled Hilbert space.}
With respect to a Gaussian reference state, one can define a set of annihilation operators for the doubled Hilbert space~ \cite{arunachalam2026quantum}.
More specifically, for any $1\leq r\leq n$,  
 for any  operator $X$, 
 define
\begin{align}
\widehat c_r(X)
=a(uv_r)X- PXP \,a(wv_r),\qquad
\widehat c_{n+r}(X)
=a^\dagger(wv_r)X-PXP \,a^\dagger(uv_r).
\label{eq:thermal_anni}
\end{align}

We use $\widehat c_r^\dagger$ and $\widehat c_r^\dagger$  to denote the adjoint maps w.r.t. the Hilbert-Schmidt inner product, where 
\begin{align}
\widehat c_r^\dagger(X)
=a^\dagger(uv_r) X + PX P a^\dagger(wv_r) ,\qquad
\widehat c_{n+r}^\dagger (X)
=a(wv_r)X + PXP a(uv_r).
\end{align}
Here we use the fact that $a^\dagger(wv_r)$ and $a(uv_r)$ have odd fermion parity and thus anticommute with the fermion parity operator $P$.

\begin{lemma}[Thermal annihilation operators]
The maps $\{\widehat c_j\}_{j=1}^{2n}$ form a set of 
fermionic annihilation operators on the doubled Hilbert space. In
particular, they satisfy the CAR
\[
\{\widehat c_j,\widehat c_k\}=0,
\qquad
\{\widehat c_j,\widehat c_k^\dagger\}
=\delta_{jk}I.
\]
Moreover, the canonical purification of the Gaussian reference state
is their common vacuum:
\[
\widehat c_j(\sqrt{\sigma})=0,\qquad 1\leq j\leq 2n.
\]
\end{lemma}
\begin{proof}
Since $a(z)$ has odd fermion parity, we have
\[
Pa(z)P=-a(z),\qquad Pa^\dagger(z)P=-a^\dagger(z).
\]

Using the  CAR of the physical operators, we obtain, for $1\leq r,s\leq n$,
\begin{align*}
&\{\widehat c_r,\widehat c_s^\dagger\}(X)
=
\bigl(
\langle uv_r,uv_s\rangle
+\langle wv_r,wv_s\rangle
\bigr)X=
\langle v_r,(u^2+w^2)v_s\rangle X
=\delta_{rs}X,\\
&\{\widehat c_r,\widehat c_{n+s}\}(X)
=
\bigl(
\langle uv_r,wv_s\rangle
-\langle wv_r,uv_s\rangle
\bigr)X
=0,
\end{align*}
where the first equation uses $u^2 + w^2=1$.  
 The second equation uses $uw=wu$. Similarly, one can directly check the other 
 anticommutator relations for the CAR.

Since the Gaussian reference state has even parity,
$P\sqrt{\sigma}P=\sqrt{\sigma}$. The thermal identities
\[
a(z)\sqrt{\sigma}
=\sqrt{\sigma}\,a(e^{-\beta E_{\mathrm{ref}}/2}z),
\qquad
a^\dagger(z)\sqrt{\sigma}
=\sqrt{\sigma}\,a^\dagger(e^{\beta E_{\mathrm{ref}}/2}z),
\]
together with
$e^{-\beta E_{\mathrm{ref}}/2}u=w$ and
$e^{\beta E_{\mathrm{ref}}/2}w=u$, give
\[
\widehat c_r(\sqrt{\sigma})
=
\widehat c_{n+r}(\sqrt{\sigma})
=0.
\]
Thus $\sqrt{\sigma}$ is the normalized common vacuum.
\end{proof}

The operators $\{\widehat c_j\}_{j=1}^{2n}$, with respect to the Gaussian
reference state $\sigma$, define a fermionic Fock space on
the doubled Hilbert space, with
$\lvert\sqrt{\sigma}\rangle\!\rangle$ as its vacuum. For any purification of a quantum state $\oket{\sqrt{\rho}}$, 
the occupation measures the departure from the reference state and is denoted by 
\begin{align}
    \| \widehat{c}_j (\sqrt{\rho}) \|_2^2 = tr\left( \widehat{c}^\dagger_j \widehat{c}_j  \oket{\sqrt{\rho}}\obra{\sqrt{\rho}} \right).
\end{align}

\subsection{Thermal-to-physical operator comparison}
\label{sec:thermal2physical}

As mentioned in the introduction, our spectral analysis of the
classical chain relies on studying related quantities expressed
in terms of thermal operators on the doubled Hilbert space.
In this section, we relate quantities in the physical Hilbert
space to their counterparts in the doubled Hilbert space.
To keep focus, here we only list the statements, and the proofs are deferred to Appendix \ref{app:thermal_physics_compare}.

For notational convenience, set
\[
p_{n+r}=p_r,\qquad 1\leq r\leq n.
\]
Let $D$ be the diagonal operator on the physical single-particle space
defined by
\[
Dv_r=p_r^{-1/2}v_r.
\]

We first bound the weighted sum of squared overlaps involving
thermal creation and annihilation operators in terms of those
involving left and right multiplication by physical Majorana
operators.

\begin{restatable}[Weighted one-step comparison]{lemma}{weightedcomparison}
\label{lem:weighted-one-step-comparison}
There exists a constant $C$ such that for any operators
$A$ and $B$ of definite fermion parity, we have
\begin{align*}
\sum_{j=1}^{2n}p_j
\left(
|\langle A,\widehat c_j(B)\rangle|^2
+
|\langle A,\widehat c_j^\dagger(B)\rangle|^2
\right)
\leq
C\sum_{r=1}^np_r
\sum_{\nu\in\{2r-1,2r\}}
\left(
|\Tr(A^\dagger\gamma_\nu B)|^2
+
|\Tr(A^\dagger B\gamma_\nu)|^2
\right).
\end{align*}
\end{restatable}

\paragraph{Thermal global unitary in the thermal coordinate}
Since our algorithm uses global jumps, we accordingly define the thermal global unitary in the doubled Hilbert space. 
Let $C_0$ be the constant in Lemma \ref{lem:weighted-one-step-comparison}.
Fix
\[
\kappa=\min\{1,(8C_0)^{-1}\},
\qquad
q_j=\kappa^2p_j^2.
\]
Define the thermal Pauli operators 
\[
\widehat \gamma_{2j-1}=\widehat c_j+\widehat c_j^\dagger,
\qquad
\widehat \gamma_{2j}=i(\widehat c_j^\dagger-\widehat c_j).
\]

For each $j$, independently set $\widehat W_j$ to $I$ with probability $1-3q_j/4$, and to each of $\widehat \gamma_{2j-1},\widehat \gamma_{2j},-i \widehat \gamma_{2j-1}\widehat \gamma_{2j}$ with probability $q_j/4$.
Set the thermal global word as
\[
\widehat W=\widehat W_1\cdots\widehat W_{2n}.
\]

\begin{restatable}
[Physical-to-thermal random overlap comparison]{lemma}{weightedcomparisonglobal}
\label{lem:physical-to-thermal-word-comparison}
Let $A,B$ be operators of  even fermion parity with $\|A\|_2=\|B\|_2=1$, and let $J,K$
be independent global words sampled according to the second
part of the proposal in Section~\ref{sec:physical-proposal}.
There exists a numerical constant $C>0$ such that
\begin{equation}
\mathbb E_{J,K}
\left|
\Tr\!\left(A^\dagger\,JB\,K\right)
\right|^2
\geq
\exp\!\left(-C\sum_{r=1}^np_r\right)
\mathbb E_{\widehat W}
\left|
\left\langle
A,\widehat W(B)
\right\rangle
\right|^2.
\label{eq:physical-to-thermal-word-comparison}
\end{equation}
\end{restatable}

\section{Spectral gap of the quantum Gibbs sampler}
\label{sec:spectral_gap}

In this section, we prove a lower bound on the spectral gap of
our algorithm. In Sections~\ref{sec:thermal_occupation_base_conductance}--\ref{sec:gap_classical},
we first analyze the spectral gap of the classical chain appearing
in Theorem~\ref{thm:qtoc}. Then, in Section~\ref{sec:gap_quantum},
we apply Theorem~\ref{thm:qtoc} to obtain a lower bound on the
spectral gap of the quantum chain.

To define the corresponding classical chain, recall the conditions in Theorem \ref{thm:qtoc} and
consider any Hermitian operator \(K\) satisfying Eq.~(\ref{eq:robust})
\[
[K,H]=0,\qquad [K,P]=0,\qquad
\beta\|K-H\|\leq\delta_0.
\]
Fix any common orthonormal eigenbasis $\{\ket{x}\}_x$ of $K$ and $P$. Each projector $\ket{x}\bra{x}$ is an operator of even fermion parity. 
 In this basis,
\begin{align}
   \overline H:= K:=\sum_x\overline E_x|x\rangle\langle x|,\qquad \deltaH:=\|H-\overline{H}\| \leq
\frac{\delta_0}{\beta},\qquad [P,\overline{H}]=0.\label{eq:overline_H_property}
\end{align}

\paragraph{Conductance.}
For a classical Metropolis chain with transition matrix $P_K$ and a nonempty proper subset $T$
of its state space, write
\[
\pi(T)=\sum_{x\in T}\pi_x,\qquad
Q(T)=\sum_{\substack{x\in T\\y\in T^c}}\pi_xP_K(x,y),\qquad
\Phi(T)=\frac{Q(T)}{\min\{\pi(T),\pi(T^c)\}}.
\]
The conductance of the chain is $\Phi=\min_T\Phi(T)$.
By Cheeger's inequality, the spectral gap of $P_K$ satisfies
\[
\frac{\Phi^2}{2}\leq\gap(P_K)\leq2\Phi.
\]

We write $\Phi_{\mathrm{base}}(T)$ and $\Phi_{\mathrm{global}}(T)$
for the conductances of the cut $T$ in the Metropolis chains
using only base jumps and only global jumps, respectively.

\subsection{Thermal occupation of conditioned state and base-jump chain conductance}\label{sec:thermal_occupation_base_conductance}

Recall that all thermal modes are unoccupied in
$\oket{\sqrt{\sigma}}$. Intuitively, in
$\oket{\sqrt{\rho_\beta}}$, thermal modes far from the impurity
have small occupation probabilities. Using slightly different
thermal annihilation operators, Ref.~\cite{arunachalam2026quantum}
showed that these occupation probabilities decay exponentially
with depth.

Our goal is to relate the thermal occupation to the conductance of the base-jump chain.
 Instead of considering $\sqrt{\rho_\beta}$, we consider its conditioned counterpart. 
 For any subset  $T\subseteq \{x\}_x$,
 the  \textit{restricted thermal state} is defined as 
\begin{equation}
\rho_T=\frac1{\pi(T)}\sum_{x\in T}\pi_x|x\rangle\langle x|, \qquad \pi_x = e^{-\beta \overline{E}_x}/\Tr(e^{-\beta\overline H}), \qquad \pi(T)= \sum_{x\in T} \pi_x.
\label{eq:conditioned-states}
\end{equation}

We prove that the thermal-mode occupations of these restricted states are controlled by
the conductance of the base-jump chain across the cut, together with a term
that decays with depth and a small rounding error.

\begin{lemma}
    Since
$\sqrt{\rho_T}$ is Hermitian, we have
$\bigl\|\widehat c_{n+r}(\sqrt{\rho_T})\bigr\|_2
=
\bigl\|\widehat c_r(\sqrt{\rho_T})\bigr\|_2.$
\end{lemma}
\begin{proof}
     For every Hermitian operator $Y$, the definitions and
$Pa^\dagger(z)P=-a^\dagger(z)$ give
\begin{align}
\widehat c_{n+r}(Y)
=
P \left(\widehat c_r(Y)\right)^\dagger P.
\end{align}
Thus, we prove $\bigl\|\widehat c_{n+r}(\sqrt{\rho_T})\bigr\|_2
=
\bigl\|\widehat c_r(\sqrt{\rho_T})\bigr\|_2.$ 
\end{proof}
\begin{lemma}[Conditioned thermal occupation decay]\label{lem:condition_decay} Consider any cut $T$. There exists a constant $C$ such that 
for every band $\ell$ and every depth $d\geq 1$,
\begin{equation}
\sum_{v_r\in W_{\ell,d}}
\bigl\|\widehat c_r(\sqrt{\rho_T})\bigr\|_2^2
\leq
\frac{C r_\ell}{(d+1)^6}
+
Cn\Phi_{\mathrm{base}}(T)
+
Cn\beta^2\deltaH^2.
\label{eq:conditioned-occupation-decay}
\end{equation}
In addition, for every integer $d_*\ge0$,
\begin{equation}
\sum_{r:d_r>d_*}
\|\widehat c_r(\sqrt{\rho_T})\|_2^2
\le
\frac{C\ms}{5(d_*+1)^5}
+Cn\Phi_{\mathrm{base}}(T)+Cn\beta^2\deltaH^2.
\label{eq:conditioned-occupation-tail}
\end{equation}
\end{lemma}

To ease the proof of Lemma \ref{eq:conditioned-occupation-decay}, we define some notations. Write
\begin{align}
    \Delta=\overline H-H,
\qquad
\tau=\overline E_x-\overline E_y,
\qquad
\sqrt{\rho_T}
=
\sum_x\alpha_x|x\rangle\langle x|,\qquad \alpha_x=
\mathbf 1_{\{x\in T\}}\sqrt{ \frac{\pi_x}{\pi(T)}}.
\end{align}

By the definition of $\widehat c_r$, the central quantity is the matrix element $\langle y|a(f(E_\ell)v_r)|x\rangle$. If the boundary coupling matrix $T_\ell=0$ and the rounding error $\Delta=0$, then the commutator identity \eqref{eq:boundary} will imply that $\langle y|a(f(E_\ell)v_r)|x\rangle=f(\tau)\langle y|a(v_r)|x\rangle$. We call $f(\tau)\langle y|a(v_r)|x\rangle$ the \textit{ideal free term}. When $T_\ell$ and $\Delta$ are nonzero, there will be a \textit{boundary correction term} and a \textit{rounding correction term}.

\begin{lemma}

\begin{align}
\langle y|a(f(E_\ell)v_r)|x\rangle
=
\underbrace{
f(\tau)\langle y|a(v_r)|x\rangle
}_{\text{ideal free term}}
-
\underbrace{
\langle y|b(T_\ell h_\tau^f(E_\ell)v_r)|x\rangle
}_{\text{boundary correction}}+
\underbrace{
\langle y|[\Delta,a(h_\tau^f(E_\ell)v_r)]|x\rangle
}_{\text{rounding correction}}.\label{eq:afE}
\end{align}

\end{lemma} 
\begin{proof} 
 By the definition of $\tau$, the following identity holds for $\tau$:
\begin{align}
	-\tau
\langle y|a(h_\tau^f(E_\ell)v_r)|x\rangle= \langle y|[\overline H,a(h_\tau^f(E_\ell)v_r)]|x\rangle. \label{eq:overlineH}
\end{align}
	To obtain \eqref{eq:afE}, first apply
$$f(E_\ell)=	f(\tau)I +(-\tau+E_\ell)h_\tau^f(E_\ell)$$  by the definition of $h_t^f$. Then apply Eq.~(\ref{eq:overlineH}), substitute  $\overline H=H+\Delta$ and apply the commutator identity  \eqref{eq:boundary}.
\end{proof}

We now prove the lemma.
\begin{proof}[Proof of Lemma \ref{lem:condition_decay}]
Throughout the proof, $C$ denotes a sufficiently large universal
constant whose value can be made explicit by tracing the estimates.

Fix $v_r\in W_{\ell,d}$. Applying \eqref{eq:afE} with
$f=u$ and $f=w$ in the definition of
$\widehat c_r(\sqrt{\rho_T})$ gives

\begin{equation}
\begin{aligned}
\langle y|\widehat c_r(\sqrt{\rho_T})|x\rangle
={}&
\underbrace{
\bigl(\alpha_xu(\tau)-\alpha_yw(\tau)\bigr)
\langle y|a(v_r)|x\rangle
}_{\mathrm{(I_{xy})}\ \text{ideal free term}}\\
&+
\underbrace{
\Bigl(
-\alpha_x
\langle y|b(T_\ell h_\tau^u(E_\ell)v_r)|x\rangle
+\alpha_y
\langle y|b(T_\ell h_\tau^w(E_\ell)v_r)|x\rangle
\Bigr)
}_{\mathrm{(II_{xy})}\ \text{boundary correction}}\\
&+
\underbrace{
\Bigl(
\alpha_x
\langle y|[\Delta,a(h_\tau^u(E_\ell)v_r)]|x\rangle
-\alpha_y
\langle y|[\Delta,a(h_\tau^w(E_\ell)v_r)]|x\rangle
\Bigr)
}_{\mathrm{(III_{xy})}\ \text{rounding correction}}.
\end{aligned}
\label{eq:conditioned-decomposition}
\end{equation}

We refer to the three lines on the right-hand side as the ideal free term, the boundary correction, and the rounding
correction, respectively.

\underline{Estimates for $\mathrm{(I_{xy})}$.}
For the ideal free term $\mathrm{(I_{xy})}$, the detailed balance condition gives
\[
\sqrt{\pi_x}\,u(\tau)
=
\sqrt{\pi_y}\,w(\tau)
=
\sqrt{\frac{\pi_x\pi_y}{\pi_x+\pi_y}}.
\]
Thus this term vanishes when $x$ and $y$ are both in $T$ or both
outside $T$. 
When they lie on opposite sides of the cut,  the squared
absolute value of the ideal free term is at most
\[ |\mathrm{(I_{xy})}|^2\leq 
\frac{\min\{\pi_x,\pi_y\}}{\pi(T)}
|\langle y|a(v_r)|x\rangle|^2.
\]
where we use $\frac{\pi_x\pi_y}{\pi_x+\pi_y}\leq \min \{\pi_x,\pi_y\}$.
Moreover, since $|z+z'|^2\leq 2(|z|^2+|z'|^2)$ for all scalar $z,z'$, we have
$$\sum_{r:d_r\geq 1}
|\langle y|a(v_r)|x\rangle|^2
\leq
\frac12\sum_{\nu=1}^{2n}
|\langle y|\gamma_\nu|x\rangle|^2.$$
Therefore, by the definition of the base-jump transition flow,  the total squared
contribution of the ideal free term satisfies
\begin{align}
\sum_{r:d_r\geq 1}\sum_{x,y}
|\mathrm{(I_{xy})}|^2
\leq
\frac{Cn}{\pi(T)}Q_{\mathrm{base}}(T)
\leq
Cn\Phi_{\mathrm{base}}(T).
\end{align}

\underline{Estimates for $\mathrm{(II_{xy})}$.}
For the boundary correction $(\mathrm{II_{xy}})$, let
$e_1,\ldots,e_{r_\ell}$ be an orthonormal basis of
$\operatorname{ran}T_\ell$. Corollary
\ref{cor:divided-difference-source-decay} gives, uniformly in
$\tau$ and $f\in\{u,w\}$, 
\[
\sum_{v_r\in W_{\ell,d}}
\left|
\langle y|b(T_\ell h_\tau^f(E_\ell)v_r)|x\rangle
\right|^2
\leq
\frac{C}{(d+1)^6}
\sum_{q=1}^{r_\ell}
|\langle y|b(e_q)|x\rangle|^2.
\]
Furthermore, note that
$\sum_{x,y}\alpha_x^2
|\langle y|b(e_q)|x\rangle|^2
=
\Tr\!\left(\rho_Tb(e_q)^\dagger b(e_q)\right)
\leq1$, and similarly for the inequality when changing $a_x$ to $a_y$.
Using $|z_1+z_2|^2\leq2(|z_1|^2+|z_2|^2)$, the total squared
contribution of the boundary correction is therefore at most
\[
  \sum_{v_r\in W_{\ell,d}} \sum_{x,y} |\mathrm{(II_{xy})}|^2  \leq  \frac{Cr_\ell}{(d+1)^6}.
\]

\underline{Estimates for $\mathrm{(III_{xy})}$.}
Recall that $E=\bigoplus_\ell E_\ell$ on the residual-bath single-particle space.
For $v_r\in W_{\ell,d}$, we have
$h_\tau^f(E)v_r=h_\tau^f(E_\ell)v_r$.
For the rounding correction $(\mathrm{III_{xy}})$, the mean-value theorem gives, uniformly
in $\tau$ and $f\in\{u,w\}$,
\[
\|h_\tau^f(E)\|
\leq
\sup_{s\in\mathbb R}|f'(s)|
\leq
C\beta.
\]
Since $a(z)$ is linear in $z$, expanding
$h_\tau^f(E)v_r$ in the orthonormal basis
$v_1,\ldots,v_n$ and using
$\|h_\tau^f(E)\|\leq C\beta$ gives
\[
\sum_{r:d_r\geq 1}
\left|
\langle y|[\Delta,a(h_\tau^f(E)v_r)]|x\rangle
\right|^2
\leq
C^2\beta^2
\sum_{j=1}^n
|\langle y|[\Delta,a(v_j)]|x\rangle|^2.
\]

For each $j$,
\[
\|[\Delta,a(v_j)]\|
\leq
2\|\Delta\|
\leq
2\deltaH.
\]

Consequently,
\[
\sum_{x,y}\alpha_x^2
|\langle y|[\Delta,a(v_j)]|x\rangle|^2
\leq4\deltaH^2,
\qquad
\sum_{x,y}\alpha_y^2
|\langle y|[\Delta,a(v_j)]|x\rangle|^2
\leq4\deltaH^2.
\]
It follows that the total squared contribution of the rounding
correction is 
\[
\sum_{r:d_r\geq 1 }\sum_{xy} |(\mathrm{III}_{xy})|^2\leq 
C' n\beta^2\deltaH^2.
\]
for some other constant $C'$.

\underline{Combining the estimates.} Combining these three estimates with
$|z_1+z_2+z_3|^2
\leq
3\bigl(|z_1|^2+|z_2|^2+|z_3|^2\bigr)$,
and summing over $x,y$ and $v_r\in W_{\ell,d}$ proves
\eqref{eq:conditioned-occupation-decay}.

Moreover, \[
\begin{aligned}
\sum_{r:d_r>d_*}
\|\widehat c_r(\sqrt{\rho_T})\|_2^2
&\le
3\sum_{r:d_r\ge1}\sum_{x,y}
\left(
|\mathrm{I}_{xy}|^2+|\mathrm{III}_{xy}|^2
\right)
+3\sum_{r:d_r>d_*}\sum_{x,y}
|\mathrm{II}_{xy}|^2\\
&\le
Cn\Phi_{\mathrm{base}}(T)+Cn\beta^2\deltaH^2
+C\sum_{\ell=0}^Lr_\ell
\sum_{d>d_*}(d+1)^{-6}\\
&\le
Cn\Phi_{\mathrm{base}}(T)+Cn\beta^2\deltaH^2
+\frac{C\ms}{5(d_*+1)^5}.
\end{aligned}
\]
Here we used $\sum_\ell r_\ell\le\ms$ and
$\sum_{d>d_*}(d+1)^{-6}
\le\int_{d_*+1}^{\infty}t^{-6}\,dt$.
This proves \eqref{eq:conditioned-occupation-tail}.

\end{proof}

\subsection{Connecting global-jump conductance to thermal occupation}

We first relate thermal occupation to overlaps under thermal
global unitaries. We then transfer these bounds to physical
global jumps and use them to lower-bound the global-jump
conductance.

\begin{restatable}[Thermal-word overlap from thermal occupations]{lemma}{thermalwordoverlap}
\label{lem:thermal-word-overlap}
Let $A$ and $B$ be  operators satisfying $\|A\|_2 =\|B\|_2=1$.  Let $\widehat{W}$ be the random thermal global unitary defined in Section \ref{sec:thermal2physical}. 
There exists a numerical constant $C>0$ such that
\begin{equation}
\begin{aligned}
\mathbb E_{\widehat W}
\left|
\langle A,\widehat W(B) \rangle
\right|^2
\geq
\exp\!\Bigg[
-C\sum_{j=1}^{2n}
\bigg(
p_j^2
+\log(1/p_j)
\left(
\|\widehat c_j(A)\|_2
+\|\widehat c_j(B)\|_2
\right)
\bigg)
\Bigg].
\end{aligned}
\label{eq:thermal-word-overlap}
\end{equation}
\end{restatable}

The proof is given in Appendix \ref{app:proof-thermal-word-overlap}. Here
we illustrate the intuition behind the proof in a special case: suppose $A$ and $B$ correspond to thermal occupation-basis vectors in the doubled Hilbert space. That is, denote the respective sets of occupied modes by $S,T\subseteq [2n]$ and write $\oket{A},\oket{B}$ as $\oket{S},\oket{T}$. Let \(\triangle\) denote the symmetric difference of two sets:
\[
S\triangle T=(S\setminus T)\cup(T\setminus S).
\]

Then one can check that
\[
\mathbb E_{\widehat W}
|\langle\!\langle S|\widehat W|T\rangle\!\rangle|^2
=
\prod_{j\in S\triangle T}\frac{q_j}{2}
\prod_{j\notin S\triangle T}\left(1-\frac{q_j}{2}\right).
\]
Indeed, at a mode where their occupations differ, the word must choose \(\widehat \gamma_{2j-1}\) or \(\widehat \gamma_{2j}\), whose total probability is \(q_j/2\). Where their occupations agree, it must choose \(I\) or \(-i\widehat \gamma_{2j-1}\widehat \gamma_{2j}\).

Taking the negative logarithm and using \(-\log(1-x)\leq2x\) for \(0\leq x\leq1/2\), we find that there exists a constant $C$ such that
\[
\begin{aligned}
-\log\mathbb E_{\widehat W}
|\langle\!\langle S|\widehat W|T\rangle\!\rangle|^2
&=
-\sum_{j\notin S\triangle T}\log(1-q_j/2)+\sum_{j\in S\triangle T}\log(2/q_j)\\
&\leq
C\sum_jp_j^2 +  C\sum_{j\in S\triangle T}\log(1/p_j)\\
\end{aligned}
\]
Thus, we prove Lemma \ref{lem:thermal-word-overlap} in this special case since $\|\widehat{c}_j(A)\|_2 = \|\widehat{c}_j \oket{S}\|$ equals $1$ if $j\in S$ and $0$ otherwise.

We next show that small thermal occupations imply a large
average overlap under physical global jumps.

\begin{lemma}[Physical-word overlap from thermal occupations]
\label{lem:two-sided-global-overlap}
Let $A$ and $B$ be  operators of even parity satisfying $\|A\|_2 =\|B\|_2=1$.   Let $J$ and $K$ be independent global words
sampled according to the second part of the proposal in
Section~\ref{sec:physical-proposal}. Then there exists a constant $C$
such that
\begin{equation}
\mathbb E_{J,K}
\left|\Tr(A^\dagger JBK)\right|^2
\geq
\exp\!\left[
-C\sum_{j=1}^{2n}\log(1/p_j)
\left(
\sqrt{p_j}
+\|\widehat c_j(A)\|_2
+\|\widehat c_j(B)\|_2
\right)
\right].
\label{eq:two-sided-global-overlap}
\end{equation}
\end{lemma}

\begin{proof}
Combining Lemmas~\ref{lem:physical-to-thermal-word-comparison}
and~\ref{lem:thermal-word-overlap}, there exists a constant $C$ such that
\[
\begin{aligned}
\mathbb E_{J,K}
\left|\Tr(A^\dagger JBK)\right|^2
\geq
\exp\!\Bigg[
&-C\sum_{r=1}^np_r
-C\sum_{j=1}^{2n}p_j^2 -C\sum_{j=1}^{2n}\log(1/p_j)
\left(
\|\widehat c_j(A)\|_2
+\|\widehat c_j(B)\|_2
\right)
\Bigg].
\end{aligned}
\]
Since $p_{n+r}=p_r$ and $0<p_j\leq1/2$, we have 
$\sum_{r=1}^np_r+\sum_{j=1}^{2n}p_j^2
\leq
2\sum_{j=1}^{2n}\sqrt{p_j}\log(1/p_j).$
Substituting this estimate into the preceding inequality and replacing $C$ with $2C$ proves
\eqref{eq:two-sided-global-overlap}.
\end{proof}

Finally, we connect physical global overlap to the global-jump conductance.

\begin{lemma}[Physical overlap controls cut flow]
\label{lem:physical-overlap-cut-flow}
Let $J$ and $K$ be independent global words
sampled according to the second part of the proposal in
Section~\ref{sec:physical-proposal}. Then
\[
\Phi_{\mathrm{global}}(T)
\ge\frac12\,
\mathbb E_{J,K}
\left|
\Tr\!\left(\sqrt{\rho_T}J\sqrt{\rho_{T^c}}K\right)
\right|^2.
\]
\end{lemma}

\begin{proof}
For fixed unitaries $J,K$, expansion in the energy eigenbasis gives
\[
\sqrt{\pi(T)\pi(T^c)}
\Tr\!\left(\sqrt{\rho_T}J\sqrt{\rho_{T^c}}K\right)
=
\sum_{\substack{x\in T, y\in T^c}}
\sqrt{\pi_x\pi_y}\,
\langle x|J|y\rangle\langle y|K|x\rangle.
\]
Using the Cauchy--Schwarz inequality, we have
\[
\begin{aligned}
\pi(T)\pi(T^c)
\left|\Tr\!\left(\sqrt{\rho_T}J\sqrt{\rho_{T^c}}K\right)\right|^2
&\le
\left(
\sum_{\substack{x\in T\\y\in T^c}}
\min\{\pi_x,\pi_y\}|\langle x|J|y\rangle|^2
\right)
\left(
\sum_{\substack{x\in T\\y\in T^c}}
\max\{\pi_x,\pi_y\}|\langle y|K|x\rangle|^2
\right)\\
&\quad\le
\sum_{\substack{x\in T\\y\in T^c}}
\min\{\pi_x,\pi_y\}|\langle x|J|y\rangle|^2.
\end{aligned}
\]
where the second factor in the first inequality is at most
$\pi(T)+\pi(T^c)=1$ since $K$ is a unitary. Averaging over $J,K$ gives
\begin{align}
    E_{J,K} \left|\Tr\!\left(\sqrt{\rho_T}J\sqrt{\rho_{T^c}}K\right)\right|^2 \leq \frac{1}{\pi(T)\pi(T^c)} Q_{\mathrm{global}}(T) \leq 2  \Phi_{\mathrm{global}}(T)
\end{align}
since $\max\{\pi(T),\pi(T^c)\}\ge1/2$.
\end{proof}

\subsection{Small base-jump conductance implies large global-jump conductance}

Recall that the size of the enlarged impurity is
\[
\ms=m+\sum_{\ell=0}^L s_{\ell,0}
\le3m(L+1)
\le9m\log(2+\beta \es),
\]
where the last inequality follows from the dyadic band count
$L+1\le3\log(2+\beta \es)$.

The key lemma in this section states that, for every
fixed cut $T$, small base-jump conductance across $T$ implies large global-jump conductance across the same
cut.

\begin{lemma}[Base--global cut complementarity]
\label{lem:base-global-cut-complementarity}
There exists a numerical constant $C>0$ such that, 
if $\overline H$ has even fermion parity and
\[
\beta\deltaH\le n^{-1/2}e^{-8C\ms},
\]
then, for every nonempty proper cut $T$,
\[
\Phi_{\mathrm{base}}(T)<\frac1n e^{-8C\ms}
\quad\Longrightarrow\quad
\Phi_{\mathrm{global}}(T)\ge\frac12e^{-C\ms}.
\]
\end{lemma}

\begin{proof}
Let \(C_1\) and \(C_2\) be the constants in
Lemmas~\ref{lem:condition_decay}
and~\ref{lem:two-sided-global-overlap}, respectively. Set
\[
C=\max\left\{
2,\ C_1,\ 1+(21+16\sqrt{C_1})C_2
\right\}.
\]

Fix a nonempty proper cut \(T\) satisfying the stated bound on
\(\Phi_{\mathrm{base}}(T)\). Write
\[
X=\sqrt{\rho_T},
\qquad
Y=\sqrt{\rho_{T^c}},
\qquad
d_*=\lceil e^{C\ms}\rceil,
\]
where $d_*$ is a depth cutoff. Define
\[
\Pi_*=
\prod_{r:d_r>d_*}
(I-\widehat c_r^\dagger\widehat c_r)
(I-\widehat c_{n+r}^\dagger\widehat c_{n+r}),
\qquad
\eta=\|(I-\Pi_*)X\|_2^2+\|(I-\Pi_*)Y\|_2^2.
\]

\paragraph{Step 1: Control the cutoff error.}
We first estimate the cut-off error:
\begin{equation}
\eta\leq\frac1{16}e^{-2C\ms},
\qquad
4\sqrt{\eta}\leq e^{-C\ms}.
\label{eq:cutoff-error-target}
\end{equation}
To prove this,
recall that
$\|\widehat c_{n+r}(X)\|_2=\|\widehat c_r(X)\|_2,$
and similarly for \(Y\). Since
\[
I-\Pi_*
\leq
\sum_{r:d_r>d_*}
\left(
\widehat c_r^\dagger\widehat c_r+
\widehat c_{n+r}^\dagger\widehat c_{n+r}
\right),
\]
applying the tail estimate
\eqref{eq:conditioned-occupation-tail} in
Lemma~\ref{lem:condition_decay} to \(T\) and \(T^c\), and using
\(\Phi_{\mathrm{base}}(T^c)=\Phi_{\mathrm{base}}(T)\), gives
\[
\begin{aligned}
\eta
\leq
4C_1\left[
\frac{\ms}{5(d_*+1)^5}
+n\Phi_{\mathrm{base}}(T)
+n\beta^2\deltaH^2
\right]\leq
4C_1\left[
\frac{\ms}{5}e^{-5C\ms}
+e^{-8C\ms}
+e^{-16C\ms}
\right].
\end{aligned}
\]
Since \(\ms\geq1\), \(C\geq2\), and \(C_1\leq C\), the last
inequality implies
\[
16e^{2C\ms}\eta
\leq
64C_1\left[
\frac{\ms}{5}e^{-3C\ms}
+e^{-6C\ms}
+e^{-14C\ms}
\right]
<1.
\]
This proves \eqref{eq:cutoff-error-target}. In particular,
\(\eta<1/4\).

\paragraph{Step 2: Bound the thermal occupations after the cutoff.}
Define
\[
X_*=\frac{\Pi_*X}{\|\Pi_*X\|_2},
\qquad
Y_*=\frac{\Pi_*Y}{\|\Pi_*Y\|_2}.
\]
The goal of this step is to prove
\begin{equation}
\begin{aligned}
\sum_{j=1}^{2n}\log(1/p_j)
\left(
\|\widehat c_j(X_*)\|_2+
\|\widehat c_j(Y_*)\|_2
\right)
\leq
(4+16\sqrt{C_1})\ms.
\end{aligned}
\label{eq:cutoff-occupation-target}
\end{equation}

The operators \(X_*,Y_*\) are of even fermion parity and are normalized with respect to $\|\cdot\|_2$. By definition of $\Pi_*$, their thermal
occupations vanish for modes of depth greater than \(d_*\).
Moreover, since
\begin{align}
\|\Pi_*X\|_2^2,\|\Pi_*Y\|_2^2\geq1-\eta\geq\frac34,\label{eq:norm}
\end{align}

we have for \(d_r\leq d_*\),
\begin{equation}
\|\widehat c_r(X_*)\|_2
\leq2\|\widehat c_r(X)\|_2,
\qquad
\|\widehat c_r(Y_*)\|_2
\leq2\|\widehat c_r(Y)\|_2.
\label{eq:scaleeta}
\end{equation}

The hypotheses on \(\Phi_{\mathrm{base}}(T)\) and
\(\beta\deltaH\), together with \(d_*+1\leq3e^{C\ms}\), give
\begin{equation}
\sqrt n\,(d_*+1)^2
\left(
\sqrt{\Phi_{\mathrm{base}}(T)}
+\beta\deltaH
\right)
\leq
9\left(e^{-2C\ms}+e^{-6C\ms}\right)
<1.
\label{eq:small-base-rounding-term}
\end{equation}

Since the number of depth-zero modes is at most \(\ms\),
\eqref{eq:scaleeta} gives
\begin{align}
\sum_{j=1}^{2n}\log(1/p_j)
\left(
\|\widehat c_j(X_*)\|_2+
\|\widehat c_j(Y_*)\|_2
\right)\leq
4\ms+
\sum_{j:1\leq d_j\leq d_*}
2\log(1/p_j)
\left(
\|\widehat c_j(X)\|_2+
\|\widehat c_j(Y)\|_2
\right).
\label{eq:7733}
\end{align}

Recall that \(\log(1/p_j)=d_j+1\). 
We have
\[
\begin{aligned}
\sum_{j:1\leq d_j\leq d_*}
\log(1/p_j)
\left(
\|\widehat c_j(X)\|_2+
\|\widehat c_j(Y)\|_2
\right)=2\sum_{\ell=0}^L\sum_{d=1}^{d_*}(d+1)
\sum_{v_r\in W_{\ell,d}}
\left(
\|\widehat c_r(X)\|_2+
\|\widehat c_r(Y)\|_2
\right).
\end{aligned}
\]
Consider the part for $X$ first. Using the Cauchy--Schwarz inequality within \(W_{\ell,d}\), the bound
\(\dim W_{\ell,d}\leq s_{\ell,0}\), 
Lemma~\ref{lem:condition_decay}, and then $\sqrt{a+b+c}\leq \sqrt{a}+\sqrt{b}+\sqrt{c}$ for non-negative $a,b,c$, we have 
\[
\begin{aligned}
\sum_{v_r\in W_{\ell,d}}
\|\widehat c_r(X)\|_2
\leq
\sqrt{C_1s_{\ell,0}}
\left[
\frac{\sqrt{r_\ell}}{(d+1)^3}
+\sqrt{n\Phi_{\mathrm{base}}(T)}
+\sqrt n\,\beta\deltaH
\right].
\end{aligned}
\]
The same inequality holds when replacing $X=\sqrt{\rho_T}$ with $Y=\sqrt{\rho_{T^c}}$ since
\(\Phi_{\mathrm{base}}(T^c)=\Phi_{\mathrm{base}}(T)\).

Consequently,
\[
\begin{aligned}
&\sum_{j:1\leq d_j\leq d_*}
\log(1/p_j)
\left(
\|\widehat c_j(X)\|_2+
\|\widehat c_j(Y)\|_2
\right)\\
&\quad\leq
4\sqrt{C_1}
\sum_{\ell=0}^L\sqrt{s_{\ell,0}}
\sum_{d=1}^{d_*}
\left[
\frac{\sqrt{r_\ell}}{(d+1)^2}
+
(d+1)\sqrt n
\left(
\sqrt{\Phi_{\mathrm{base}}(T)}
+\beta\deltaH
\right)
\right].
\end{aligned}
\]
Using
\[
r_\ell\leq s_{\ell,0},
\qquad
\sum_{\ell=0}^Ls_{\ell,0}\leq\ms,
\qquad
\sum_{\ell=0}^L\sqrt{s_{\ell,0}}\leq\ms,
\]
together with
\[
\sum_{d=1}^{\infty}\frac1{(d+1)^2}\leq2,
\qquad
\sum_{d=1}^{d_*}(d+1)\leq(d_*+1)^2,
\]
we obtain
\[
\begin{aligned}
\sum_{j:1\leq d_j\leq d_*}
\log(1/p_j)
\left(
\|\widehat c_j(X)\|_2+
\|\widehat c_j(Y)\|_2
\right)
&\leq
4\sqrt{C_1}\,\ms
\left[
1+
\sqrt n\,(d_*+1)^2
\left(
\sqrt{\Phi_{\mathrm{base}}(T)}
+\beta\deltaH
\right)
\right]\\
&\quad\leq 8\sqrt{C_1}\,\ms,
\end{aligned}
\]
where the last inequality follows from
\eqref{eq:small-base-rounding-term}. Combining this with
\eqref{eq:7733} proves
\eqref{eq:cutoff-occupation-target}.

\paragraph{Step 3: Lower-bound the global overlap after the cutoff.}
Recall that  \(p_{n+r}=p_r\).
At depth $0$, there are $2(m+\sum_{\ell=0}^L s_{\ell,0})=2m$ thermal modes. At
 each depth $d\geq 1$, the number of thermal modes that have depth $d$ is at most  $$2\sum_{\ell=0}^L\dim W_{\ell,d}
\leq 2\sum_{\ell=0}^Ls_{\ell,0}\leq2\ms.$$ 
Thus
\[
\begin{aligned}
\sum_{j=1}^{2n}\sqrt{p_j}\log(1/p_j)
<17\ms.
\end{aligned}
\]

Combining this with \eqref{eq:cutoff-occupation-target} and
applying Lemma~\ref{lem:two-sided-global-overlap} gives
\begin{align}
\begin{aligned}
\mathbb E_{J,K}
\left|
\Tr(X_*^\dagger JY_*K)
\right|^2
&\geq
e^{-(21+16\sqrt{C_1})C_2\ms}\geq2e^{-C\ms}. \label{eq:overlap_cutoff}
\end{aligned}
\end{align}

\paragraph{Step 4: Remove the cutoff.}
Fix unitaries $J,K$. Let 
\[
a=\Tr(X^\dagger JYK),\qquad
b=\Tr(X_*^\dagger JY_*K).
\]
Since \(J,K\) are unitary and \(X,Y,X_*,Y_*\) have unit Hilbert–Schmidt norm,
$|a|,|b|\leq1.$ Therefore,
\[
\begin{aligned}
\bigl||a|^2-|b|^2\bigr|
\leq (|a|+|b|)|a-b|
\leq
2\bigl(\|X-X_*\|_2+\|Y-Y_*\|_2\bigr) \leq 4 \sqrt{\eta}
\end{aligned}
\]

Thus 
\[
\left|
|\Tr(X^\dagger JYK)|^2
-
|\Tr(X_*^\dagger JY_*K)|^2
\right|
\leq
4\sqrt{\eta}
\leq e^{-C\ms},
\]
where the last inequality follows from
\eqref{eq:cutoff-error-target}. Combining this with
\eqref{eq:overlap_cutoff}, averaging over
\(J,K\), and using Lemma~\ref{lem:physical-overlap-cut-flow}, we conclude 
\[
\Phi_{\mathrm{global}}(T)
\geq
\frac12
\mathbb E_{J,K}
\left|
\Tr(XJYK)
\right|^2
\geq
\frac12e^{-C\ms}.
\]
\end{proof}

\subsection{Spectral gap of  the classical chain}\label{sec:gap_classical}

\begin{lemma}[Uniform classical gap]
\label{lem:uniform-classical-gap}
There exists a numerical constant $C\ge1$ such that if
\[
\deltaH=\|\overline H-H\|
\le
\frac{1}{\beta\sqrt n}e^{-C\ms} 
\]
then in every common orthonormal eigenbasis of $\overline{H}$ and $P$, the classical chain with respect to $\overline{H}$ using the equal mixture
of the base and global jumps has spectral gap
\[
\lambda_{\mathrm{cl}}(\overline H)
\ge
\frac{1}{Cn^2}e^{-C\ms} \geq 
\frac{1}{Cn^2} \left(2+\beta \es\right)^{-9Cm}
\]
\end{lemma}
\begin{proof}
Let \(C_0\) be the constant in
Lemma~\ref{lem:base-global-cut-complementarity}, and choose
\(C\ge\max\{16C_0,8\}\). The assumed bound on \(\deltaH\)
implies \(\beta\deltaH\le n^{-1/2}e^{-8C_0\ms}\), so that
Lemma~\ref{lem:base-global-cut-complementarity} applies.

Fix a nonempty proper cut \(T\). If
\(\Phi_{\mathrm{base}}(T)\ge n^{-1}e^{-8C_0\ms}\), then the
equal-mixture chain has conductance $\Phi$ at least
\(\frac12\Phi_{\mathrm{base}}(T)\ge
\frac1{2n}e^{-8C_0\ms}\) across this cut. Otherwise,
Lemma~\ref{lem:base-global-cut-complementarity} gives
\(\Phi_{\mathrm{global}}(T)\ge\frac12e^{-C_0\ms}\), and hence
the equal-mixture chain has conductance at least
$$\frac12\Phi_{\mathrm{global}}(T)\ge
\frac14e^{-C_0\ms}\ge\frac1{2n}e^{-8C_0\ms}.$$

Thus the conductance of the equal-mixture chain is at least
\(\frac1{2n}e^{-8C_0\ms}\). Cheeger's inequality then proves the lemma:
$$
\frac{\Phi^2}{2}
\le
\lambda_{\mathrm{cl}}(\overline{H})
\le
2\Phi.
$$
\end{proof}

\subsection{Spectral gap of  the quantum chain}\label{sec:gap_quantum}

\begin{corollary}[Quantum gap and mixing time]
\label{cor:impurity-quantum-gap}
Let $C\ge1$ be the constant in
Lemma~\ref{lem:uniform-classical-gap}, and set
\[
\lambda_0=\frac{e^{-C\ms}}{Cn^2},
\qquad
\delta_0=\frac{e^{-C\ms}}{\sqrt n},
\qquad
\tau=\frac{\beta}{\delta_0}.
\]
Consider the quantum Gibbs sampler $\cL$ in
\eqref{eq:total-physical}, with these parameters and the cutoff
$R$ specified in Theorem~\ref{thm:qtoc},
Eq.~\eqref{eq:dephasing-cutoff}.
There exist universal constants $c,C'>0$ such that its KMS
spectral gap on even operators satisfies
\begin{equation}
\lambda_{\mathrm{even}}
:=\gap_{\mathrm{even}}(\cL)
\ge \frac{c}{n^2}e^{-C\ms}
\ge \frac{c}{n^2(2+\beta\es)^{C'm}}.
\label{eq:lambda_even}
\end{equation}
For every $\delta\in(0,1)$, the mixing time to trace distance
at most $\delta$ from $\rho_\beta$, starting from any even
density operator, satisfies
\[
t_{\mathrm{mix}}(\delta)
\le
C'n^2(2+\beta\es)^{C'm}
\bigl[n+\beta\|H\|+\log(1/\delta)\bigr].
\]
\end{corollary}

\begin{proof}
The chosen parameters satisfy $0<\lambda_0,\delta_0\le1$.
For any Hamiltonian $K$ satisfying \eqref{eq:robust},
\[
\|K-H\|
\le \frac{\delta_0}{\beta}
=\frac{e^{-C\ms}}{\beta\sqrt n}.
\]
Applying Lemma~\ref{lem:uniform-classical-gap} with
$\overline H=K$ therefore gives a classical spectral gap
of at least $\lambda_0$ in every common orthonormal eigenbasis
of $K$ and $P$.
Since $H$ is even and every jump operator has definite parity,
Theorem~\ref{thm:qtoc} gives
\[
\lambda_{\mathrm{even}}
\ge \frac{c}{n^2}e^{-C\ms}
\]
for a universal constant $c>0$.
The bounds \eqref{eq:enlarged-impurity-size} and
\eqref{eq:bath-band-count} imply
$\ms=O(m\log(2+\beta\es))$.
This proves the second inequality in \eqref{eq:lambda_even}
for a sufficiently large universal constant $C'$.

For every even initial density operator $\rho_0$,
KMS self-adjointness and the even spectral gap give
\[
\left\|e^{t\cL^\dagger}(\rho_0)-\rho_\beta\right\|_1
\le
\lambda_{\min}(\rho_\beta)^{-1/2}
e^{-\lambda_{\mathrm{even}}t}.
\]
Since
\[
\lambda_{\min}(\rho_\beta)
\ge 2^{-n}e^{-2\beta\|H\|},
\]
the right-hand side is at most $\delta$ whenever
\[
t\ge
\lambda_{\mathrm{even}}^{-1}
\left[
\frac n2\ln2+\beta\|H\|+\ln(1/\delta)
\right].
\]
Substituting \eqref{eq:lambda_even} proves the mixing-time bound,
after increasing $C'$ if necessary.
\end{proof}

Combining Corollary~\ref{cor:impurity-quantum-gap} with the
Gibbs sampler implementation of \cite[Theorem~I.2]{chen2023efficient}
and the random Hamiltonian evolutions implementing
the dephasing term in~\eqref{eq:total-physical} gives the following.

\begin{corollary}[Circuit complexity of thermal state preparation]
\label{cor:impurity-gibbs-preparation}
Use the parameters of Corollary~\ref{cor:impurity-quantum-gap}.
For $\delta\in(0,1)$, set
\[
T=C\lambda_0^{-1}
\bigl[n+\beta\|H\|+\log(1/\delta)\bigr]
\]
for a sufficiently large numerical constant $C$.
There is a quantum algorithm that prepares $\rho_\beta$ to
trace distance at most $\delta$ using $\widetilde O(n)$
qubits,
$\widetilde O(\beta T/(\delta_0\lambda_0))$
total Hamiltonian evolution time, including controlled
evolutions, and $\widetilde O(nT)$ additional one- and
two-qubit gates.
In particular, the Hamiltonian evolution time is
\[
\widetilde O\!\left(
\beta n^{9/2}(2+\beta\es)^{Cm}
[n+\beta\|H\|+\log(1/\delta)]\right).
\]
If $H$ has a spectral gap $\Delta_{\mathrm{gs}}>0$ above its
ground space, choose
$\beta=\Delta_{\mathrm{gs}}^{-1}[n\ln2+\ln(4/\delta)]$.
Applying this corollary with accuracy $\delta/2$ then prepares
the uniform mixture over the ground space
to trace distance at most $\delta$, with the resource bounds
above evaluated at this choice of $\beta$.
\end{corollary}
\begin{proof}
By Corollary~\ref{cor:impurity-quantum-gap}, the chosen $T$
satisfies
$\|e^{T\cL^\dagger}(I/2^n)-\rho_\beta\|_1\le\delta/2$.
It therefore suffices to simulate this evolution to
accuracy $\delta/2$.

We first discretize the phase-jump integral on $[-R,R]$.
Let $N$ be the smallest power of two with
$N\ge\max\{1,8RT\|H\|/\delta\}$, and set $h=2R/N$.
For $k=0,\ldots,N-1$, choose
\[
u_k=-R+(k+\tfrac12)h,
\qquad
w_k=\int_{-R+kh}^{-R+(k+1)h}p_\tau(u)\,du.
\]
Let $\cL_N$ be obtained by replacing the phase-jump
integral in~\eqref{eq:total-physical} by
$\sum_k w_k(e^{iu_kH}Xe^{-iu_kH}-X)$.
Because unitary conjugation by $e^{-iuH}$ is
$2\|H\|$-Lipschitz in $u$ in diamond norm, the midpoint
discretization gives
\[
\|\cL_N^\dagger-\cL^\dagger\|_\diamond
\le h\|H\|\le\frac{\delta}{4T}.
\]
Duhamel's formula then implies
$\|e^{T\cL_N^\dagger}-e^{T\cL^\dagger}\|_\diamond\le\delta/4$.

Next, we express $\cL_N$ as a CKG generator with additional
jump operators. Since $e^{-iu_kH}$ commutes with $H$,
its filtered jump is
$\varphi_\beta(\omega)e^{-iu_kH}$ and its coherent
correction vanishes. Its CKG contribution is therefore
$\alpha_{00}(e^{iu_kH}Xe^{-iu_kH}-X)$, where
\[
\alpha_{00}
=\int_{\mathbb R}
\gamma_\beta(\omega)\varphi_\beta(\omega)^2\,d\omega
=\Theta(1).
\]
Here $\alpha_{00}$ is bounded below by a constant independent of $\beta$, as can be seen
by substituting $x=\beta\omega$.
Thus we obtain $\cL_N$ by adding the weighted jumps
\[
A_k=\sqrt{w_k/\alpha_{00}}\,e^{-iu_kH}
\]
to the original weighted CKG jump family.
The symmetric grid preserves closure under adjoints.
Moreover, $\sum_k A_k^\dagger A_k
=\alpha_{00}^{-1}(\sum_k w_k)I=O(1)I$.
Hence a constant rescaling of the jumps and evolution
time gives the normalization required by the CKG theorem.

To apply that theorem, we also account for the jump
encoding. The weights $w_k$ can be prepared coherently
by recursive interval bisection: each conditional
probability is computed from the explicit Cauchy
interval probabilities, which are differences of
arctangents. This uses $\log N$ levels and
polylogarithmic arithmetic and precision overhead,
without enumerating the grid points.
Since $u_k$ depends linearly on $k$, its binary
representation implements the selected evolution
$e^{-iu_kH}$ using $O(R)$ total Hamiltonian evolution
time. Combining this with the original jump encoding,
which uses $\widetilde O(n)$ gates, gives an enlarged
jump encoding with $O(R)$ evolution time and
$\widetilde O(n)$ additional gates per call.

By~\cite[Theorem~I.2]{chen2023efficient}, a simulation
of $e^{T\cL_N^\dagger}$ to accuracy $\delta/4$, including
implementation errors, uses $\widetilde O(T)$ calls
to this encoding and $\widetilde O(\beta T)$ further
Hamiltonian evolution time.
The total costs are therefore
\[
\widetilde O((\beta+R)T)
=
\widetilde O\!\left(\frac{\beta T}{\delta_0\lambda_0}\right)
\]
Hamiltonian evolution time,
$\widetilde O(nT)$ additional gates, and
$\widetilde O(n)$ qubits.
The mixing, discretization, and simulation errors sum
to at most $\delta$.
Substituting the parameters and using
$\ms=O(m\log(2+\beta\es))$ gives the explicit bound
in the statement.

Finally, the Gibbs weight outside the ground space
is at most $2^n e^{-\beta\Delta_{\mathrm{gs}}}$.
At the stated $\beta$, this is at most $\delta/4$,
so $\rho_\beta$ is within trace norm $\delta/2$ of
the uniform ground-space mixture.
Preparing $\rho_\beta$ to accuracy $\delta/2$ and
applying the triangle inequality proves the claim.
\end{proof}

\section{Observable estimation with Heisenberg scaling}\label{sec:observable}

We first observe that there is an efficiently computable Gaussian state on the doubled Hilbert space with large overlap with the purified thermal state. More precisely, define a positive quadratic operator on this space by
\begin{equation}
\widehat{Q}=
\sum_{1\leq r\leq n,\,\,d_r\geq1}
3\ln(d_r+1)
\bigl(\widehat N_r+\widehat N_{n+r}\bigr),\qquad \widehat N_j=\widehat c_j^\dagger\widehat c_j.
\label{eq:dominating-Q}
\end{equation}
Define the Gaussian state $\widehat\sigma_{\mathrm{dom}}$ on the doubled Hilbert space by
\begin{equation}
\widehat\sigma_{\mathrm{dom}}
=\frac{e^{-\widehat{Q}}}{\Tr(e^{-\widehat{Q}})}.
\label{eq:doubled-dominating-state}
\end{equation}

\begin{lemma}[High overlap with purified thermal state] There exists a constant $C'$ such that
\label{lem:dominating-reference}
\begin{align}
\Tr\left(\widehat\sigma_{\mathrm{dom}} \oket{\sqrt{\rho_\beta}}\obra{\sqrt{\rho_\beta}}\right) \geq \left(\frac{1}{2+\beta\es}\right)^{C'm}. \label{eq:dominate_Gaussian}
\end{align}
\end{lemma}

\begin{proof}
Consider $d\geq 1$.
For every  vector $v_r\in W_{\ell,d}$, put
\[
\eta_r=
\left(
\int_0^{\beta/2}
\left\|
T_\ell e^{-tE_\ell}u(E_\ell)v_r
\right\|\,dt
\right)^2.
\]
By Corollary~\ref{cor:integrated-source-decay}, there is a constant \(C\) such that
\begin{equation}
\sum_{v_r\in W_{\ell,d}}\eta_r
\leq
\frac{Cr_\ell}{(d+1)^6}.
\label{eq:eta-layer-bound}
\end{equation}

Let $\oket{\Psi_\beta}=\oket{\sqrt{\rho_\beta}}$. 
For every subset $I\subseteq\bigcup_{r:\,d_r\geq 1}\{r,n+r\}$,
the CAR and Lemma~\ref{lem:multiparticle-pull-through} give
\begin{equation}
\obra{\Psi_\beta}
\prod_{j\in I}\widehat N_j
\oket{\Psi_\beta}
\leq
\prod_{j\in I}\eta_{r(j)},
\label{eq:joint-defect-bound}
\end{equation}
where $r(j)=j$ for $j\leq n$ and $r(j)=j-n$ for $j>n$.
Since the number operators commute and are projections,
\[
e^{\widehat{Q}}
=
\prod_{1\leq r\leq n,\,\,d_r\geq1}
\left[\I+\bigl((d_r+1)^3-1\bigr)\widehat N_r\right]
\left[\I+\bigl((d_r+1)^3-1\bigr)\widehat N_{n+r}\right].
\]
Expanding this product and applying \eqref{eq:joint-defect-bound}, and then applying \eqref{eq:eta-layer-bound},
\begin{align}
\obra{\Psi_\beta}e^{\widehat{Q}}\oket{\Psi_\beta}
&\leq
\prod_{1\leq r\leq n,\,\,d_r\geq1}
\left[1+\bigl((d_r+1)^3-1\bigr)\eta_r\right]^2\nonumber\\
&\leq
\exp\left(
2\sum_{1\leq r\leq n,\,\,d_r\geq1}
(d_r+1)^3\eta_r
\right)\leq
\exp\left(
2C\sum_{\ell=0}^L r_\ell
\sum_{d\geq1}(d+1)^{-3}
\right)
\leq e^{C\ms}.\label{eq:psi_beta_Q}
\end{align}
where we use \(\sum_{d\geq1}(d+1)^{-3}\leq\int_1^\infty x^{-3}\,dx=\tfrac12\).
Next,
\[
\Tr(e^{-\widehat{Q}})
=
2^{2\ms}
\prod_{\substack{1\leq r\leq n, \,\, d_r\geq1}}
\left(1+(d_r+1)^{-3}\right)^2.
\]
Here $2^{2\ms}$ comes from the thermal modes that have depth $0$.
Using $\dim W_{\ell,d}\leq s_{\ell,0}$, we have
\begin{align}
\ln\Tr(e^{-\widehat{Q}})
\leq
2\ms\ln2+
2\sum_{\ell=0}^L s_{\ell,0}
\sum_{d\geq1}(d+1)^{-3}\leq 2\ms.\label{eq:lneQ}
\end{align}

For any positive definite state $\tau$ and any vector $\ket{\psi}$,
\[
\ket{\psi}\bra{\psi}
\preceq
\bra{\psi}\tau^{-1}\ket{\psi}\,\tau.
\]
Applying this with $\tau=\widehat\sigma_{\mathrm{dom}}$ and using \eqref{eq:lneQ} and
\eqref{eq:psi_beta_Q}, we have
\[
\oket{\Psi_\beta}\obra{\Psi_\beta}
\preceq
\Tr(e^{-\widehat{Q}})
\obra{\Psi_\beta}e^{\widehat{Q}}\oket{\Psi_\beta}\,
\widehat\sigma_{\mathrm{dom}}
\preceq
e^{C'\ms}\widehat\sigma_{\mathrm{dom}}.
\]
Here $C'=C+2$ is another constant. Multiplying by $\obra{\Psi_\beta}$ on the left and $\oket{\Psi_\beta}$ on the right, we obtain \[
1
\preceq
e^{C'\ms}\widehat tr\left( \sigma_{\mathrm{dom}}\oket{\Psi_\beta}\obra{\Psi_\beta} \right).
\]
\end{proof}

We now show that thermal expectation values can be estimated with
Heisenberg scaling in the target precision.
An $(\alpha,k,\delta')$-block encoding of an operator $O$ is a
unitary $U$ on the system and $k$ ancilla qubits such that
\[
\left\|
O-\alpha(\langle0^k|\otimes I)U(|0^k\rangle\otimes I)
\right\|\leq\delta'.
\]
Note that we allow $O$
 to be non-Hermitian.
In particular, taking $O=e^{itH}Ae^{-itH}B$ allows us to
estimate thermal dynamical correlation functions. If $A$ and $B$ are each linear combinations of $O(1)$ Pauli
strings, as for local spin observables and fermionic creation
and annihilation operators, encoding $O$ into a unitary requires only $O(1)$ ancillas.

\begin{corollary}[Thermal observable estimation with Heisenberg scaling]
\label{cor:thermal-observable-estimation}
Use the parameters of Corollary~\ref{cor:impurity-quantum-gap},
and let $\lambda_{\mathrm{even}}$ and $R$ be as in
\eqref{eq:lambda_even} and~\eqref{eq:dephasing-cutoff}.
For every $\delta\in(0,1)$, given a circuit $U$ for an
$(\alpha,k,\delta')$-block encoding of an operator $O$ with
$\alpha\geq1$ and $\delta'\leq\delta/4$, there is a quantum
algorithm that outputs a complex number $z$ satisfying
\[
\Pr\!\left(|z-\operatorname{Tr}(O\rho_\beta)|\leq\delta\right)
\geq 2/3.
\]
For a universal constant $C>0$, the algorithm uses
$\widetilde O(n+k)$ qubits and
\begin{itemize}
\item
$\widetilde O\!\left(
(\beta+R)\lambda_{\mathrm{even}}^{-1}
\bigl[(2+\beta\es)^{Cm}+\alpha\delta^{-1}\bigr]
\right)$
Hamiltonian evolution time, including controlled evolutions,
\item
$\widetilde O(\alpha\delta^{-1})$ calls to $U$ and
$U^\dagger$, including their controlled versions, and
\item
$\widetilde O\!\left(
n^2(2+\beta\es)^{Cm}
+n\lambda_{\mathrm{even}}^{-1}
\bigl[(2+\beta\es)^{Cm}+\alpha\delta^{-1}\bigr]
+k\alpha\delta^{-1}
\right)$
additional one- and two-qubit gates.
\end{itemize}
The Hamiltonian evolution time excludes the cost of implementing
the calls to $U$ and $U^\dagger$.

The purified thermal state $\oket{\sqrt{\rho_\beta}}$ can also
be prepared to trace-distance error $\delta/\alpha$ using
$\widetilde O(n)$ qubits and the same time and gate bounds
with the terms containing $\alpha\delta^{-1}$ omitted.
This preparation does not use $U$, and its dependence on
$\alpha/\delta$ is only polylogarithmic.
\end{corollary}

\begin{proof}
The KMS similarity transform of $-\cL$, together with the
odd-parity projector $(I-P\otimes P^{\mathsf T})/2$, gives
a positive parent Hamiltonian $\widehat{\mathcal H}$ whose
unique ground state is $\oket{\sqrt{\rho_\beta}}$.
Its norm is $O(1)$ and its spectral gap is at least
$\min\{\lambda_{\mathrm{even}},1\}
=\Omega(\lambda_{\mathrm{even}})$.
By~\cite[Theorem~I.3]{chen2023efficient}, using the phase-jump
discretization and jump encoding from the proof of
Corollary~\ref{cor:impurity-gibbs-preparation}, this parent
Hamiltonian has a block encoding with constant normalization
and cost $\widetilde O(\beta+R)$ Hamiltonian evolution time
and $\widetilde O(n)$ additional gates per call.
The transposed circuits required for the doubled space have
the same costs, and the parity projector adds $O(n)$ gates.
The dependence on the inverse block-encoding error is
polylogarithmic.

Set $M=(2+\beta\es)^{C'm}$, with $C'$ as in
Lemma~\ref{lem:dominating-reference}.
The Gaussian state $\widehat\sigma_{\mathrm{dom}}$ can
be prepared with $\widetilde O(n^2)$ gates by preparing its
independent mode occupations and applying a Gaussian basis
change~\cite[Section~IV]{jiang2018quantum}.
By Lemma~\ref{lem:dominating-reference},
$\widehat\sigma_{\mathrm{dom}}$ has overlap at least $M^{-1}$
with $\oket{\sqrt{\rho_\beta}}$.
This overlap bound, together with the ground-state projection
result of~\cite[Lemma~5 and Section~3]{lin2020near},
allows us to prepare $\oket{\sqrt{\rho_\beta}}$
to trace-distance error $\delta/\alpha$ using
$\widetilde O((\beta+R)M\lambda_{\mathrm{even}}^{-1})$
Hamiltonian evolution time and
$\widetilde O(n^2M+nM\lambda_{\mathrm{even}}^{-1})$
additional one- and two-qubit gates.
The dependence on $\alpha/\delta$ is only polylogarithmic.

Let $O_U=\alpha(\langle0^k|\otimes I)U(|0^k\rangle\otimes I)$,
so that $\|O-O_U\|\leq\delta'$.
The real and imaginary parts of
$\operatorname{Tr}(O_U\rho_\beta)/\alpha$ are the expectations
of the Hermitian operators
\[
\frac{O_U+O_U^\dagger}{2\alpha},
\qquad
\frac{O_U-O_U^\dagger}{2i\alpha}.
\]
Both have norm at most one and can be block encoded using
a constant number of applications of $U$ and
$U^\dagger$~\cite{gilyen2019quantum}.
Applying catalytic tomography~\cite[Algorithm~III.1]{chen2025catalytic}
to these operators on the physical register of
$\oket{\sqrt{\rho_\beta}}$ estimates their expectations to
error $\delta/(4\alpha)$, using one prepared copy per estimate.
Combining the two estimates and multiplying by $\alpha$
gives a complex number $z$ satisfying
\[
|z-\operatorname{Tr}(O\rho_\beta)|
\leq \frac\delta4+\frac\delta4+\|O-O_U\|
\leq\frac{3\delta}{4}.
\]
Choosing sufficiently small failure probabilities and
sufficiently accurate state preparation and circuit
implementation gives an overall success probability of
at least $2/3$, with only polylogarithmic overhead.

This estimation uses $\widetilde O(\alpha\delta^{-1})$
applications of $U$ and $U^\dagger$.
Using~\cite{gilyen2019quantum} to simulate evolution under
$\widehat{\mathcal H}$ requires
$\widetilde O(\alpha\lambda_{\mathrm{even}}^{-1}\delta^{-1})$
applications of the unitary that encodes $\widehat{\mathcal H}$.

Preparing the purified thermal state and simulating
$\widehat{\mathcal H}$ use $\widetilde O(n)$ qubits.
Observable estimation adds the $k$ ancillas of $U$ and
logarithmic-size registers~\cite[Appendix~A]{chen2025catalytic},
which are reused between the two estimates.
Operations on the ancillas of $U$ require
$\widetilde O(k\alpha\delta^{-1})$ additional gates.
Adding the preparation and estimation costs proves the
claimed bounds, after increasing $C$ if necessary.

\end{proof}

\appendix
\appendix
\addtocontents{toc}{\protect\setcounter{tocdepth}{1}}

\section{Details of the quantum-to-classical reduction for the spectral gap} \label{app:gap_reduction}
We organize the proof through several sections. Section~\ref{sec:ckg-dirichlet} introduce the Dirichlet form.
Section~\ref{sec:classical-comparison} establishes bounds for this form and filtered commutators used in subsequent perturbation arguments.
Section~\ref{sec:commuting-perturbations} compares Dirichlet forms under commuting perturbations.
Finally, Section~\ref{sec:dephasing-gap} combines these bounds with spectral pinching to prove Theorem~\ref{thm:qtoc}.

\subsection{The  Dirichlet form}
\label{sec:ckg-dirichlet}

\paragraph{Dirichlet form of the combined generator.}

The Dirichlet form of the Metropolis CKG generator takes a standard form as follows. Define
\begin{equation}
 \cE_K(X,Y):=-\ip{X}{\cL_{\mathrm{CKG},K}(Y)}{K},\qquad
 \cE_K(X):=\cE_K(X,X).
 \label{eq:form}
\end{equation}

\begin{lemma}[Dirichlet-form identity]\label{lem:sos}
Let
\begin{equation}
 g_\beta(t):=\frac1{\beta\cosh(2\pi t/\beta)},\qquad
 B_{K,J}(t,\omega):=\sum_\nu e^{it\nu+\beta\nu/4}\varphi_\beta(\omega-\nu)J_\nu^K.
 \label{eq:B}
\end{equation}
Then, for all operators $X,Y$,
\begin{equation}
 \cE_K(X,Y)=\E_J\int_{\R^2}g_\beta(t)\gamma_\beta(\omega)
 \ip{[B_{K,J}(t,\omega),X]}{[B_{K,J}(t,\omega),Y]}{K}\dd\omega\dd t.
 \label{eq:sos}
\end{equation}
Consequently $\cL_{\mathrm{CKG},K}$ satisfies KMS detailed balance and $\rho_{\beta,K}$ is stationary.
\end{lemma}
The identity follows from~\cite[Theorem~3 and Appendix~A.4.2]{leng2026accelerating}, using the coefficients in~\eqref{eq:filtered-jumps}.
The cited theorem assumes Hermitian jump operators, but it can be directly generalized to the case where the set of jump operators is closed under taking adjoints.

The Metropolis CKG generator with an extra dephasing term as in Eq.~\eqref{eq:total-physical} takes the following Dirichlet form.

\begin{claim} \label{claim:totaldirichlet}
    The generator in Eq.~\eqref{eq:total-physical} satisfies KMS detailed balance, preserves the thermal state, and maps even operators to even operators. Its Dirichlet form is
    \begin{equation}
        \cF_H(X)=\cE_H(X)+\cE_{\mathrm{deph}}(X),
        \label{eq:dephasing-form}
    \end{equation}
    where, in an eigenbasis of $H$ with energies $H_i$ and Gibbs weights $\pi_i$,
    \begin{equation}
        \cE_{\mathrm{deph}}(X):=
        \sum_{i,j}\sqrt{\pi_i\pi_j}|X_{ij}|^2
        \int_{-R}^R p_\tau(u)\bigl(1-\cos(u(H_i-H_j))\bigr)\dd u.
        \label{eq:dephasing-entries}
    \end{equation}
\end{claim}
\begin{proof}
    Since $e^{iuH}$ commutes with $\rho_\beta$ and $P$, the dephasing term in the generator preserves the thermal state and maps even operators to even operators.
    By symmetry of $p_\tau$,
    \begin{align*}
        (\cL(Y)-\cL_{\mathrm{CKG},H}(Y))_{ij}
        &=Y_{ij}\int_{-R}^R p_\tau(u)
        \bigl(\cos(u(H_i-H_j))-1\bigr)\dd u\\
        &\quad+iY_{ij}\underbrace{\int_{-R}^R p_\tau(u)
        \sin(u(H_i-H_j))\dd u}_{=0}\\
        &=-Y_{ij}\int_{-R}^R p_\tau(u)
        \bigl(1-\cos(u(H_i-H_j))\bigr)\dd u.
    \end{align*}
    Thus, for arbitrary operators $X,Y$,
    \begin{align*}
        -\ip{X}{\cL(Y)}H
        &=\cE_H(X,Y)+\sum_{i,j}\sqrt{\pi_i\pi_j}\,\overline{X_{ij}}Y_{ij}
          \int_{-R}^R p_\tau(u)\bigl(1-\cos(u(H_i-H_j))\bigr)\dd u\\
        &=-\ip{\cL(X)}{Y}H.
    \end{align*}
    Here the second equality uses that the integral is real and $\cL_{\mathrm{CKG},H}$ satisfies KMS detailed balance. Thus $\cL$ satisfies KMS detailed balance. Taking $Y=X$ proves~\eqref{eq:dephasing-form}--\eqref{eq:dephasing-entries}.
\end{proof}

\subsection{The classical comparison and uniform bounds}
\label{sec:classical-comparison}

Here we establish two bounds required for the proof of the theorem. First, we show that the spectral gap of the classical Metropolis chain controls the Dirichlet form of the generator under commuting observables. Second, we establish a pair of useful bounds for our perturbation arguments.

\begin{lemma}[Comparison on commuting observables]\label{lemma:classicalcomp}
    Assume $[K,P]=0$ and the classical Metropolis chain for $K$ has spectral gap at least $\lambda_0$ in every common  eigenbasis of $K$ and $P$. For every even Hermitian $Y$ with $[Y,K]=0$,
    \begin{equation}
    \cE_K(Y)\ge a_0\lambda_0\Var_K(Y),
    \label{eq:classical-comparison}
    \end{equation}
    where $a_0>0$ is a numerical constant.
\end{lemma}
\begin{proof}
    Choose an orthonormal basis simultaneously diagonalizing $K$, $Y$, and $P$, and write $\{K_x\}_x,\{y_x\}_x$ for their eigenvalues. Define
    \[
    r_\beta(\nu):=\int_\R\gamma_\beta(\omega)\varphi_\beta(\omega-\nu)^2\dd\omega,
    \]
    where $\gamma_\beta$ and $\varphi_\beta$ are given by Eq.~\eqref{eq:filters}.
    Set $z=\omega-\nu$. For $|z|\le1/\beta$, $\max\{\beta(\nu+z)+1/2,0\}\le\max\{\beta\nu,0\}+3/2$. Restricting the integral to $z\in[-1/\beta,1/\beta]$, equivalently $\omega\in[\nu-1/\beta,\nu+1/\beta]$, gives
    \begin{equation}
    \begin{aligned}
    r_\beta(\nu)
    &\ge\int_{-1/\beta}^{1/\beta}\gamma_\beta(\nu+z)\varphi_\beta(z)^2\dd z\\
    &\ge e^{-3/2}\min\{1,e^{-\beta\nu}\}
    \int_{-1/\beta}^{1/\beta}\frac{\beta}{\sqrt{2\pi}}e^{-\beta^2z^2/2}\dd z\\
    &=\underbrace{\frac{e^{-3/2}}{\sqrt{2\pi}}\int_{-1}^1e^{-x^2/2}\dd x}_{a_0>0}
    \min\{1,e^{-\beta\nu}\},
    \end{aligned}
    \label{eq:rate-domination}
    \end{equation}
    where $x=\beta z$ in the Gaussian integral.
    Let $\pi_x=e^{-\beta K_x}/\sum_z e^{-\beta K_z}$ and $q_{xy}=\E_J|\langle y|J|x\rangle|^2$.
    Since $Y$ is diagonal and only $\nu=K_y-K_x$ contributes to the $(y,x)$ entry of $B_{K,J}$,
    \begin{align*}
    \langle y|[B_{K,J}(t,\omega),Y]|x\rangle
    &=(y_x-y_y)\langle y|B_{K,J}(t,\omega)|x\rangle\\
    &=(y_x-y_y)e^{it(K_y-K_x)+\beta(K_y-K_x)/4}
    \varphi_\beta(\omega-K_y+K_x)\langle y|J|x\rangle.
    \end{align*}
    Taking the squared modulus, averaging over $J$, and evaluating~\eqref{eq:sos} entrywise gives
    \begin{equation}
    \begin{aligned}
    \cE_K(Y)
    &=\E_J\int_{\R^2}g_\beta(t)\gamma_\beta(\omega)
      \sum_{x,y}\sqrt{\pi_x\pi_y}\,
      \abs{\langle y|[B_{K,J}(t,\omega),Y]|x\rangle}^2\dd\omega\dd t\\
    &=\sum_{x,y}\sqrt{\pi_x\pi_y}\,e^{\beta(K_y-K_x)/2}
      q_{xy}(y_x-y_y)^2\int_{\R^2}g_\beta(t)\gamma_\beta(\omega)
      \underbrace{\abs{e^{it(K_y-K_x)}}^2}_{=1}
      \varphi_\beta(\omega-K_y+K_x)^2\dd\omega\dd t\\
    &=\frac12\sum_{x,y}\sqrt{\pi_x\pi_y}\,e^{\beta(K_y-K_x)/2}
      q_{xy}(y_x-y_y)^2
      \int_\R\gamma_\beta(\omega)\varphi_\beta(\omega-K_y+K_x)^2\dd\omega\\
    &=\frac12\sum_{x,y}\pi_xq_{xy}r_\beta(K_y-K_x)(y_x-y_y)^2.
    \end{aligned}
    \label{eq:diagonal-form}
    \end{equation}
    Here $\int g_\beta=1/2$, and the KMS matrix-entry weight satisfies
    $\sqrt{\pi_x\pi_y}\,e^{\beta(K_y-K_x)/2}=\pi_x$. Apply~\eqref{eq:rate-domination} and the classical Poincar\'e inequality~\eqref{eq:classical-poincare}, with $f_x=y_x$, to~\eqref{eq:diagonal-form}. This proves the lemma, including when $K$ has degenerate eigenvalues.
\end{proof}

We require a pair of bounds in our perturbation arguments. The first bound below controls the error when replacing $X$ by its pinched approximation $Y$ in the proof of Theorem~\ref{thm:qtoc}, by applying it to $Z=X-Y$.
The second controls the error when replacing the Hamiltonian in Lemma~\ref{lemma:comparison}.
There, changing the Hamiltonian changes the energy differences $\nu$ in~\eqref{eq:B}: differentiating $e^{(it+\beta/4)\nu}\varphi_\beta(\omega-\nu)$ with respect to $\nu$ produces $it+\beta/4$ from the exponential and $-\partial_\omega$ from the Gaussian, giving the operator $\mathsf D_t$ below.

\begin{lemma}[Uniform bounds for the filtered commutators]\label{lemma:uniform} 
    For every $K$ and every operator $Z$,
    \begin{equation}
        \cE_K(Z)\le2\wnorm ZK^2.
        \label{eq:bounded-form}
    \end{equation}
    Define the differential operator
    $\mathsf D_t:=-\partial_\omega+it+\tfrac\beta4$, acting on the $\omega$ variable. Then
    \begin{equation}
        \E_J\int_{\R^2}g_\beta(t)\gamma_\beta(\omega)
        \wnorm{[\mathsf D_tB_{K,J}(t,\omega),Z]}{K}^{2}\dd\omega\dd t
        \le3\beta^2\wnorm ZK^2.
        \label{eq:derivative-bound}
    \end{equation}
    The bounds are independent of $K$, its spectral width, and the Hilbert-space dimension.
\end{lemma}

To avoid distraction, the proof of Lemma~\ref{lemma:uniform} is delayed to Section~\ref{app:uniform-bounds}.

\subsection{Comparison under commuting perturbations}
\label{sec:commuting-perturbations}

In this section, we establish our main perturbation lemma (Lemma~\ref{lemma:comparison}). Before doing so, we need the following claim.

\begin{claim}[Gibbs-weight comparison]\label{claim:gibbs-comparison}
Let $A,B$ be commuting Hamiltonians with $\norm{A-B}\le\varepsilon$. Then
\[
 e^{-2\beta\varepsilon}\rho_{\beta,B}\preceq\rho_{\beta,A}\preceq e^{2\beta\varepsilon}\rho_{\beta,B}.
\]
For every operator $X$,
\begin{equation}
 e^{-2\beta\varepsilon}\wnorm XB^2\le\wnorm XA^2\le e^{2\beta\varepsilon}\wnorm XB^2.
 \label{eq:norm-comparison}
\end{equation}
For every Hermitian $X$, moreover,
\[
 e^{-2\beta\varepsilon}\Var_B(X)\le\Var_A(X)\le e^{2\beta\varepsilon}\Var_B(X).
\]
\end{claim}
\begin{proof}
Choose a common orthonormal eigenbasis of $A$ and $B$. Since $|A_i-B_i|\le\varepsilon$,
\[
 e^{-\beta\varepsilon}e^{-\beta B_i}\le e^{-\beta A_i}\le e^{\beta\varepsilon}e^{-\beta B_i},\qquad
 e^{-\beta\varepsilon}\Tr(e^{-\beta B})\le\Tr(e^{-\beta A})\le e^{\beta\varepsilon}\Tr(e^{-\beta B}).
\]
Thus their normalized Gibbs weights satisfy
$e^{-2\beta\varepsilon}\pi_i^{(B)}\le\pi_i^{(A)}\le e^{2\beta\varepsilon}\pi_i^{(B)}$,
proving the operator inequalities. The norm comparison follows from
\[
 \wnorm XA^2=\sum_{i,j}\sqrt{\pi_i^{(A)}\pi_j^{(A)}}\,|X_{ij}|^2.
\]
Finally, apply~\eqref{eq:norm-comparison} to $X-cI$ and take the infimum over $c\in\R$ to obtain the variance comparison.
\end{proof}

\begin{lemma}\label{lemma:comparison}
    If $[H,K]=0$ and $\beta\norm{H-K}\le1$, then, for every operator $Y$,
    \begin{equation}
        \boxed{\sqrt{\cE_K(Y)}
        \le C\left(\sqrt{\cE_H(Y)}+\beta\wnorm{[H-K,Y]}H\right),}
        \label{eq:comparison}
    \end{equation}
    where $C > 0$ is a universal numerical constant.
    In particular, if $[K,Y]=0$,
    \begin{equation}
        \cE_K(Y)\le C\left(\cE_H(Y)+\beta^2\wnorm{[H,Y]}H^2\right).
        \label{eq:comparison-commuting}
    \end{equation}
\end{lemma}
\begin{proof}
By~\eqref{eq:sos},
\[
\cE_K(Y)=\E_J\int_{\R^2}g_\beta(t)\gamma_\beta(\omega)
\wnorm{[B_{K,J}(t,\omega),Y]}K^2\dd\omega\dd t.
\]
We compare this expression with $\cE_H(Y)$ by controlling the change in the filtered commutator along a path from $K$ to $H$.

Put $\Delta:=H-K$ and $K_r:=K+r\Delta$ for $0\le r\le1$. All $K_r$ commute, and $\beta\norm{K_r-K_s}\le1$ for $r,s\in[0,1]$. Claim~\ref{claim:gibbs-comparison} therefore gives
\begin{equation}
 e^{-2}\wnorm{A}{K_s}^2\le\wnorm{A}{K_r}^2\le e^2\wnorm{A}{K_s}^2
 \qquad(r,s\in[0,1])
 \label{eq:path-norm-comparison}
\end{equation}
for every operator $A$.

Write $B_r:=B_{K_r,J}(t,\omega)$ and $V_r:=[B_r,Y]$.
Use the weighted norm
\begin{equation}
 \norm F_*^2:=\E_J\int_{\R^2}g_\beta(t)\gamma_\beta(\omega)\wnorm{F_J(t,\omega)}K^2\dd\omega\dd t
 \label{eq:aux-norm}
\end{equation}
for a family of matrices $F_J(t,\omega)$, with the fixed KMS norm associated with $K$.
By~\eqref{eq:sos} and~\eqref{eq:path-norm-comparison},
\[
 \norm{V_0}_*=\sqrt{\cE_K(Y)},\qquad
 \norm{V_1}_*\le e\sqrt{\cE_H(Y)}.
\]
Thus we want to bound $\norm{V_0}_*$ in terms of $\norm{V_1}_*$ and $\beta\wnorm{[H-K,Y]}H$. The identity
\[
 V_0=V_1-\int_0^1\partial_rV_r\dd r
\]
motivates studying $\partial_rV_r$. We first compute this derivative, then cancel its frequency-derivative term before bounding the remaining error.

Fix a common eigenbasis of $H$ and $K$ and write
$K_r\ket{i}=\kappa_i(r)\ket{i}$ and $\Delta\ket{i}=\Delta_i\ket{i}$.
Put $\nu_r=\kappa_i(r)-\kappa_j(r)$, so that
$\partial_r\nu_r=\Delta_i-\Delta_j$. Differentiating a matrix entry of~\eqref{eq:B} gives
\begin{align*}
 (B_r)_{ij}
 &=e^{(it+\beta/4)\nu_r}\varphi_\beta(\omega-\nu_r)J_{ij},\\
 \partial_r(B_r)_{ij}
 &=(\Delta_i-\Delta_j)e^{(it+\beta/4)\nu_r}
 \left[(it+\tfrac\beta4)\varphi_\beta(\omega-\nu_r)-\varphi_\beta'(\omega-\nu_r)\right]J_{ij}\\
 &=(\Delta_i-\Delta_j)(\mathsf D_tB_r)_{ij}
 = [\Delta,\mathsf D_tB_r]_{ij}.
\end{align*}
Thus
\begin{equation}
 \partial_r B_r=[\Delta,\mathsf D_tB_r].
 \label{eq:Br-derivative}
\end{equation}
Put $Z:=[\Delta,Y]$. Jacobi's identity yields
\begin{equation}
 \begin{aligned}
 \partial_r V_r
 &=[\partial_rB_r,Y]=[[\Delta,\mathsf D_tB_r],Y]\\
 &=[\Delta,[\mathsf D_tB_r,Y]]-[\mathsf D_tB_r,[\Delta,Y]]\\
 &=\underbrace{[\Delta,\mathsf D_tV_r]}_{\text{first term}}
 -\underbrace{[\mathsf D_tB_r,Z]}_{\text{second term}}.
 \end{aligned}
 \label{eq:Vr-derivative}
\end{equation}

Here $[\mathsf D_tB_r,Y]=\mathsf D_t[B_r,Y]$ because $Y$ is independent of $\omega$. 

\paragraph{First term.}
We cancel $[\Delta,\mathsf D_tV_r]$ in~\eqref{eq:Vr-derivative} by shifting the frequency and multiplying by an exponential.
Put $d_{ij}=\Delta_i-\Delta_j$ and define
\begin{equation}
 (W_r)_{ij}(t,\omega)
 :=e^{-rd_{ij}(it+\beta/4)}(V_r)_{ij}(t,\omega+rd_{ij}).
 \label{eq:shifted-commutator}
\end{equation}
The product and chain rules, together with~\eqref{eq:Vr-derivative}, give
\begin{equation}
 \begin{aligned}
 \partial_r(W_r)_{ij}(t,\omega)
 &=e^{-rd_{ij}(it+\beta/4)}
 \left(\partial_rV_r+d_{ij}\partial_\omega V_r
       -d_{ij}(it+\tfrac\beta4)V_r\right)_{ij}(t,\omega+rd_{ij})\\
 &=-e^{-rd_{ij}(it+\beta/4)}
 [\mathsf D_tB_r,Z]_{ij}(t,\omega+rd_{ij}).
 \end{aligned}
 \label{eq:shifted-derivative}
\end{equation}
In the last equality, the two terms involving $V_r$ cancel, leaving the term involving $Z$. We next show that the transformation from $V_r$ to $W_r$ increases the weighted norm by at most a constant factor. Since $\ln\gamma_\beta$ is $\beta$-Lipschitz,
\[
 \ln\gamma_\beta(s-rd_{ij})-\ln\gamma_\beta(s)\le \beta|rd_{ij}|,
 \qquad
 \gamma_\beta(s-rd_{ij})\le e^{\beta|rd_{ij}|}\gamma_\beta(s).
\]
Together with $\beta|rd_{ij}|\le2$, this gives
\[
 e^{-\beta rd_{ij}/2}\gamma_\beta(s-rd_{ij})
 \le e^{\beta|rd_{ij}|/2}e^{\beta|rd_{ij}|}\gamma_\beta(s)
 \le e^3\gamma_\beta(s).
\]
Thus, changing variables by $s=\omega+rd_{ij}$ gives, for any $F$,
\begin{align*}
 &\int_\R\gamma_\beta(\omega)
 \abs{e^{-rd_{ij}(it+\beta/4)}F_{ij}(t,\omega+rd_{ij})}^2\dd\omega\\
 &\qquad=\int_\R e^{-\beta rd_{ij}/2}\gamma_\beta(s-rd_{ij})\abs{F_{ij}(t,s)}^2\dd s\\
 &\qquad\le e^3\int_\R\gamma_\beta(s)\abs{F_{ij}(t,s)}^2\dd s.
\end{align*}
Let $\pi_i$ be the Gibbs weights of $K$ in the chosen basis. Since
$\wnorm AK^2=\sum_{i,j}\sqrt{\pi_i\pi_j}\,|A_{ij}|^2$, the definition~\eqref{eq:aux-norm} gives
\begin{align*}
 \norm{W_r}_*^2
 &=\E_J\int_\R g_\beta(t)\sum_{i,j}\sqrt{\pi_i\pi_j}
   \int_\R\gamma_\beta(\omega)|(W_r)_{ij}(t,\omega)|^2\dd\omega\dd t\\
 &=\E_J\int_\R g_\beta(t)\sum_{i,j}\sqrt{\pi_i\pi_j}
   \int_\R\gamma_\beta(\omega)
   \abs{e^{-rd_{ij}(it+\beta/4)}(V_r)_{ij}(t,\omega+rd_{ij})}^2\dd\omega\dd t\\
 &\le e^3\E_J\int_\R g_\beta(t)\sum_{i,j}\sqrt{\pi_i\pi_j}
   \int_\R\gamma_\beta(\omega)|(V_r)_{ij}(t,\omega)|^2\dd\omega\dd t\\
 &=e^3\norm{V_r}_*^2.
\end{align*}
Taking square roots and applying the same estimate to~\eqref{eq:shifted-derivative} gives
\begin{equation}
 \norm{W_1}_*\le e^{3/2}\norm{V_1}_*,\qquad
 \norm{\partial_rW_r}_*\le e^{3/2}\norm{[\mathsf D_tB_r,Z]}_*.
 \label{eq:shifted-bounds}
\end{equation}
We can now integrate along the path. Since $W_0=V_0$, the fundamental theorem of calculus and the triangle inequality imply
\begin{equation}
 \begin{aligned}
 \norm{V_0}_*
 &\le\norm{W_1}_*+\int_0^1\norm{\partial_rW_r}_*\dd r\\
 &\le e^{3/2}\left(\norm{V_1}_*
       +\int_0^1\norm{[\mathsf D_tB_r,Z]}_*\dd r\right).
 \end{aligned}
 \label{eq:variation}
\end{equation}

\paragraph{Second term.}
It remains to bound $[\mathsf D_tB_r,Z]$, where $Z=[H-K,Y]$. Lemma~\ref{lemma:uniform} and two applications of~\eqref{eq:path-norm-comparison} give
\begin{align*}
 \norm{[\mathsf D_tB_r,Z]}_*^2
 &\le e^2\E_J\int g_\beta(t)\gamma_\beta(\omega)
 \wnorm{[\mathsf D_tB_{K_r,J}(t,\omega),Z]}{K_r}^2\dd\omega\dd t\\
 &\le3\beta^2e^2\wnorm Z{K_r}^2\le3\beta^2e^4\wnorm ZH^2.
\end{align*}
Substituting this bound and the endpoint estimates into~\eqref{eq:variation} gives
\[
 \sqrt{\cE_K(Y)}\le e^{5/2}\sqrt{\cE_H(Y)}
       +\sqrt3\,e^{7/2}\beta\wnorm{[H-K,Y]}H,
\]
which proves~\eqref{eq:comparison}. If $[K,Y]=0$, then $[H-K,Y]=[H,Y]$. Squaring both sides proves~\eqref{eq:comparison-commuting}.
\end{proof}

\subsection{Dephasing, rounding, and the gap}
\label{sec:dephasing-gap}

To lower bound the full Dirichlet form of observable $X$ under $H$, we first construct a nearby Hamiltonian $K$ and a pinched observable $Y$ commuting with $K$, with $\wnorm{X-Y}H^2$ controlled by $\cE_{\mathrm{deph}}(X)$ and the Cauchy tail beyond $R$.
We then combine this construction with the preceding comparison bounds and the dephasing term to prove Theorem~\ref{thm:qtoc}.

Write
\begin{equation}
    q_R:=\int_{|u|>R}p_\tau(u)\dd u
    =\frac2\pi\arctan\frac\tau R
    \le\frac{2\tau}{\pi R}.
    \label{eq:cauchy-tail}
\end{equation}

\begin{lemma}[Spectral pinching] \label{lemma:spectralpinching}
   Let $X=X^\dagger$ satisfy $\Tr(\rho_\beta X)=0$ and $\wnorm XH=1$. Set $\delta=\delta_0/\beta=1/\tau$. There exist a Hamiltonian $K$ and a Hermitian $Y$ such that
    \begin{gather}
        [K,H]=0,\qquad \norm{K-H}\le\delta/2,\qquad [K,Y]=0,\label{eq:pinch-algebra}\\
        d:=\wnorm{X-Y}H,\qquad d^2\le2\cE_{\mathrm{deph}}(X)+4q_R,\qquad
        \Tr(\rho_\beta Y)=0,\qquad \wnorm YH^2=1-d^2,\label{eq:pinch-distance}\\
        \beta^2\wnorm{[H,Y]}H^2\le2\delta_0^2\bigl(\cE_{\mathrm{deph}}(X)+2q_R\bigr).\label{eq:pinch-comm}
    \end{gather}
    If $H$ and $X$ are even, then $K$ and $Y$ can also be chosen even.
\end{lemma}
\begin{proof}
    For $\theta\in[0,\delta)$, divide the energy axis into intervals
    $[\theta+j\delta,\theta+(j+1)\delta)$, $j\in\mathbb Z$. Let $P_{j,\theta}$ be the spectral projection of $H$ onto the $j$th interval, and set
    \[
        K_\theta:=\sum_j\bigl(\theta+(j+1/2)\delta\bigr)P_{j,\theta},\qquad
        Y_\theta:=\sum_jP_{j,\theta}XP_{j,\theta}.
    \]
    If $[H,P]=[X,P]=0$, then each $P_{j,\theta}$ commutes with $P$, so $[K_\theta,P]=[Y_\theta,P]=0$.
    The properties in~\eqref{eq:pinch-algebra} are immediate. The pinching map is an orthogonal projection in the KMS inner product, because each $P_{j,\theta}$ commutes with $\rho_\beta$. Thus the Pythagorean theorem gives
    \[
        \wnorm{Y_\theta}H^2=\wnorm XH^2-\wnorm{X-Y_\theta}H^2
        =1-\wnorm{X-Y_\theta}H^2.
    \]
    It also preserves the expectation in the thermal state: using $[P_{j,\theta},\rho_\beta]=0$ and $\sum_jP_{j,\theta}=I$,
    \[
        \Tr(\rho_\beta Y_\theta)
        =\sum_j\Tr(\rho_\beta P_{j,\theta}XP_{j,\theta})
        =\Tr(\rho_\beta X)=0.
    \]
    These are the last two identities in~\eqref{eq:pinch-distance}, valid for every $\theta$.

    The Fourier transform of the Cauchy density is
    $\int_\R p_\tau(u)e^{iu\nu}\dd u=e^{-\tau|\nu|}$. Hence
    \begin{align}
        1-e^{-\tau|\nu|}
        &=\int_\R p_\tau(u)(1-\cos(u\nu))\dd u\nonumber\\
        &\le\int_{-R}^R p_\tau(u)(1-\cos(u\nu))\dd u+2q_R,
        \label{eq:dephasing-tail-comparison}
    \end{align}
    since $0\le1-\cos(u\nu)\le2$.

    Write $H_i$ and $\pi_i$ for the energies and Gibbs weights of $H$ in an eigenbasis. A uniform shift $\theta$ puts $H_i,H_j$ in different intervals with probability $\min\{1,|H_i-H_j|/\delta\}$.
    
    Since $\min\{1,x\}\le2(1-e^{-x})$ for $x\ge0$,
    \begin{align*}
        \E_\theta\wnorm{X-Y_\theta}H^2
        &=\sum_{i,j}\sqrt{\pi_i\pi_j}|X_{ij}|^2
          \min\{1,\tau|H_i-H_j|\}\\
        &\le2\sum_{i,j}\sqrt{\pi_i\pi_j}|X_{ij}|^2
          (1-e^{-\tau|H_i-H_j|})\\
        &\le2\sum_{i,j}\sqrt{\pi_i\pi_j}|X_{ij}|^2
          \left[\int_{-R}^R p_\tau(u)\bigl(1-\cos(u(H_i-H_j))\bigr)\dd u+2q_R\right]\\
        &=2\cE_{\mathrm{deph}}(X)+4q_R\wnorm XH^2
        =2\cE_{\mathrm{deph}}(X)+4q_R.
    \end{align*}
    The last two lines use~\eqref{eq:dephasing-tail-comparison} and~\eqref{eq:dephasing-entries}, respectively, together with $\wnorm XH=1$. Some shift attains this upper bound. Choose that shift to obtain $K$ and $Y$.
    Any remaining entry $Y_{ij}=X_{ij}$ has $|H_i-H_j|<\delta$.
    Thus, using $x^2\le2(1-e^{-x})$ for $0\le x\le1$, we obtain
    \begin{align*}
        \beta^2\wnorm{[H,Y]}H^2
        &=\delta_0^2\sum_{i,j}\sqrt{\pi_i\pi_j}
          (\tau|H_i-H_j|)^2|Y_{ij}|^2\\
        &\le2\delta_0^2\sum_{i,j}\sqrt{\pi_i\pi_j}
          (1-e^{-\tau|H_i-H_j|})|Y_{ij}|^2\\
        &\le2\delta_0^2\sum_{i,j}\sqrt{\pi_i\pi_j}
          (1-e^{-\tau|H_i-H_j|})|X_{ij}|^2\\
        &\le2\delta_0^2\sum_{i,j}\sqrt{\pi_i\pi_j}|X_{ij}|^2
          \left[\int_{-R}^R p_\tau(u)\bigl(1-\cos(u(H_i-H_j))\bigr)\dd u+2q_R\right]\\
        &=2\delta_0^2\bigl(\cE_{\mathrm{deph}}(X)+2q_R\bigr).
    \end{align*}
    Again, the last two lines use~\eqref{eq:dephasing-tail-comparison} and~\eqref{eq:dephasing-entries}, respectively, with $\wnorm XH=1$.

\end{proof}

\begin{proof}[Proof of Theorem~\ref{thm:qtoc}] 
By Claim~\ref{claim:totaldirichlet}, the total Dirichlet form is
$\cF_H(X)=\cE_H(X)+\cE_{\mathrm{deph}}(X)$. Take any even Hermitian $X$ with $\Tr(\rho_\beta X) = 0$ and $\wnorm{X}{H}=1$, and define $\eta := \cF_H(X)+2q_R$. Then
\begin{equation}
     \cE_H(X)\le\eta,\qquad \cE_{\mathrm{deph}}(X)+2q_R\le\eta.
 \label{eq:slow}
\end{equation}

Apply Lemma~\ref{lemma:spectralpinching} with $\delta=\delta_0/\beta$, and obtain operators $K,Y$ and scalar $d$ with the guarantee
\begin{equation} \label{eq:dsquared}
    d^2\le2\eta,\qquad \beta^2\wnorm{[H,Y]}H^2\le2\delta_0^2\eta.
\end{equation}

The pinching construction gives $[K,P]=[Y,P]=0$. The robustness hypothesis~\eqref{eq:robust} applies to $K$, because $[K,H]=[K,P]=0$ and $\beta\norm{K-H}\le\delta_0/2$. Since $[Y,K]=0$, Lemma~\ref{lemma:classicalcomp} gives
\[
    \cE_K(Y) \geq a_0 \lambda_0 \Var_K(Y),
\]
for a constant $a_0 > 0$. 

Claim~\ref{claim:gibbs-comparison}, using $\beta\norm{K-H}\le\delta_0/2\le1$, and Eq.~\eqref{eq:pinch-distance} give
\[
    \Var_K(Y) \geq e^{-2} \Var_H(Y) = e^{-2} (1-d^2).
\]

Therefore the Dirichlet form of $Y$ w.r.t. $K$ satisfies
\begin{equation} \label{eq:lower}
    \cE_K(Y) \geq a \lambda_0 (1-d^2),
\end{equation}
for some numerical constant $a>0$.

Conversely, Lemma~\ref{lemma:comparison}, $[Y,K] = 0$, and Eq.~\eqref{eq:pinch-comm} imply

\begin{align}
    \cE_K(Y)
    &\le C'\bigl(\cE_H(Y)+\beta^2\wnorm{[H,Y]}H^2\bigr)\nonumber\\
    &\le C'\bigl(2\cE_H(X)+2\cE_H(Y-X)+\beta^2\wnorm{[H,Y]}H^2\bigr)\nonumber\\
    &\le C'\bigl(2\eta+4d^2+2\delta_0^2\eta\bigr)
    \le C''\eta.
    \label{eq:upper}
\end{align}
Here the second line uses the triangle inequality for $\sqrt{\cE_H(\cdot)}$ and $(a+b)^2\le2a^2+2b^2$; the last line uses Lemma~\ref{lemma:uniform}, Eqs.~\eqref{eq:slow}--\eqref{eq:dsquared}, and $\delta_0\le1$.

Combining~\eqref{eq:lower} and~\eqref{eq:upper}, using $\lambda_0\le1$, gives
\[
    a\lambda_0
    \le \cE_K(Y)+a\lambda_0d^2
    \le C''\eta+2a\lambda_0\eta
    \le(C''+2a)\eta.
\]
The constants $a,C',C''$ are independent of $R$. Therefore, using~\eqref{eq:cauchy-tail} and $R=C\tau/\lambda_0$,
\[
    \cF_H(X)=\eta-2q_R
    \ge\left(\frac{a}{C''+2a}-\frac{4}{\pi C}\right)\lambda_0
    \ge c\lambda_0
\]
for a sufficiently large numerical constant $C$ and a sufficiently small numerical constant $c>0$. Taking the infimum over centered, normalized, even Hermitian $X$ proves~\eqref{eq:main}.

\end{proof}

\subsection{Proof of Lemma~\ref{lemma:uniform}}
\label{app:uniform-bounds}

\begin{proof}
Bound~\eqref{eq:bounded-form} follows from~\cite[Theorem~I.3]{chen2023efficient}: the discriminant has operator norm at most $2$, so $\cE_K(Z)\le2\wnorm ZK^2$. To prove~\eqref{eq:derivative-bound}, we use the following estimates.
The definition of the filtered jump and the Gaussian Fourier-transform identity give
\begin{equation}
 \widehat J_K(\omega)=\frac1{\sqrt{2\pi}}\int_\R f_\beta(u)e^{iuK}Je^{-iuK}e^{-iu\omega}\dd u,
 \qquad f_\beta(u):=\beta^{-1/2}(2/\pi)^{1/4}e^{-u^2/\beta^2}.
 \label{eq:OFT}
\end{equation}
Apply the operator Parseval identity~\cite[Proposition~A.2]{chen2023thermalpreparation} with filters $f_\beta(u)$ and $-iu f_\beta(u)$, respectively; the latter gives $\partial_\omega\widehat J_K$. Since $\int|f_\beta|^2=1$ and $\int u^2|f_\beta(u)|^2\dd u=\beta^2/4$, the normalization~\eqref{eq:boundedjump} yields
\begin{align}
 \E_J\int_\R\widehat J_K(\omega)^\dagger\widehat J_K(\omega)\dd\omega&\preceq I,
 &\E_J\int_\R\widehat J_K(\omega)\widehat J_K(\omega)^\dagger\dd\omega&\preceq I,\label{eq:parseval-base}\\
 \E_J\int_\R(\partial_\omega\widehat J_K)^\dagger(\partial_\omega\widehat J_K)\dd\omega&\preceq\tfrac{\beta^2}4I,
 &\E_J\int_\R(\partial_\omega\widehat J_K)(\partial_\omega\widehat J_K)^\dagger\dd\omega&\preceq\tfrac{\beta^2}4I.
 \label{eq:parseval-derivative}
\end{align}

For this proof only, abbreviate $B=B_{K,J}(t,\omega)$ and set
\[
 B^\sharp:=\rho_{\beta,K}^{1/4}B\rho_{\beta,K}^{-1/4},\quad
 B^\flat:=\rho_{\beta,K}^{-1/4}B\rho_{\beta,K}^{1/4},\quad
 \widetilde Z:=\rho_{\beta,K}^{1/4}Z\rho_{\beta,K}^{1/4}.
\]
Use the Gibbs relation $\rho_{\beta,K}^aJ_\nu^K\rho_{\beta,K}^{-a}=e^{-a\beta\nu}J_\nu^K$. Applying the Gaussian conjugation identity~\cite[Lemma~IX.2]{chen2025locallymarkovian} with conjugation $e^{\beta K/2}$ and frequency width $1/\beta$ then gives
\begin{align}
 B^\sharp&=e^{itK}\widehat J_K(\omega)e^{-itK},\label{eq:sharp}\\
 B^\flat&=e^{\beta\omega/2+1/4}e^{itK}\widehat J_K(\omega+1/\beta)e^{-itK}.
 \label{eq:flat}
\end{align}
For fixed $t$, write $U=e^{itK}$. Since $\gamma_\beta\le1$ and
\[
\gamma_\beta(\omega)e^{\beta\omega+1/2}
=e^{-\max\{\beta\omega+1/2,0\}+\beta\omega+1/2}
=e^{-\max\{-\beta\omega-1/2,0\}}\le1,
\]
equations~\eqref{eq:sharp}--\eqref{eq:flat} give
\begin{equation}
 \begin{aligned}
 \E_J\int\gamma_\beta(\omega)(B^\sharp)^\dagger B^\sharp\dd\omega
 &\preceq U\left(\E_J\int\widehat J_K(\omega)^\dagger\widehat J_K(\omega)\dd\omega\right)U^\dagger\preceq I,\\
 \E_J\int\gamma_\beta(\omega)B^\flat(B^\flat)^\dagger\dd\omega
 &\preceq U\left(\E_J\int\widehat J_K(\omega+1/\beta)\widehat J_K(\omega+1/\beta)^\dagger\dd\omega\right)U^\dagger\preceq I.
 \end{aligned}
 \label{eq:row-column}
\end{equation}
The final inequalities follow from~\eqref{eq:parseval-base}.
The KMS norm is related to the Hilbert--Schmidt norm by
\[
\wnorm{Z}{K}^2
=\Tr(Z^\dagger\rho_{\beta,K}^{1/2}Z\rho_{\beta,K}^{1/2})
=\wnorm{\widetilde Z}{2}^2.
\]
Moreover, inserting $\rho_{\beta,K}^{-1/4}\rho_{\beta,K}^{1/4}=I$ between the factors gives
\begin{equation}
 \rho_{\beta,K}^{1/4}[B,Z]\rho_{\beta,K}^{1/4}
 =B^\sharp\widetilde Z-\widetilde ZB^\flat.
 \label{eq:commutator-weight}
\end{equation}
By~\eqref{eq:row-column} and cyclicity of the trace,
\begin{align*}
 \E_J\int\gamma_\beta(\omega)\wnorm{B^\sharp\widetilde Z}{2}^2\dd\omega
 &=\Tr\!\left[\widetilde Z^\dagger
 \left(\E_J\int\gamma_\beta(\omega)(B^\sharp)^\dagger B^\sharp\dd\omega\right)
 \widetilde Z\right]\le\wnorm{\widetilde Z}{2}^2,\\
 \E_J\int\gamma_\beta(\omega)\wnorm{\widetilde ZB^\flat}{2}^2\dd\omega
 &=\Tr\!\left[\widetilde Z^\dagger\widetilde Z
 \left(\E_J\int\gamma_\beta(\omega)B^\flat(B^\flat)^\dagger\dd\omega\right)
 \right]\le\wnorm{\widetilde Z}{2}^2.
\end{align*}
Applying $\wnorm{A-C}{2}^2\le2\wnorm A2^2+2\wnorm C2^2$
to~\eqref{eq:commutator-weight} and using
$\wnorm{\widetilde Z}{2}^2=\wnorm ZK^2$ therefore gives
\begin{equation}
 \E_J\int\gamma_\beta(\omega)\wnorm{[B,Z]}K^2\dd\omega\le4\wnorm ZK^2.
 \label{eq:fixed-t}
\end{equation}

For the derivative, differentiating~\eqref{eq:sharp}--\eqref{eq:flat} gives
\begin{align*}
 \partial_\omega B^\sharp
 &=U(\partial_\omega\widehat J_K(\omega))U^\dagger,\\
 \partial_\omega B^\flat
 &=e^{\beta\omega/2+1/4}U
 \left(\partial_\omega\widehat J_K(\omega+1/\beta)+\tfrac\beta2\widehat J_K(\omega+1/\beta)\right)U^\dagger.
\end{align*}
Using $\gamma_\beta\le1$ and~\eqref{eq:parseval-derivative}, we obtain
\[
 \E_J\int\gamma_\beta(\omega)(\partial_\omega B^\sharp)^\dagger
 (\partial_\omega B^\sharp)\dd\omega
 \preceq U\left(\E_J\int(\partial_\omega\widehat J_K)^\dagger
 (\partial_\omega\widehat J_K)\dd\omega\right)U^\dagger
 \preceq\frac{\beta^2}{4}I.
\]
For $B^\flat$, use $\gamma_\beta(\omega)e^{\beta\omega+1/2}\le1$ and
$(A+C)(A+C)^\dagger\preceq2AA^\dagger+2CC^\dagger$, then change variables from $\omega+1/\beta$ to $\omega$. Equations~\eqref{eq:parseval-base}--\eqref{eq:parseval-derivative} give
\begin{align*}
 &\E_J\int\gamma_\beta(\omega)(\partial_\omega B^\flat)
 (\partial_\omega B^\flat)^\dagger\dd\omega\\
 &\qquad\preceq U\left(\E_J\int\left[
 2(\partial_\omega\widehat J_K)(\partial_\omega\widehat J_K)^\dagger
 +\frac{\beta^2}{2}\widehat J_K\widehat J_K^\dagger
 \right]\dd\omega\right)U^\dagger\\
 &\qquad\preceq\left(2\cdot\frac{\beta^2}{4}+\frac{\beta^2}{2}\right)I
 =\beta^2I.
\end{align*}
Since $\rho_{\beta,K}$ and $Z$ are independent of $\omega$, differentiating~\eqref{eq:commutator-weight} gives
\[
 \rho_{\beta,K}^{1/4}[\partial_\omega B,Z]\rho_{\beta,K}^{1/4}
 =(\partial_\omega B^\sharp)\widetilde Z-\widetilde Z(\partial_\omega B^\flat).
\]
Applying the same trace calculation as above to the two derivative bounds yields
\begin{align*}
 \E_J\int\gamma_\beta(\omega)\wnorm{[\partial_\omega B,Z]}K^2\dd\omega
 &\le2\E_J\int\gamma_\beta(\omega)
 \left(\wnorm{(\partial_\omega B^\sharp)\widetilde Z}{2}^2
 +\wnorm{\widetilde Z(\partial_\omega B^\flat)}{2}^2\right)\dd\omega\\
 &\le2\left(\frac{\beta^2}{4}+\beta^2\right)\wnorm{\widetilde Z}{2}^2
 =\frac{5\beta^2}{2}\wnorm ZK^2.
\end{align*}
Now $[\mathsf D_tB,Z]=-[\partial_\omega B,Z]+(it+\beta/4)[B,Z]$, so~\eqref{eq:fixed-t} gives
\begin{align*}
 \E_J\int\gamma_\beta(\omega)\wnorm{[\mathsf D_tB,Z]}K^2\dd\omega
 &\le2\E_J\int\gamma_\beta(\omega)\wnorm{[\partial_\omega B,Z]}K^2\dd\omega\\
 &\quad+2(t^2+\beta^2/16)\E_J\int\gamma_\beta(\omega)\wnorm{[B,Z]}K^2\dd\omega\\
 &\le[5\beta^2+8(t^2+\beta^2/16)]\wnorm ZK^2.
\end{align*}
Finally, the Fourier identity
\[
 \int_\R g_\beta(t)e^{itx}\dd t=\frac1{2\cosh(\beta x/4)}
\]
and its second derivative at $x=0$ give
\[
 \int_\R g_\beta(t)\dd t=\frac12,\qquad
 \int_\R t^2g_\beta(t)\dd t
 =-\left.\frac{\dd^2}{\dd x^2}\frac1{2\cosh(\beta x/4)}\right|_{x=0}
 =\frac{\beta^2}{32}.
\]
Thus
\begin{align*}
 \E_J\int_{\R^2}g_\beta(t)\gamma_\beta(\omega)
 \wnorm{[\mathsf D_tB,Z]}K^2\dd\omega\dd t
 &\le\left[\left(5\beta^2+\frac{\beta^2}{2}\right)\frac12
 +8\frac{\beta^2}{32}\right]\wnorm ZK^2\\
 &=3\beta^2\wnorm ZK^2,
\end{align*}
proving~\eqref{eq:derivative-bound}.
\end{proof}

\section{Polynomial approximation}

\subsection{Polynomial approximation}

For the low-energy band, let $I_\ell=[0,\omega_\ell]$, where
$\omega_\ell=1/\beta$. For every other band $\ell$, let $I_\ell=[\omega_\ell,2\omega_\ell]$,
where $\omega_\ell$ is the lower endpoint of that band. Recall that
\[
u(t)=(1+e^{-\beta t})^{-1/2}\qquad
w(t)=(1+e^{\beta t})^{-1/2}.
\]

For $f\in\{u,w\}$ and $\tau\in\mathbb R$, define the divided
difference
\begin{align}
    h_\tau^f(z)=
\begin{cases}
\dfrac{f(z)-f(\tau)}{z-\tau},&z\neq\tau,\\[1ex]
f'(\tau),&z=\tau.
\end{cases}
\label{eq:divided_diff}
\end{align}

\begin{lemma}[Polynomial approximation on an interval]
\label{lem:polynomial-approximation}
Let $I=[a,b]\subseteq\mathbb R$ with $a<b$, and let
$f\in C^4(I)$. For every integer $k\geq0$, there exists a
polynomial $q_k^f$ of degree at most $k$ such that
\begin{equation}
\sup_{z\in I}|f(z)-q_k^f(z)|
\leq
\frac{8}{3(k+1)^3}
\sum_{j=1}^4
(b-a)^j\sup_{z\in I}|f^{(j)}(z)|.
\label{eq:polynomial-approximation}
\end{equation}
The numerical constant $8/3$ is universal; all dependence on
the interval $I$ and the function $f$ is displayed explicitly
on the right-hand side.
\end{lemma}

\begin{proof}
Put
\[
c=\frac{a+b}{2},
\qquad
r=\frac{b-a}{2},
\qquad
\widetilde f(\theta)=f(c+r\cos\theta).
\]
Define
\[
A_j=\frac{2}{\pi}\int_0^\pi
\widetilde f(\theta)\cos(j\theta)\,d\theta, \qquad
q_k^f(z)
=
\frac{A_0}{2}
+
\sum_{j=1}^k
A_jT_j\left(\frac{z-c}{r}\right),
\]
where $T_j$ is the degree-$j$ Chebyshev polynomial. Thus
$q_k^f$ has degree at most $k$.

For $j\geq1$, four integrations by parts give
\[
A_j
=
\frac{2}{\pi j^4}
\int_0^\pi
\widetilde f^{(4)}(\theta)\cos(j\theta)\,d\theta.
\]
The boundary terms vanish because
$\widetilde f'(0)=\widetilde f'(\pi)
=\widetilde f'''(0)=\widetilde f'''(\pi)=0.$
It follows that
\[
|A_j|
\leq
\frac{2}{j^4}
\sup_{\theta\in[0,\pi]}
|\widetilde f^{(4)}(\theta)|.
\]
Writing $z=c+r\cos\theta$, direct differentiation gives
\begin{align*}
\widetilde f^{(4)}(\theta)
={}&
r\cos\theta\,f'(z)
+r^2(3\cos^2\theta-4\sin^2\theta)f''(z)-6r^3\sin^2\theta\cos\theta\,f^{(3)}(z)
+r^4\sin^4\theta\,f^{(4)}(z).
\end{align*}
Since
$|3\cos^2\theta-4\sin^2\theta|\leq4,$
we obtain
\[
\begin{aligned}
\sup_\theta|\widetilde f^{(4)}(\theta)|
\leq{}&
r\sup_I|f'|
+4r^2\sup_I|f''|
+6r^3\sup_I|f^{(3)}|
+r^4\sup_I|f^{(4)}|\\
\leq{}&
\sum_{p=1}^4
(b-a)^p\sup_{z\in I}|f^{(p)}(z)|,
\end{aligned}
\]
where the second inequality uses \(r=(b-a)/2\).
Since $|T_j(x)|\leq1$ for $x\in[-1,1]$, we have
\[
\sup_{z\in I}|f(z)-q_k^f(z)|
\leq
\sum_{j=k+1}^\infty |A_j|.
\]
Moreover,
$\sum_{j=k+1}^\infty\frac1{j^4}
\leq
\frac{4}{3(k+1)^3}.$
Combining these estimates gives
\[
\begin{aligned}
\sup_{z\in I}|f(z)-q_k^f(z)|
\leq
2\sup_\theta|\widetilde f^{(4)}(\theta)|
\sum_{j=k+1}^\infty\frac1{j^4}\leq
\frac{8}{3(k+1)^3}
\sum_{p=1}^4
(b-a)^p\sup_{z\in I}|f^{(p)}(z)|.
\end{aligned}
\]
\end{proof}

\begin{corollary}[Polynomial approximation of thermal divided differences]
\label{cor:thermal-polynomial-approximation}
For every band $\ell$, every integer $k\geq0$, every
$f\in\{u,w\}$, and every $\tau\in\mathbb R$, define the divided
difference $h_\tau^f(z)$ as in Eq.~\eqref{eq:divided_diff}.
Then there exists a polynomial $q_{\tau,k}^f$ of degree at most
$k$ such that
\begin{equation}
\sup_{z\in I_\ell}
\left|
h_\tau^f(z)-q_{\tau,k}^f(z)
\right|
\leq
\frac{100}{\omega_\ell(k+1)^3}.
\label{eq:divided-difference-approximation}
\end{equation}
The numerical constant $100$ is universal; in particular, it is
independent of $\beta$, $\ell$, $k$, $\tau$, and
$f\in\{u,w\}$.
\end{corollary}

\begin{proof}
We show that, for every $f\in\{u,w\}$ and every $\tau\in\mathbb R$,
\begin{equation}
\sup_{z\in I_\ell}
\left|\frac{d^j}{dz^j}h_\tau^f(z)\right|
\leq \frac{j!}{\omega_\ell^{j+1}},
\qquad 0\leq j\leq4.
\label{eq:divided-difference-derivative-bound}
\end{equation}

First consider the low band, where $\omega_\ell=1/\beta$.
Both $u$ and $w$ extend holomorphically to a neighborhood of the
strip $|\operatorname{Im}z|\leq1/\beta$ and have modulus at most
one there, since $\operatorname{Re}(1+e^{\pm\beta z})\geq1$ on
the strip. Cauchy's estimate on the circle of radius $1/\beta$
about any real point $y$ therefore gives
\[
|f^{(j+1)}(y)|\leq (j+1)!\,\beta^{j+1}.
\]
Using the integral formula for the divided difference, including
at $z=\tau$, we obtain
\[
\left|\frac{d^j}{dz^j}h_\tau^f(z)\right|
=
\left|\int_0^1 s^j
f^{(j+1)}\bigl(\tau+s(z-\tau)\bigr)\,ds\right|
\leq j!\,\beta^{j+1}.
\]
This proves \eqref{eq:divided-difference-derivative-bound} on the
low band.

Now consider a dyadic band. We will prove the stronger bound
$|(h_\tau^f)^{(j)}(z)|\leq j!/z^{j+1}$ for every $z>0$.
For $0\leq p\leq1$, put $L_p(z)=(1+p e^{-\beta z})^{-1}$.
The substitution $v=\tan\theta$ gives
\[
u(z)=\frac{2}{\pi}\int_0^{\pi/2}L_{\sin^2\theta}(z)\,d\theta.
\]
Write $h_\tau^{L_p}$ for the divided difference of $L_p$.
For $z>0$, expand $L_p(z)$ as a geometric series and use
\[
\frac{e^{-\beta\tau}-e^{-\beta z}}{z-\tau}
=\int_0^\beta e^{-sz-(\beta-s)\tau}\,ds,
\]
with the continuous value at $z=\tau$. This gives
\[
h_\tau^{L_p}(z)
=\sum_{n=0}^{\infty}\int_0^\beta
e^{-(n\beta+s)z}(-p)^n
\frac{p\,e^{-(\beta-s)\tau}}{1+p e^{-\beta\tau}}\,ds.
\]
The coefficient of the exponential has absolute value at most one,
because
\[
p\,e^{-(\beta-s)\tau}
\leq \max\{p,p e^{-\beta\tau}\}
\leq 1+p e^{-\beta\tau}
\qquad (0\leq s\leq\beta).
\]
Thus the series is absolutely convergent. Setting $r=n\beta+s$
and averaging over $\theta$ gives a representation
\[
h_\tau^u(z)=\int_0^\infty e^{-rz}K_\tau^u(r)\,dr,
\qquad |K_\tau^u(r)|\leq1.
\]

Since $w(z)=e^{-\beta z/2}u(z)$, its divided difference satisfies
\[
h_\tau^w(z)
=e^{-\beta z/2}h_\tau^u(z)
-\int_0^{\beta/2}e^{-rz}u(\tau)e^{-(\beta/2-r)\tau}\,dr.
\]
The first term shifts the kernel $K_\tau^u$ to $[\beta/2,\infty)$.
On $[0,\beta/2]$, the kernel in the second term has absolute value
at most $\max\{u(\tau),w(\tau)\}\leq1$. Hence $h_\tau^w$ also has
a Laplace integral representation with a kernel bounded by one.
For either $f=u$ or $f=w$, differentiation under the integral gives
\[
\left|\frac{d^j}{dz^j}h_\tau^f(z)\right|
\leq\int_0^\infty r^j e^{-rz}\,dr
=\frac{j!}{z^{j+1}}.
\]
Since $z\geq\omega_\ell$ on a dyadic band, this proves
\eqref{eq:divided-difference-derivative-bound} in the remaining case.

Finally, apply Lemma~\ref{lem:polynomial-approximation} to
$h_\tau^f$ on $I_\ell$. Since $|I_\ell|\leq\omega_\ell$, there
exists a polynomial $q_{\tau,k}^f$ of degree at most $k$ such that
\[
\sup_{z\in I_\ell}|h_\tau^f(z)-q_{\tau,k}^f(z)|
\leq \frac{8}{3\omega_\ell(k+1)^3}\sum_{j=1}^4 j!
=\frac{88}{\omega_\ell(k+1)^3}
<\frac{100}{\omega_\ell(k+1)^3}.
\]
\end{proof}

\begin{lemma}[Integrated polynomial approximation of thermal sources]
\label{lem:integrated-source-polynomial-approximation}
Define
\[
g_t(z)=e^{-tz}u(z),
\qquad 0\leq t\leq\beta/2.
\]
For every band $\ell$ and every integer $k\geq0$, there exists,
for each $t\in[0,\beta/2]$, a polynomial $q_{t,k}$ of degree at
most $k$ such that
\begin{equation}
\int_0^{\beta/2}
\sup_{z\in I_\ell}
|g_t(z)-q_{t,k}(z)|\,dt
\leq
\frac{100}{\omega_\ell(k+1)^3}.
\label{eq:integrated-source-approximation}
\end{equation}
The numerical constant $100$ is universal; in particular, it is
independent of $\beta$, $\ell$, $k$, and the interval $I_\ell$.
\end{lemma}

\begin{proof}
We first prove that, for $1\leq j\leq4$,
\begin{equation}
\int_0^{\beta/2}
\sup_{z\in I_\ell}|g_t^{(j)}(z)|\,dt
\leq
\frac{C_j}{\omega_\ell^{j+1}},
\qquad
C_j\leq j!+1.
\label{eq:integrated-source-derivative-bound}
\end{equation}

Consider a dyadic band. By Leibniz's rule,
\[
g_t^{(j)}(z)
=
\sum_{p=0}^j
\binom jp
(-t)^p e^{-tz}u^{(j-p)}(z).
\]

Direct differentiation of $u(z)=(1+e^{-\beta z})^{-1/2}$ gives
\[
|u^{(s)}(z)|\leq \beta^s e^{-\beta z},
\qquad z\geq 0,\quad 1\leq s\leq 4.
\]
On a dyadic band, $z\geq\omega_\ell$. Using $u(z)\leq 1$ for
the term $p=j$ and the derivative bound for $p<j$, we obtain
\[
\sup_{z\in I_\ell}|g_t^{(j)}(z)|
\leq t^j e^{-t\omega_\ell}
+(t+\beta)^j e^{-(t+\beta)\omega_\ell}.
\]
Integrating the two terms over $0\leq t\leq\beta/2$ gives
integrals of $s^j e^{-s\omega_\ell}$ over the disjoint intervals
$[0,\beta/2]$ and $[\beta,3\beta/2]$. Their sum is therefore at most
\[
\int_0^\infty s^j e^{-s\omega_\ell}\,ds
=\frac{j!}{\omega_\ell^{j+1}}.
\]

On the low band, Leibniz's rule gives
$|g_t^{(j)}(z)|\leq(t+\beta)^j\leq(3\beta/2)^j$.
Since $\omega_\ell=1/\beta$ and $(3/2)^j/2\leq j!$ for
$1\leq j\leq4$, integration over $0\leq t\leq\beta/2$
gives the same bound. Thus \eqref{eq:integrated-source-derivative-bound}
holds on every band with $C_j=j!$.

For each fixed $t$, apply
Lemma~\ref{lem:polynomial-approximation} to $g_t$ and denote the
resulting polynomial by $q_{t,k}$. It has degree at most $k$ and
satisfies
\[
\sup_{z\in I_\ell}|g_t(z)-q_{t,k}(z)|
\leq
\frac{8}{3(k+1)^3}
\sum_{j=1}^4
|I_\ell|^j
\sup_{z\in I_\ell}|g_t^{(j)}(z)|.
\]
Integrating this inequality, using
\eqref{eq:integrated-source-derivative-bound} and
$|I_\ell|\leq\omega_\ell$, gives
\[
\begin{aligned}
\int_0^{\beta/2}
\sup_{z\in I_\ell}|g_t(z)-q_{t,k}(z)|\,dt
&\leq
\frac{8}{3(k+1)^3}
\sum_{j=1}^4
|I_\ell|^j
\int_0^{\beta/2}
\sup_{z\in I_\ell}|g_t^{(j)}(z)|\,dt\\
&\leq
\frac{8}{3\omega_\ell(k+1)^3}
\sum_{j=1}^4 C_j\\
&\leq
\frac{296}{3\omega_\ell(k+1)^3}\\
&<
\frac{100}{\omega_\ell(k+1)^3}.
\end{aligned}
\]
\end{proof}

\subsection{Some simple polynomial approximation}

\begin{lemma}[Exponential approximation of the thermal functions]
\label{lem:thermal-function-exponential-approximation}
For every band $I_\ell$, every $f\in\{u,w\}$, and every integer
$k\geq0$, there exists a polynomial $q_{\ell,k}^f$ of degree at most
$k$ such that
\begin{equation}
\sup_{t\in I_\ell}
|f(t)-q_{\ell,k}^f(t)|
\leq C2^{-k},
\label{eq:thermal-function-exponential-approximation}
\end{equation}
where $C$ is a numerical constant independent of $\beta$, $\ell$,
$k$, and $f$.
\end{lemma}

\begin{proof}
First consider a dyadic band $I_\ell=[\omega_\ell,2\omega_\ell]$ and write
$c_\ell=\frac32 \omega_\ell$. We use
\[
u(t)=\sum_{j\geq0}
(-1)^j 4^{-j}\binom{2j}{j}e^{-j\beta t},
\qquad
w(t)=\sum_{j\geq0}
(-1)^j 4^{-j}\binom{2j}{j}e^{-(j+1/2)\beta t}.
\]
Using these expansions and
$4^{-j}\binom{2j}{j}\leq1$, for every $s\geq1$ we have

\begin{align}
\frac{|u^{(s)}(c_\ell)|}{s!}
&\leq
\omega_\ell^{-s}
\sum_{j\geq1}
\frac{(j\beta\omega_\ell)^s}{s!}
e^{-\frac32 j \beta\omega_\ell}\nonumber\\
&\leq
\omega_\ell^{-s}
\sum_{j \geq1}e^{-\frac12 j\beta\omega_\ell}
\leq 2\omega_\ell^{-s}, \label{eq:u_derivative}
\end{align}

where 
the last inequality follows from $\beta\omega_\ell\geq1$. The same
argument, with $j$ in the formulas replaced by $j+\frac12$, gives 
$\frac{|w^{(s)}(c_\ell)|}{s!}
\leq 2\omega_\ell^{-s}.$
The claim for $s=0$ follows from $|u|,|w|\leq1$.

Define the degree-$k$ polynomial
\[
q_{\ell,k}^f(t)
:=
\sum_{s=0}^k
\frac{f^{(s)}(c_\ell)}{s!}(t-c_\ell)^s.
\]
Since $|t-c_\ell|\leq \omega_\ell/2$ for
$t\in[\omega_\ell,2\omega_\ell]$, Eq.~(\ref{eq:u_derivative}) and the Taylor-series remainder give
\[
|f(t)-q_{\ell,k}^f(t)|
\leq
2\sum_{s=k+1}^{\infty}
\left(\frac{|t-c_\ell|}{\omega_\ell}\right)^s
\leq
2\sum_{s=k+1}^{\infty}2^{-s}
\leq 2\cdot2^{-k}.
\]

It remains to consider the low-energy band
$t\in I_0=[0,1/\beta]$. Set $x=\beta t$. The Taylor expansions about
$x=0$ of
$(1+e^{-x})^{-1/2}$
and
$(1+e^x)^{-1/2}$
converge for $|x|\leq2$. Therefore, their degree-$k$ truncations have
error at most $C2^{-k}$ for $x\in[0,1]$. Replacing $x$ by $\beta t$
gives the required polynomials on $I_0$.
\end{proof}

\begin{lemma}\label{lem:spectral_norm_test}
Let $\mathcal H=\bigoplus_{d\geq0}\mathcal H_d$, and let $P_d$ be the
projection onto $\mathcal H_d$. If
\[
\|P_dAP_{d'}\|\leq C_0 \,2^{-|d-d'|/2},
\] 
for some constant $C_0$, then there exists another constant $C$ such that $\|A\|\leq C$.  \end{lemma}

\begin{proof}
Consider a vector $\phi=\sum_{d'}\phi_{d'}$ with $\phi_{d'}=P_d' \phi$. The assumption gives
\[
\|A\phi\|^2
\leq C_0^2\sum_d
\left(\sum_{d'}2^{-|d-d'|/2}\|\phi_{d'}\|\right)^2
\leq
C_0 ^2\left(\sum_{r\in\mathbb Z}2^{-|r|/2}\right)^2\|\phi\|^2,
\]
where the last inequality follows from Cauchy--Schwarz.
\end{proof}

\section{Bandwise decay of thermal boundary sources}

\subsection{Decay bound}

\begin{corollary}[Divided-difference source decay]
\label{cor:divided-difference-source-decay} 
For every band $\ell$, depth $d\geq1$, $f\in\{u,w\}$, and
$\tau\in\mathbb R$,
\begin{equation}
\left\|
\left.T_\ell h_\tau^f(E_\ell)\right|_{W_{\ell,d}}
\right\|
\leq
\frac{C}{(d+1)^3}.
\label{eq:divided-difference-source-decay}
\end{equation}
The constant $C$ is universal; in particular, it is independent of
$\beta,\ell,d,\tau$, and $f\in\{u,w\}$.
\end{corollary}

\begin{proof}
Suppose first that $d\geq2$. By
Corollary~\ref{cor:thermal-polynomial-approximation}, applied with
$k=d-2$, there exists a polynomial $q_{\tau,d-2}^f$ of degree at
most $d-2$ such that
\[
\sup_{z\in I_\ell}
\left|
h_\tau^f(z)-q_{\tau,d-2}^f(z)
\right|
\leq
\frac{100}{\omega_\ell(d-1)^3}.
\]
The Krylov cancellation \eqref{eq:krylov} gives
\[
T_\ell q_{\tau,d-2}^f(E_\ell)v=0,
\qquad v\in W_{\ell,d}.
\]
Consequently, for every unit vector $v\in W_{\ell,d}$,
\[
\begin{aligned}
\|T_\ell h_\tau^f(E_\ell)v\|
&=
\left\|
T_\ell
\bigl(h_\tau^f(E_\ell)-q_{\tau,d-2}^f(E_\ell)\bigr)v
\right\|\\
&\leq
\|T_\ell\|
\sup_{z\in I_\ell}
\left|h_\tau^f(z)-q_{\tau,d-2}^f(z)\right|\leq
\frac{200}{(d-1)^3}
\leq
\frac{C}{(d+1)^3},
\end{aligned}
\]
where we used $\|T_\ell\|\leq2\omega_\ell$ and
$(d+1)/(d-1)\leq3$ for $d\geq2$.
For $d=1$, the direct estimates
$\|T_\ell\|\leq2\omega_\ell,$
$\|h_\tau^f(E_\ell)\|\leq\frac{2}{\omega_\ell}$
give
$\|T_\ell h_\tau^f(E_\ell)\|\leq4,$
which is also bounded by $C/(d+1)^3$ for a sufficiently large
universal $C$.
\end{proof}

\begin{corollary}[Integrated thermal-source decay]
\label{cor:integrated-source-decay}
For every band $\ell$ and depth $d\geq1$,
\begin{equation}
\sum_{v_r\in W_{\ell,d}}
\left(
\int_0^{\beta/2}
\left\|
T_\ell e^{-tE_\ell}u(E_\ell)v_r
\right\|\,dt
\right)^2
\leq
\frac{Cr_\ell}{(d+1)^6}.
\label{eq:integrated-source-decay}
\end{equation}
The sum is over any orthonormal basis of $W_{\ell,d}$.
The constant $C$ is universal; in particular, it is independent
of $\beta,\ell$, and $d$.
\end{corollary}

\begin{proof}
Let $V$ be the isometry whose columns form the chosen orthonormal
basis of $W_{\ell,d}$, and write
$g_t(z)=e^{-tz}u(z)$. Minkowski's inequality gives
\begin{equation}
\begin{aligned}
\left[
\sum_{v_r\in W_{\ell,d}}
\left(
\int_0^{\beta/2}
\|T_\ell g_t(E_\ell)v_r\|\,dt
\right)^2
\right]^{1/2}\leq
\int_0^{\beta/2}
\|T_\ell g_t(E_\ell)V\|_{\mathrm F}\,dt.
\end{aligned}
\label{eq:integrated-source-minkowski}
\end{equation}
where $ \|A\|_{\mathrm F}$ is the Frobenius norm. 

Suppose first that $d\geq2$. By
Lemma~\ref{lem:integrated-source-polynomial-approximation}, applied
with $k=d-2$, there exists, for every $t\in[0,\beta/2]$, a
polynomial $q_{t,d-2}$ of degree at most $d-2$ such that
\[
\int_0^{\beta/2}
\sup_{z\in I_\ell}
|g_t(z)-q_{t,d-2}(z)|\,dt
\leq
\frac{100}{\omega_\ell(d-1)^3}.
\]
The Krylov cancellation \eqref{eq:krylov} gives
\[
T_\ell q_{t,d-2}(E_\ell)V=0.
\]
Note that for any operator $A$, we have $\|A\|_{\mathrm{F}}^2\leq rank(A) \|A\|^2$.
Since the rank of
$T_\ell(g_t(E_\ell)-q_{t,d-2}(E_\ell))V$
is at most $r_\ell$, we obtain
\[
\begin{aligned}
\int_0^{\beta/2}
\|T_\ell g_t(E_\ell)V\|_{\mathrm F}\,dt
&=
\int_0^{\beta/2}
\left\|
T_\ell
\bigl(g_t(E_\ell)-q_{t,d-2}(E_\ell)\bigr)V
\right\|_{\mathrm F}\,dt\\
&\leq
\sqrt{r_\ell}\,\|T_\ell\|
\int_0^{\beta/2}
\sup_{z\in I_\ell}
|g_t(z)-q_{t,d-2}(z)|\,dt\\
&\leq
\frac{200\sqrt{r_\ell}}{(d-1)^3},
\end{aligned}
\]
where we used $\|T_\ell\|\leq2\omega_\ell$.
For $d=1$, we use directly
\begin{align}
    \int_0^{\beta/2}
\|T_\ell g_t(E_\ell)V\|_{\mathrm F}\,dt
&\leq
\sqrt{r_\ell}\,\|T_\ell\|
\int_0^{\beta/2}
\|e^{-tE_\ell}u(E_\ell)\|\,dt \\
&\leq 2\omega_\ell\sqrt{r_\ell}
\int_0^\infty e^{-t\omega_\ell}\,dt
=
2\sqrt{r_\ell}
\end{align}
Combining these estimates with
$(d+1)/(d-1)\leq3$ for $d\geq2$, we conclude that
\[
\int_0^{\beta/2}
\|T_\ell g_t(E_\ell)V\|_{\mathrm F}\,dt
\leq
\frac{\sqrt{Cr_\ell}}{(d+1)^3}
\]
for every $d\geq1$, with a universal $C$.
Substituting into \eqref{eq:integrated-source-minkowski} and
squaring proves \eqref{eq:integrated-source-decay}.
\end{proof}

\subsection{A dominating Gaussian reference state}\label{app:dominate_Gaussian_State}

 \textbf{Notation.}
 Fix residual basis vectors $v_{r_1},\ldots,v_{r_k}$ with
$d_{r_i}\geq1$. For each $i$, let $\ell_i$ be the band index such
that
$v_{r_i}\in W_{\ell_i,d_{r_i}},$
and independently choose the  index
$j_i\in\{r_i,n+r_i\}.$
For $0\leq t\leq\beta/2$, define
\begin{align}
	O_i(t) &=
\begin{cases}
a(e^{-tE_{\ell_i}}u(E_{\ell_i})v_{r_i}), & j_i=r_i,\\
a^\dagger(e^{-(\beta/2-t)E_{\ell_i}}u(E_{\ell_i})v_{r_i}),
    & j_i=n+r_i,
\end{cases}
\\
C_i(t)
&:=[H,O_i(t)]-\frac{d}{dt}O_i(t)
=
\begin{cases}
-b(T_{\ell_i}e^{-tE_{\ell_i}}u(E_{\ell_i})v_{r_i}),
    & j_i=r_i,\\
b^\dagger(T_{\ell_i}e^{-(\beta/2-t)E_{\ell_i}}
u(E_{\ell_i})v_{r_i}), & j_i=n+r_i,
\end{cases}
\end{align}

where the second equality follows from \eqref{eq:boundary}.
For each $i=1,\ldots,k$ and each permutation
$\pi\in S_i$, let
\begin{align}
\Delta_\pi
&=
\left\{
(t_1,\ldots,t_i):
0<t_{\pi(1)}<\cdots<t_{\pi(i)}<\beta/2
\right\},\\
X_\pi(t_1,\ldots,t_i)
=&e^{-t_{\pi(1)}H}
C_{\pi(1)}(t_{\pi(1)})
e^{-(t_{\pi(2)}-t_{\pi(1)})H}
\cdots
C_{\pi(i)}(t_{\pi(i)})
e^{-(\beta/2-t_{\pi(i)})H}.
\end{align}

\begin{lemma}[Time-ordered expansion]
\label{lem:thermal-annihilation-expansion}
For every $i=1,\ldots,k$,
\begin{equation}
\widehat c_{j_i}\cdots\widehat c_{j_1}(e^{-\beta H/2})
=
\sum_{\pi\in S_i}\varepsilon_\pi
\int_{\Delta_\pi}
X_\pi(t_1,\ldots,t_i)\,dt_1\cdots dt_i,
\label{eq:thermal-annihilation-expansion}
\end{equation}
where each $\varepsilon_\pi\in\{-1,1\}$.
\end{lemma}

\begin{proof}
Since $E_{\mathrm{ref}}$ equals $E_{\ell_i}$ on
$W_{\ell_i}\ominus W_{\ell_i,0}$, we have
\[
u(E_{\ell_i})v_{r_i}=uv_{r_i},
\qquad
e^{-\beta E_{\ell_i}/2}u(E_{\ell_i})v_{r_i}=wv_{r_i}.
\]
Evaluating $O_i$ at $0$ and $\beta/2$ and using
\eqref{eq:thermal_anni} therefore gives
\[
\widehat c_{j_i}(X)
=
O_i(0)X-(PXP)O_i(\beta/2).
\]
This identity covers both choices
$j_i=r_i$ and $j_i=n+r_i$.

For fixed $0\leq t\leq t'\leq\beta/2$, differentiating
$-e^{-(\tau-t)H}O_i(\tau)e^{-(t'-\tau)H}$ with respect to $\tau$
and integrating over $\tau\in[t,t']$ gives
\begin{align}
	O_i(t)e^{-(t'-t)H}-e^{-(t'-t)H}O_i(t')
=
\int_t^{t'}
e^{-(\tau-t)H}C_i(\tau)e^{-(t'-\tau)H}\,d\tau.\label{eq:xxx}
\end{align}

We prove \eqref{eq:thermal-annihilation-expansion} by induction on
$i$. For $i=0$, the right-hand side is understood as
$e^{-\beta H/2}$. Suppose the formula holds for $i-1$. By linearity and the definition
of $\widehat c_{j_i}$,

\begin{align}
\widehat c_{j_i}\cdots\widehat c_{j_1}(e^{-\beta H/2})
=
\sum_{\pi\in S_{i-1}}\varepsilon_\pi
\int_{\Delta_\pi}
\left(
O_i(0)X_\pi(t_1,\ldots,t_{i-1})
-
(-1)^{i-1}
X_\pi(t_1,\ldots,t_{i-1})O_i(\beta/2)
\right)
\,dt_1\cdots dt_{i-1},\label{eq:reduction}
\end{align}
where we used
$PX_\pi(t_1,\ldots,t_{i-1})P
=
(-1)^{i-1}X_\pi(t_1,\ldots,t_{i-1}).$

Define \(t_{\pi(0)}=0\) and \(t_{\pi(i)}=\beta/2\).
Applying the preceding
integral identity Eq.~(\ref{eq:xxx}) to the $i$ exponential factors in $X_\pi$ gives
\begin{align}
&O_i(0)X_\pi(t_1,\ldots,t_{i-1})
-(-1)^{i-1}X_\pi(t_1,\ldots,t_{i-1})O_i(\beta/2)\nonumber\\
&=
\sum_{q=0}^{i-1}(-1)^q
\int_{t_{\pi(q)}}^{t_{\pi(q+1)}}
\left[
\prod_{a=1}^{q}
e^{-(t_{\pi(a)}-t_{\pi(a-1)})H}
C_{\pi(a)}(t_{\pi(a)})
\right]\times
e^{-(t_i-t_{\pi(q)})H}\,C_i(t_i)\,
e^{-(t_{\pi(q+1)}-t_i)H}\nonumber\\[-2pt]
&\qquad\qquad{}\times
\left[
\prod_{a=q+1}^{i-1}
C_{\pi(a)}(t_{\pi(a)})
e^{-(t_{\pi(a+1)}-t_{\pi(a)})H}
\right]\,dt_i.
\end{align}
Here we use the fact that $O_i(t)$ anticommutes with
every $C_j(t')$ by the CAR.
Substituting this identity into Eq.~(\ref{eq:reduction}) and regrouping the pairs
$(\pi,q)$ gives
\begin{align}
	\widehat c_{j_i}\cdots\widehat c_{j_1}(e^{-\beta H/2})
=
\sum_{\sigma\in S_i}\varepsilon_\sigma
\int_{\Delta_\sigma}
X_\sigma(t_1,\ldots,t_i)\,dt_1\cdots dt_i,
\end{align}

because every $\sigma\in S_i$ is obtained from a unique pair
$(\pi,q)$. This proves the induction step.
\end{proof}

\begin{lemma}[Joint thermal occupation bound]
\label{lem:multiparticle-pull-through}
With the notation at the beginning of Section \ref{app:dominate_Gaussian_State},
\begin{equation}
\left\|
\widehat c_{j_k}\cdots\widehat c_{j_1}(\sqrt{\rho_\beta})
\right\|_2
\leq
\prod_{i=1}^k
\int_0^{\beta/2}
\left\|
T_{\ell_i}e^{-tE_{\ell_i}}u(E_{\ell_i})v_{r_i}
\right\|\,dt.
\label{eq:multiparticle-pull-through}
\end{equation}
\end{lemma}
\begin{proof}
This is the joint occupation bound of
\cite[Lemma~4.2]{arunachalam2026quantum}, specialized to our residual
modes. We give the proof in our notation.

Apply Lemma~\ref{lem:thermal-annihilation-expansion} with $i=k$.
For $(t_1,\ldots,t_k)\in\Delta_\pi$, use $t_{\pi(0)}=0$ and
$t_{\pi(k+1)}=\beta/2$. Since
\[
\sum_{a=0}^k \left(t_{\pi(a+1)}-t_{\pi(a)}\right)/{\beta}
=\frac12,
\]
Schatten H\"older gives
\begin{align*}
\|X_\pi(t_1,\ldots,t_k)\|_2
&\leq
\prod_{a=0}^k
\left\|e^{-(t_{\pi(a+1)}-t_{\pi(a)})H}\right\|_{
\beta/(t_{\pi(a+1)}-t_{\pi(a)})}
\prod_{i=1}^k\|C_i(t_i)\|
&=
\|e^{-\beta H/2}\|_2
\prod_{i=1}^k\|C_i(t_i)\|.
\end{align*}
In the last equality, each exponential contributes a factor
$\bigl(\Tr(e^{-\beta H})\bigr)^{(t_{\pi(a+1)}-t_{\pi(a)})/\beta}$,
and the powers sum to $1/2$.

Using $\sqrt{\rho_\beta}=e^{-\beta H/2}/\|e^{-\beta H/2}\|_2$,  Lemma~\ref{lem:thermal-annihilation-expansion}, 
and the triangle inequality, we obtain
\begin{align*}
\left\|
\widehat c_{j_k}\cdots\widehat c_{j_1}(\sqrt{\rho_\beta})
\right\|_2
&\leq
\sum_{\pi\in S_k}\int_{\Delta_\pi}
\prod_{i=1}^k\|C_i(t_i)\|\,dt_1\cdots dt_k\\
&=
\int_{[0,\beta/2]^k}
\prod_{i=1}^k\|C_i(t_i)\|\,dt_1\cdots dt_k=
\prod_{i=1}^k\int_0^{\beta/2}\|C_i(t)\|\,dt.
\end{align*}
Finally, the CAR give $\|b(y)\|=\|b^\dagger(y)\|=\|y\|$.
Taking norms in the formula for $C_i(t)$ and changing variables
$t\mapsto\beta/2-t$ when $j_i=n+r_i$ gives
\eqref{eq:multiparticle-pull-through}.
\end{proof}

\section{Details of comparing physical overlap to thermal overlap}\label{app:thermal_physics_compare}

\subsection{Comparing physical operators to thermal operators}

In this section, we connect the overlap in the physical coordinate to the overlap in the thermal coordinate, as stated in Lemma \ref{lem:weighted-one-step-comparison} and Lemma \ref{lem:physical-to-thermal-word-comparison}.
Recall that
\[
p_{n+r}=p_r,\qquad 1\leq r\leq n.
\]
Let $D$ be the diagonal operator on the physical single-particle space
defined by
\[
Dv_r=p_r^{-1/2}v_r.
\]

\begin{lemma}[Depth-weighted stability of the thermal functions]
\label{lem:thermal-depth-stability}
There exists a  constant $C$ such that, for every
$f\in\{u,w\}$,
\begin{equation}
\|D^{-1}f(E_{\mathrm{ref}})D\|
+
\|Df(E_{\mathrm{ref}})D^{-1}\|
\leq C.
\label{eq:thermal-depth-weight}
\end{equation}
\end{lemma}

\begin{proof}
Let $\Pi_{\ell,d}$ denote the projection onto $W_{\ell,d}$. Since
$E_\ell$ is block tridiagonal in the depth decomposition, a
polynomial of degree $k$ changes the depth by at most $k$. 
For $d\neq d'$,
Lemma~\ref{lem:thermal-function-exponential-approximation}, applied with
$k=|d-d'|-1$, shows that there exists a constant $C_0$ such that 
\[
\|\Pi_{\ell,d}f(E_\ell)\Pi_{\ell,d'}\|
\leq
C_0 2^{-|d-d'|}.
\]
The case $d=d'$ follows directly from $\|f\|_\infty\leq1$.

Recall that $p_r=2^{-(d_r+1)}$, the restriction of $D$ to
$W_{\ell,d}$ corresponds to multiplication by $2^{(d+1)/2}$. Consequently,
\[
\left\|
\Pi_{\ell,d}D^{\pm1}f(E_\ell)D^{\mp1}\Pi_{\ell,d'}
\right\|
\leq
C_0 2^{-|d-d'|/2}.
\]

Then, by Lemma \ref{lem:spectral_norm_test}, there exists another constant $C$ such that
\[
\|D^{-1}f(E_\ell)D\|
+
\|Df(E_\ell)D^{-1}\|
\leq C
\]
uniformly in $\ell$. Finally, we conclude the proof because $E_{\mathrm{ref}}$ is block diagonal
over the bands and is zero on the boundary.
\end{proof}

\weightedcomparison*
\begin{proof}[Proof of Lemma \ref{lem:weighted-one-step-comparison}]
Fix $f\in\{u,w\}$, write $F=f(E_{\mathrm{ref}})$ and $z_r=\Tr\!\left(A^\dagger a(v_r)B\right)$. Since $F$ is Hermitian, one can check that
\[
 \Tr\!\left(A^\dagger a(Fv_j)B\right) 
=
( F^Tz)_j.
\]

Therefore, by Lemma~\ref{lem:thermal-depth-stability}, there exists a constant $C'$ such that

\begin{align}
\sum_{j=1}^np_j
\left|
\Tr\!\left(A^\dagger a(Fv_j)B\right)
\right|^2
=
\|D^{-1} F^Tz\|^2
\leq
C'\|D^{-1}z\|^2=
C'\sum_{r=1}^np_r
\left|
\Tr\!\left(A^\dagger a(v_r)B\right)
\right|^2,\label{eq:align}
\end{align}

The same argument applies to $a^\dagger$. 
Because $B$ has definite fermion parity, $PBP=\pm B$.  Recall that   $p_{n+r}=p_r$. Expanding the
definitions of $\widehat c_j$ and $\widehat c_j^\dagger$ as in Eq.~(\ref{eq:thermal_anni}) and using the triangle inequality, and then applying the
estimates in Eq.~(\ref{eq:align}), for some constant $C''$, we obtain

\begin{align}
\sum_{j=1}^{2n}p_j
\left(
|\langle A,\widehat c_j(B)\rangle|^2
+
|\langle A,\widehat c_j^\dagger(B)\rangle|^2
\right)
\leq
C''\sum_{r=1}^np_r
\sum_{\mathsf{a}\in\{a,a^\dagger\}}
\left(
|\Tr(A^\dagger \mathsf{a}(v_r)B)|^2
+
|\Tr(A^\dagger B \mathsf{a}(v_r))|^2
\right).
\end{align}

Finally, using
$a(v_r)=(\gamma_{2r-1}+i\gamma_{2r})/2,$ and
$a^\dagger(v_r)=(\gamma_{2r-1}-i\gamma_{2r})/2,$
we prove the lemma.
\end{proof}

\subsection{Comparing physical global operators to thermal global operators}

\begin{lemma}\label{lem:compare_hatW_hatc} Let $A$ and $B$ be  operators of even fermion parity. Then
    \begin{equation}
\begin{aligned}
&\mathbb E_{\widehat W}
\left|
\left\langle
A,\widehat W(B)
\right\rangle
\right|^2 \leq
\sum_{k=0}^{\infty}\frac{\kappa^k}{k!}
\sum_{j_1,\ldots,j_k=1}^{2n}
p_{j_1}\cdots p_{j_k}
\sum_{\substack{
\widehat{\mathsf a}_i\in
\{\widehat c_{j_i},\widehat c_{j_i}^\dagger\}\\
i=1,\ldots,k}}
\left|
\left\langle
A,
\widehat{\mathsf a}_1\cdots
\widehat{\mathsf a}_k(B)
\right\rangle
\right|^2.
\end{aligned}
\label{eq:thermal-word-expansion}
\end{equation}
\end{lemma}

\begin{proof}
For every $j$ and any operators $A,B$, the definitions of
$\widehat \gamma_{2j-1},\widehat \gamma_{2j},-i\widehat \gamma_{2j-1}\widehat \gamma_{2j}$, together with the CAR, give
\begin{align}
\mathbb E_{\widehat W_j}
\left|\langle A,\widehat W_j(B)\rangle\right|^2
=
(1-q_j)|\langle A,B\rangle|^2
+\frac{q_j}{2}
\sum_{\widehat{\mathsf{g}}\in\{
\widehat c_j,\widehat c_j^\dagger,
\widehat c_j^\dagger\widehat c_j,
\widehat c_j\widehat c_j^\dagger\}}
|\langle A,\widehat{\mathsf{g}}(B)\rangle|^2.
\end{align}

Applying the above identity successively for $j=1,...,2n$ gives

\begin{align}
\mathbb E_{\widehat W}
\left|
\left\langle
A,\widehat W(B)
\right\rangle
\right|^2= 
\sum_{\substack{
\widehat g_j\in\{I,\widehat c_j,\widehat c_j^\dagger,
\widehat c_j^\dagger\widehat c_j,\widehat c_j\widehat c_j^\dagger\}\\
\text{for each }j=1,\ldots,2n
}}.
\left(\prod_{j=1}^{2n}\lambda_j(\widehat{\mathsf g}_j)\right)
\left|
\left\langle
A,
\widehat{\mathsf g}_1\cdots
\widehat{\mathsf g}_{2n}(B)
\right\rangle
\right|^2,\label{eq:thermal_word-expansition}
\end{align}

where $\lambda_j(I)=1-q_j$  and $\lambda_j(\widehat{\mathsf g})=q_j/2$ for $\widehat{\mathsf g}\neq I$. 

For each summand, let $m_j\in\{0,1,2\}$ be the number of factors
$\widehat c_j$ or $\widehat c_j^\dagger$ in
$\widehat{\mathsf g}_j$. Since $q_j=\kappa^2p_j^2$ and
$\kappa p_j\leq1$,
\begin{equation}
\lambda_j(\widehat{\mathsf g}_j)
\leq
\frac{(\kappa p_j)^{m_j}}{m_j!}.
\label{eq:local-thermal-weight-bound}
\end{equation}

To prove Eq.~(\ref{eq:thermal-word-expansion}), observe the following:
Fix one summand in
\eqref{eq:thermal_word-expansition} and let
$k=\sum_jm_j$. Regard
$\widehat c_j^\dagger\widehat c_j$ and
$\widehat c_j\widehat c_j^\dagger$ as two-letter words $\widehat c_j^\dagger, \widehat c_j$ and
$\widehat c_j, \widehat c_j^\dagger$;  and regard $I$ as
the empty word. The right-hand side of \eqref{eq:thermal-word-expansion}
contains every word obtained by interleaving the local words
while preserving the order of the letters within each mode.
Each such word can be restored to the original order by swapping
only letters from different modes. By the CAR, these swaps
contribute only a sign, so all these words give the same squared
overlap. The number of such interleavings is
$k!/(\prod_jm_j!),$
which leads to 
\[
\frac{\kappa^k}{k!}
\frac{k!}{\prod_jm_j!}
\prod_jp_j^{m_j}
=
\prod_j\frac{(\kappa p_j)^{m_j}}{m_j!}.
\]
This dominates
$\prod_j\lambda_j(\widehat{\mathsf g}_j)$ by
\eqref{eq:local-thermal-weight-bound}. All other terms on the
right-hand side of \eqref{eq:thermal-word-expansion} are
nonnegative, proving the claim.

\end{proof}

\begin{lemma}[Physical word series comparison]
\label{lem:physical-word-series}
Let $J$ be a global word sampled according to the second part
of the proposal in Section~\ref{sec:physical-proposal}.
For any operator $O$,
\begin{equation}
\begin{aligned}
\sum_{k=0}^{\infty}\frac{1}{8^k k!}
\sum_{r_1,\ldots,r_k=1}^n
p_{r_1}\cdots p_{r_k}
\sum_{\substack{
\nu_i\in\{2r_i-1,2r_i\}\\
i=1,\ldots,k}}
\left|
\Tr\!\left(O\gamma_{\nu_1}\cdots\gamma_{\nu_k}\right)
\right|^2 \leq
\exp\!\left(\frac94\sum_{r=1}^np_r\right)
\mathbb E_J|\Tr(OJ)|^2.
\end{aligned}
\end{equation}
\end{lemma}

\begin{proof}
We first group together the terms in which each Majorana appears
the same number of times. Let $s_\nu$ be the number of times
$\gamma_\nu$ appears in a word of length $k$, so that
$k=\sum_{\nu=1}^{2n}s_\nu$. There are
$k!/(s_1!\cdots s_{2n}!)$ possible orderings with these numbers
of appearances. By the CAR, every such word agrees with
$\gamma_1^{s_1}\cdots\gamma_{2n}^{s_{2n}}$ up to a sign,
so they all give the same squared trace.
Summing over these orderings cancels the factor $k!$ in the
denominator. Summing over all lengths $k$ then gives
\begin{align*}
&\sum_{k=0}^{\infty}\frac{1}{8^k k!}
\sum_{r_1,\ldots,r_k=1}^n
p_{r_1}\cdots p_{r_k}
\sum_{\substack{
\nu_i\in\{2r_i-1,2r_i\}\\
i=1,\ldots,k}}
\left|
\Tr\!\left(O\gamma_{\nu_1}\cdots\gamma_{\nu_k}\right)
\right|^2\\
&\quad=
\sum_{s_1,\ldots,s_{2n}\geq0}
\left(
\prod_{r=1}^n
\frac{(p_r/8)^{s_{2r-1}+s_{2r}}}
{s_{2r-1}!\,s_{2r}!}
\right)
\left|
\Tr\!\left(O\gamma_1^{s_1}\cdots
\gamma_{2n}^{s_{2n}}\right)
\right|^2.
\end{align*}

In the expression on the right, the factor at mode $r$ is
$\gamma_{2r-1}^{s_{2r-1}}\gamma_{2r}^{s_{2r}}$.
Since $\gamma_\nu^2=I$, this factor depends only on whether
$s_{2r-1}$ and $s_{2r}$ are even or odd. Up to a phase,
it is one of
\[
I,\qquad
\gamma_{2r-1},\qquad
\gamma_{2r},\qquad
-i\gamma_{2r-1}\gamma_{2r},
\]
according to whether neither, only the first, only the second,
or both of $s_{2r-1},s_{2r}$ are odd.

For either $\nu\in\{2r-1,2r\}$, we have
\[
\sum_{\substack{s_\nu\geq0\\s_\nu\ {\rm even}}}
\frac{(p_r/8)^{s_\nu}}{s_\nu!}\leq e^{p_r/8},
\qquad
\sum_{\substack{s_\nu\geq0\\s_\nu\ {\rm odd}}}
\frac{(p_r/8)^{s_\nu}}{s_\nu!}
\leq\frac{p_r}{8}e^{p_r/8}.
\]
Thus, summing over $s_{2r-1}$ and $s_{2r}$ for each parity
choice gives coefficients for the four local factors above
that are bounded, in the same order, by
\[
e^{p_r/4}
\left(1,\frac{p_r}{8},\frac{p_r}{8},\frac{p_r^2}{64}\right)
\leq
e^{9p_r/4}
\left(1-\frac{3p_r}{4},
\frac{p_r}{4},\frac{p_r}{4},\frac{p_r}{4}\right).
\]
The inequality holds entry by entry, using $p_r\leq1/2$ and
$1-3p_r/4\geq1-p_r\geq e^{-2p_r}$.

The entries in the last parentheses are exactly the
probabilities of the corresponding choices of $J_r$.
Thus, we have
\begin{align*}
&\sum_{s_1,\ldots,s_{2n}\geq0}
\left(
\prod_{r=1}^n
\frac{(p_r/8)^{s_{2r-1}+s_{2r}}}
{s_{2r-1}!\,s_{2r}!}
\right)
\left|
\Tr\!\left(O\gamma_1^{s_1}\cdots
\gamma_{2n}^{s_{2n}}\right)
\right|^2\\
&\quad\leq
\exp\!\left(\frac94\sum_{r=1}^np_r\right)
\mathbb E_{J_1,\ldots,J_n}
\left|\Tr\!\left(OJ_1\cdots J_n\right)\right|^2 =
\exp\!\left(\frac94\sum_{r=1}^np_r\right)
\mathbb E_J|\Tr(OJ)|^2.
\end{align*}
The last equality follows because $A=J_1\cdots J_n$ and
$A^\dagger$ differ only by a phase, so choosing between them
does not change the squared trace. This proves the lemma.
\end{proof}

\weightedcomparisonglobal*

\begin{proof}[Proof of Lemma \ref{lem:physical-to-thermal-word-comparison}]
Starting from \eqref{eq:thermal-word-expansion} in
Lemma~\ref{lem:compare_hatW_hatc}, apply
Lemma~\ref{lem:weighted-one-step-comparison}
successively to the thermal letters, using
$\kappa C_0\leq1/8$.
All intermediate operators have definite fermion parity.
This gives
\begin{equation}
\begin{aligned}
&\mathbb E_{\widehat W}
\left|
\left\langle
A,\widehat W(B)
\right\rangle
\right|^2\\
&\quad\leq
\sum_{k=0}^{\infty}\frac{1}{8^k k!}
\sum_{\substack{s,t\geq0\\s+t=k}}
\binom{k}{s}
\sum_{r_1,\ldots,r_k=1}^n
p_{r_1}\cdots p_{r_k}
\sum_{\substack{
\nu_i\in\{2r_i-1,2r_i\}\\
i=1,\ldots,k}}
\left|
\Tr\!\left(
A^\dagger\,
\gamma_{\nu_1}\cdots\gamma_{\nu_s}
B\,
\gamma_{\nu_{s+1}}\cdots\gamma_{\nu_k}
\right)
\right|^2\\
&\quad\leq
\sum_{s,t=0}^{\infty}\frac{1}{8^{s+t}s!\,t!}
\sum_{r_1,\ldots,r_{s+t}=1}^n
p_{r_1}\cdots p_{r_{s+t}}
\sum_{\substack{
\nu_i\in\{2r_i-1,2r_i\}\\
i=1,\ldots,s+t}}
\left|
\Tr\!\left(
A^\dagger\,
\gamma_{\nu_1}\cdots\gamma_{\nu_s}
B\,
\gamma_{\nu_{s+1}}\cdots\gamma_{\nu_{s+t}}
\right)
\right|^2.
\end{aligned}
\label{eq:temp}
\end{equation}
To obtain the coefficient in \eqref{eq:temp}, note that 
$\frac{1}{8^{s+t}(s+t)!}\binom{s+t}{s}
=
\frac{1}{8^{s+t}s!\,t!}.$

Using cyclicity of the trace and
applying Lemma~\ref{lem:physical-word-series} first to the
left word $\gamma_{\nu_1}\cdots\gamma_{\nu_s}$ in
\eqref{eq:temp},
and then similarly to the right word yields
\begin{align*}
&\mathbb E_{\widehat W}
\left|
\left\langle
A,\widehat W(B)
\right\rangle
\right|^2\\
&\quad\leq
\exp\!\left(\frac94\sum_{r=1}^np_r\right)
\sum_{t=0}^{\infty}\frac{1}{8^t t!}
\sum_{r_1,\ldots,r_t=1}^n
p_{r_1}\cdots p_{r_t}
\sum_{\substack{
\nu_i\in\{2r_i-1,2r_i\}\\
i=1,\ldots,t}}
\mathbb E_J
\left|
\Tr\!\left(
A^\dagger\,J\,B\,
\gamma_{\nu_1}\cdots\gamma_{\nu_t}
\right)
\right|^2\\
&\quad\leq
\exp\!\left(\frac92\sum_{r=1}^np_r\right)
\mathbb E_{J,K}
\left|
\Tr\!\left(A^\dagger\,J\,B\,K\right)
\right|^2.
\end{align*}
Rearranging proves the claim.
\end{proof}

\section{Proof of Lemma \ref{lem:thermal-word-overlap}}\label{app:proof-thermal-word-overlap}

We first prove an auxiliary lemma. 

\begin{lemma}[Expectation comparison from thermal occupations]
\label{lem:expectation-occupation-comparison}
Let $F$ be a Hermitian operator on the doubled Hilbert space.
Suppose that, for every $j$, there exists a value $\alpha_j$ such that
\[
\|F-UFU\|\leq\alpha_j,
\qquad
U\in\{\widehat \gamma_{2j-1},\widehat \gamma_{2j},-i\widehat \gamma_{2j-1}\widehat \gamma_{2j}\}.
\]
Then, for any $A,B$ with $\|A\|_2=\|B\|_2=1$,
\[
\left|
\obra{A}F\oket{A}-\obra{B}F\oket{B}
\right|
\leq
4\sum_{j=1}^{2n}\alpha_j
\left(
\|\widehat c_j(A)\|_2+
\|\widehat c_j(B)\|_2
\right).
\]
\end{lemma}

\begin{proof}
Write $\widehat N_j=\widehat c_j^\dagger\widehat c_j$. To aid the proof, we introduce a quantum channel $\mathcal R_j$
on the doubled Hilbert space that resets thermal mode $j$
to its unoccupied state:
\begin{align}
	\mathcal R_j(\omega)
=
(I-\widehat N_j)\omega(I-\widehat N_j)
+\widehat c_j\omega\widehat c_j^\dagger.\label{eq:reset}
\end{align}

Writing $\eta_j=\Tr(\omega\widehat N_j)$, the block
decomposition into unoccupied and occupied sectors gives
\begin{align}
	\|\omega-\mathcal R_j(\omega)\|_1
\leq
2\sqrt{\eta_j(1-\eta_j)}+2\eta_j
\leq4\sqrt{\eta_j}.\label{eq:etaj}
\end{align}

Consider the averaged operator
\[
\overline F_j
=
\frac14\left(
F+\widehat \gamma_{2j-1}F\widehat \gamma_{2j-1}
+\widehat \gamma_{2j}F\widehat \gamma_{2j}
+(-i\widehat \gamma_{2j-1}\widehat \gamma_{2j})F(-i\widehat \gamma_{2j-1}\widehat \gamma_{2j})
\right)
\]

Note that 
$\widehat \gamma_{2j-1}\overline F_j\widehat \gamma_{2j-1}
=
\widehat \gamma_{2j}\overline F_j\widehat \gamma_{2j}
=
\overline F_j,$ and thus $\overline F_j$ commutes with $\widehat \gamma_{2j-1}$ and $\widehat \gamma_{2j}$.
Since
$\widehat c_j=(\widehat \gamma_{2j-1}+i\widehat \gamma_{2j})/2$
we conclude that $\overline F_j$ commutes with $\widehat c_j$ and $\widehat c_j^\dagger$.
Thus, for any density operator $\omega$, we have
\begin{align}
\operatorname{Tr}\!\left(\mathcal R_j(\omega)\overline F_j\right)
&=\operatorname{Tr}(\omega\overline F_j).\label{eq:equal}
\end{align}

Since $\|F-\overline F_j\|\leq\alpha_j$, Eqs.~\eqref{eq:equal}, \eqref{eq:reset}, and \eqref{eq:etaj} imply
\[
\left|
\Tr\bigl((\omega-\mathcal R_j(\omega))F\bigr)
\right|=\left|
\Tr\bigl((\omega-\mathcal R_j(\omega))(F-\overline{F}_j)\bigr)
\right|
\leq4\alpha_j\sqrt{\Tr(\omega\widehat N_j)}.
\]

Set $\omega^{(0)}=\oket{A}\obra{A}$ and
$\omega^{(j)}=\mathcal R_j(\omega^{(j-1)})$.
Note that $\Tr(\omega^{(j-1)}\widehat N_j)=\|\widehat c_j(A)\|_2^2$,
and $\omega^{(2n)}=\oket{\sqrt{\sigma}}\obra{\sqrt{\sigma}}$.
The preceding bound therefore gives the following telescoping
estimate:\[
\begin{aligned}
\left|\obra{A}F\oket{A}
-\obra{\sqrt{\sigma}}F\oket{\sqrt{\sigma}}\right|
&=\left|\sum_{j=1}^{2n}
\Tr\bigl((\omega^{(j-1)}-\omega^{(j)})F\bigr)\right|\\
&\leq4\sum_{j=1}^{2n}\alpha_j
\sqrt{\Tr(\omega^{(j-1)}\widehat N_j)}=4\sum_{j=1}^{2n}\alpha_j\|\widehat c_j(A)\|_2.
\end{aligned}
\]
 Applying it also to $B$ and using the triangle
inequality proves the claim.

\end{proof}

\thermalwordoverlap*

\begin{proof}
Write
\[
Z_B=
\mathbb E_{\widehat W}
\oket{\widehat W(B)}\obra{\widehat W(B)}, \qquad
\obra{A}Z_B\oket{A}
=
\mathbb E_{\widehat W}
|\langle A,\widehat W(B)\rangle|^2.
\]
First note that $Z_B$ is strictly positive since
every thermal Pauli word in $\widehat{W}$ has positive probability, and these
words span the full operator algebra. 
Keeping only the choice
$\widehat W_j=I$ at every $j$ gives
\[
Z_B\geq
\prod_{j=1}^{2n}(1-3q_j/4)\oket{B}\obra{B}.
\]
Recall $\kappa=\min\{1,(8C_0)^{-1}\},
q_j=\kappa^2p_j^2.$
Thus, we have
\begin{equation}
\obra{B}\log Z_B\oket{B}
\geq
\sum_{j=1}^{2n}\log(1-3q_j/4)
\geq
-\sum_{j=1}^{2n}p_j^2.
\label{eq:thermal-log-at-B}
\end{equation}

Fix $j$ and let
$U\in\{I,\widehat \gamma_{2j-1},\widehat \gamma_{2j},-i\widehat \gamma_{2j-1}\widehat \gamma_{2j}\}$.
Multiplying a thermal word by $U$ permutes the four choices at
mode $j$, up to a phase. Their probabilities are
$1-3q_j/4,q_j/4,q_j/4,q_j/4$. Comparing the probabilities
term by term therefore gives
\[
\frac{q_j}{4}UZ_BU^\dagger\leq Z_B\leq\frac4{q_j}UZ_BU^\dagger.
\]
By operator monotonicity of the logarithm,
\[
-\log(4/q_j)I+\log(UZ_BU^\dagger)
\leq \log Z_B
\leq \log(4/q_j)I+\log(UZ_BU^\dagger).
\]
Since $\log(UZ_BU^\dagger)=U\log Z_B U^\dagger$, there exists a constant $C'$ such that
\begin{equation}
\|\log Z_B-U(\log Z_B)U^\dagger\|
\leq\log(4/q_j)
\leq C'\log(1/p_j).
\label{eq:thermal-Pauli-log-bound}
\end{equation}

By Lemma \ref{lem:expectation-occupation-comparison}, we have
\begin{align}
\left| \obra{A}\log Z_B\oket{A}- 
\obra{B}\log Z_B\oket{B}\right|\leq 
4C'\sum_{j=1}^{2n}\log(1/p_j)
\left(
\|\widehat c_j(A)\|_2+
\|\widehat c_j(B)\|_2
\right).\label{eq:73}
\end{align}

Recall that $Z_B$ is strictly positive. Let $\lambda_k> 0$ and $\oket{v_k}$ be the eigenvalues
and elements of an orthonormal eigenbasis of $Z_B$, respectively.
The weights $w_k=|\obra{v_k}\oket{A}|^2$ satisfy
$w_k\geq0$ and $\sum_k w_k=\|A\|_2^2=1$.
Applying scalar Jensen's inequality to these weights and the
concave function $\log$ gives
\begin{align}
\log\!\left(\obra{A}Z_B\oket{A}\right)
=
\log\!\left(\sum_k w_k\lambda_k\right)
\geq
\sum_k w_k\log\lambda_k
=
\obra{A}\log Z_B\oket{A}.\label{eq:74}
\end{align}

Combining \eqref{eq:74} with \eqref{eq:73} and \eqref{eq:thermal-log-at-B} proves \eqref{eq:thermal-word-overlap}.
\end{proof}

\bibliographystyle{unsrt}
\bibliography{ref.bib}

@article{bravyi2017complexity,
  title={Complexity of quantum impurity problems},
  author={Bravyi, Sergey and Gosset, David},
  journal={Communications in Mathematical Physics},
  volume={356},
  number={2},
  pages={451--500},
  year={2017},
  publisher={Springer}
}

@article{grundner2024complex,
  title={Complex time evolution in tensor networks and time-dependent Green's functions},
  author={Grundner, M and Westhoff, P and Kugler, FB and Parcollet, O and Schollw{\"o}ck, U},
  journal={Physical Review B},
  volume={109},
  number={15},
  pages={155124},
  year={2024},
  publisher={APS}
}

@article{mejuto2019dynamical,
  title={Dynamical mean field theory simulations with the adaptive sampling configuration interaction method},
  author={Mejuto-Zaera, Carlos and Tubman, Norm M and Whaley, K Birgitta},
  journal={Physical Review B},
  volume={100},
  number={12},
  pages={125165},
  year={2019},
  publisher={APS}
}

@article{zgid2012truncated,
  title={Truncated configuration interaction expansions as solvers for correlated quantum impurity models and dynamical mean-field theory},
  author={Zgid, Dominika and Gull, Emanuel and Chan, Garnet Kin-Lic},
  journal={Physical Review B—Condensed Matter and Materials Physics},
  volume={86},
  number={16},
  pages={165128},
  year={2012},
  publisher={APS}
}

@article{chen2026thermal,
  title={Thermal expectation estimation via single-trajectory Gibbs sampling with non-destructive measurements},
  author={Chen, Hongrui and Jiang, Jiaqing and Li, Bowen and Ying, Lexing},
  journal={arXiv preprint arXiv:2603.21595},
  year={2026}
}

@article{jiang2026predicting,
  title={Predicting properties of quantum thermal states from a single trajectory},
  author={Jiang, Jiaqing and Leng, Jiaqi and Lin, Lin},
  journal={arXiv preprint arXiv:2602.12539},
  year={2026}
}

@article{lin2020near,
  title={Near-optimal ground state preparation},
  author={Lin, Lin and Tong, Yu},
  journal={Quantum},
  volume={4},
  pages={372},
  year={2020},
  publisher={Verein zur F{\"o}rderung des Open Access Publizierens in den Quantenwissenschaften}
}

@article{jiang2018quantum,
  title={Quantum algorithms to simulate many-body physics of correlated fermions},
  author={Jiang, Zhang and Sung, Kevin J and Kechedzhi, Kostyantyn and Smelyanskiy, Vadim N and Boixo, Sergio},
  journal={Physical Review Applied},
  volume={9},
  number={4},
  pages={044036},
  year={2018},
  publisher={APS}
}

@inproceedings{gilyen2019quantum,
  title={Quantum singular value transformation and beyond: exponential improvements for quantum matrix arithmetics},
  author={Gily{\'e}n, Andr{\'a}s and Su, Yuan and Low, Guang Hao and Wiebe, Nathan},
  booktitle={Proceedings of the 51st annual ACM SIGACT symposium on theory of computing},
  pages={193--204},
  year={2019}
}

@article{wouters2016practical,
  title={A practical guide to density matrix embedding theory in quantum chemistry},
  author={Wouters, Sebastian and Jim{\'e}nez-Hoyos, Carlos A and Sun, Qiming and Chan, Garnet K-L},
  journal={Journal of Chemical Theory and Computation},
  volume={12},
  number={6},
  pages={2706--2719},
  year={2016},
  publisher={ACS Publications},
  doi={10.1021/acs.jctc.6b00316}
}

@article{knizia2012density,
  title = {Density Matrix Embedding: A Simple Alternative to Dynamical Mean-Field Theory},
  author = {Knizia, Gerald and Chan, Garnet Kin-Lic},
  journal = {Physical Review Letters},
  year = {2012},
  month = nov,
  volume = {109},
  number = {18},
  pages = {186404},
  publisher = {American Physical Society (APS)},
  doi = {10.1103/physrevlett.109.186404},
  pmid = {23215304},
  url = {https://arxiv.org/pdf/1204.5783}
}

@article{RevModPhys.78.865,
  title = {Electronic structure calculations with dynamical mean-field theory},
  author = {Kotliar, G. and Savrasov, S. Y. and Haule, K. and Oudovenko, V. S. and Parcollet, O. and Marianetti, C. A.},
  journal = {Reviews of Modern Physics},
  volume = {78},
  issue = {3},
  pages = {865--951},
  numpages = {0},
  year = {2006},
  month = {Aug},
  publisher = {American Physical Society},
  doi = {10.1103/RevModPhys.78.865},
  url = {https://link.aps.org/doi/10.1103/RevModPhys.78.865}
}

@article{capone2006competition,
  title={Competition between d-wave superconductivity and antiferromagnetism in the two-dimensional Hubbard model},
  author={Capone, Massimo and Kotliar, G},
  journal={Physical Review B—Condensed Matter and Materials Physics},
  volume={74},
  number={5},
  pages={054513},
  year={2006},
  publisher={APS}
}

@article{georges1996dynamical,
  title={Dynamical mean-field theory of strongly correlated fermion systems and the limit of infinite dimensions},
  author={Georges, Antoine and Kotliar, Gabriel and Krauth, Werner and Rozenberg, Marcelo J},
  journal={Reviews of modern physics},
  volume={68},
  number={1},
  pages={13},
  year={1996},
  publisher={APS}
}

@article{georges1992hubbard,
  title={Hubbard model in infinite dimensions},
  author={Georges, Antoine and Kotliar, Gabriel},
  journal={Physical Review B},
  volume={45},
  number={12},
  pages={6479},
  year={1992},
  publisher={APS}
}

@article{chen2025catalytic,
  title={Catalytic tomography of ground states},
  author={Chen, Chi-Fang and King, Robbie},
  journal={arXiv preprint arXiv:2512.10247},
  year={2025}
}

@article{hogan2025quantum,
  title={A quantum computing approach to efficiently simulating correlated materials using impurity models and dynamical mean field theory},
  author={Hogan, Norman and K{\"o}kc{\"u}, Efekan and Steckmann, Thomas and Doak, Liam and Mejuto-Zaera, Carlos and Camps, Daan and Van Beeumen, Roel and de Jong, Wibe and Kemper, A},
  year={2025}
}

@article{rouze2026optimal,
  title={Optimal quantum algorithm for Gibbs state preparation},
  author={Rouz{\'e}, Cambyse and Stilck Fran{\c{c}}a, Daniel and Alhambra, {\'A}lvaro M},
  journal={Physical Review Letters},
  volume={136},
  number={6},
  pages={060601},
  year={2026},
  publisher={APS}
}

@inproceedings{bakshi2026dobrushin,
  title={A Dobrushin condition for quantum Markov chains: Rapid mixing and conditional mutual information at high temperature},
  author={Bakshi, Ainesh and Liu, Allen and Moitra, Ankur and Tang, Ewin},
  booktitle={Proceedings of the 58th Annual ACM Symposium on Theory of Computing},
  pages={1508--1516},
  year={2026}
}

@inproceedings{rouze2025efficient,
  title={Efficient thermalization and universal quantum computing with quantum Gibbs samplers},
  author={Rouz{\'e}, Cambyse and Fran{\c{c}}a, Daniel Stilck and Alhambra, {\'A}lvaro M},
  booktitle={Proceedings of the 57th Annual ACM Symposium on Theory of Computing},
  pages={1488--1495},
  year={2025}
}

@article{chen2025quantum,
  title={Quantum Gibbs states are locally Markovian},
  author={Chen, Chi-Fang and Rouz{\'e}, Cambyse},
  journal={arXiv preprint arXiv:2504.02208},
  year={2025}
}

@article{swendsen1987nonuniversal,
  author  = {Swendsen, Robert H. and Wang, Jian-Sheng},
  title   = {Nonuniversal critical dynamics in {Monte Carlo} simulations},
  journal = {Physical Review Letters},
  volume  = {58},
  number  = {2},
  pages   = {86--88},
  year    = {1987},
  doi     = {10.1103/PhysRevLett.58.86}
}

@article{ding2026toric,
  author  = {Ding, Zhiyan and Landau, Zeph and Li, Bowen
             and Lin, Lin and Zhang, Ruizhe},
  title   = {Polynomial-time preparation of low-temperature
             {Gibbs} states for two-dimensional toric code},
  journal = {Journal of Mathematical Physics},
  volume  = {67},
  number  = {5},
  pages   = {052201},
  year    = {2026},
  doi     = {10.1063/5.0302877},
  eprint  = {2410.01206},
  archivePrefix = {arXiv},
  primaryClass  = {quant-ph}
}

@misc{hangleiter2025code,
  author = {Hangleiter, Dominik and Ju, Nathan and Vazirani, Umesh},
  title  = {{Code Swendsen--Wang} Dynamics},
  year   = {2025},
  eprint = {2510.08446},
  archivePrefix = {arXiv},
  primaryClass  = {quant-ph},
  url    = {https://arxiv.org/abs/2510.08446}
}

@article{jamet2025anderson,
  title={Anderson impurity solver integrating tensor network methods with quantum computing},
  author={Jamet, Fran{\c{c}}ois and Lindoy, Lachlan P and Rath, Yannic and Lenihan, Connor and Agarwal, Abhishek and Fontana, Enrico and Simkovic, Fedor and Martin, Baptiste Anselme and Rungger, Ivan},
  journal={APL Quantum},
  volume={2},
  number={1},
  year={2025},
  publisher={AIP Publishing}
}

@article{mukherjee2023comparative,
  title={Comparative study of adaptive variational quantum eigensolvers for multi-orbital impurity models},
  author={Mukherjee, Anirban and Berthusen, Noah F and Getelina, Jo{\~a}o C and Orth, Peter P and Yao, Yong-Xin},
  journal={Communications Physics},
  volume={6},
  number={1},
  pages={4},
  year={2023},
  publisher={Nature Publishing Group UK, London}
}

@article{sakurai2022hybrid,
  title={Hybrid quantum-classical algorithm for computing imaginary-time correlation functions},
  author={Sakurai, Rihito and Mizukami, Wataru and Shinaoka, Hiroshi},
  journal={Physical Review Research},
  volume={4},
  number={2},
  pages={023219},
  year={2022},
  publisher={APS}
}

@article{hwang2026gibbs,
  author  = {Hwang, Yeongwoo and Jiang, Jiaqing},
  title   = {{Gibbs} state preparation for commuting {Hamiltonian}:
             Mapping to classical {Gibbs} sampling},
  journal = {Quantum},
  volume  = {10},
  pages   = {2209},
  year    = {2026},
  doi     = {10.22331/q-2026-09-18-2209},
  eprint  = {2410.04909},
  archivePrefix = {arXiv},
  primaryClass  = {quant-ph}
}

@article{ayral2025dynamical,
  title={Dynamical mean field theory with quantum computing},
  author={Ayral, Thomas},
  journal={arXiv preprint arXiv:2508.00118},
  year={2025}
}

@article{bauer2016hybrid,
  title={Hybrid quantum-classical approach to correlated materials},
  author={Bauer, Bela and Wecker, Dave and Millis, Andrew J and Hastings, Matthew B and Troyer, Matthias},
  journal={Physical Review X},
  volume={6},
  number={3},
  pages={031045},
  year={2016},
  publisher={APS}
}

@article{leng2026accelerating,
  title={Accelerating quantum {Gibbs} sampling without quantum walks},
  author={Leng, Jiaqi and Jiang, Jiaqing and Lin, Lin},
  journal={arXiv preprint arXiv:2604.22996},
  year={2026},
  note={Version 2},
  url={https://arxiv.org/abs/2604.22996v2}
}

@article{chen2023efficient,
  title={An efficient and exact noncommutative quantum {Gibbs} sampler},
  author={Chen, Chi-Fang and Kastoryano, Michael J. and Gily{\'e}n, Andr{\'a}s},
  journal={arXiv preprint arXiv:2311.09207},
  year={2023},
  url={https://arxiv.org/abs/2311.09207}
}

@article{chen2023thermalpreparation,
  title={Quantum Thermal State Preparation},
  author={Chen, Chi-Fang and Kastoryano, Michael J. and Brand{\~a}o, Fernando G. S. L. and Gily{\'e}n, Andr{\'a}s},
  journal={arXiv preprint arXiv:2303.18224},
  year={2023},
  url={https://arxiv.org/abs/2303.18224}
}

@article{chen2025locallymarkovian,
  title={Quantum {Gibbs} states are locally {Markovian}},
  author={Chen, Chi-Fang and Rouz{\'e}, Cambyse},
  journal={arXiv preprint arXiv:2504.02208},
  year={2025},
  url={https://arxiv.org/abs/2504.02208}
}

@article{jiang2026polynomial,
  title={Polynomial-time classical and quantum simulation of quantum impurity models},
  author={Jiang, Jiaqing and Ju, Nathan and Parekh, Ojas and Rayudu, Chaithanya and Zhao, Andrew},
  journal={arXiv preprint arXiv:2610.02167},
  year={2026}
}

@article{arunachalam2026quantum,
  title={Quantum impurity models: easy at equilibrium, universal in motion},
  author={Arunachalam, Srinivasan and Bravyi, Sergey and Chowdhury, Anirban and Dutt, Arkopal and Gheorghiu, Alexandru and Li, Zhi},
  journal={arXiv preprint arXiv:2610.02130},
  year={2026}
}

@article{werner2006continuous,
  title={Continuous-time solver for quantum impurity models},
  author={Werner, Philipp and Comanac, Armin and de' Medici, Luca and Troyer, Matthias and Millis, Andrew J},
  journal={Physical Review Letters},
  volume={97},
  number={7},
  pages={076405},
  year={2006},
  publisher={APS},
  doi={10.1103/PhysRevLett.97.076405}
}

@article{gull2011continuous,
  title={Continuous-time {Monte Carlo} methods for quantum impurity models},
  author={Gull, Emanuel and Millis, Andrew J and Lichtenstein, Alexander I and Rubtsov, Alexey N and Troyer, Matthias and Werner, Philipp},
  journal={Reviews of Modern Physics},
  volume={83},
  number={2},
  pages={349--404},
  year={2011},
  publisher={APS},
  doi={10.1103/RevModPhys.83.349}
}

@article{ding2026simple,
  title={Simple and efficient end-to-end quantum thermal and ground state preparation},
  author={Ding, Zhiyan and Zhan, Yongtao and Preskill, John and Lin, Lin},
  journal={Nature Physics},
  pages={1--5},
  year={2026},
  publisher={Nature Publishing Group UK, London}
}

@article{jiang2024quantum,
  title={Quantum Metropolis sampling via weak measurement},
  author={Jiang, Jiaqing and Irani, Sandy},
  journal={arXiv preprint arXiv:2406.16023},
  year={2024}
}

@article{chen2025efficient,
  title={Efficient quantum thermal simulation},
  author={Chen, Chi-Fang and Kastoryano, Michael and Brand{\~a}o, Fernando GSL and Gily{\'e}n, Andr{\'a}s},
  journal={Nature},
  volume={646},
  number={8085},
  pages={561--566},
  year={2025},
  publisher={Nature Publishing Group UK London}
}

@article{basso2025quantum,
  title={On quantum to classical comparison for Davies generators},
  author={Basso, Joao and Ganguly, Shirshendu and Sinclair, Alistair and Srivastava, Nikhil and Stier, Zachary and Vuong, Thuy-Duong},
  journal={arXiv preprint arXiv:2510.07267},
  year={2025}
}

@article{temme2013lower,
  title={Lower bounds to the spectral gap of Davies generators},
  author={Temme, Kristan},
  journal={Journal of Mathematical Physics},
  volume={54},
  number={12},
  year={2013},
  publisher={AIP Publishing}
}

@article{tong2025fast,
  title={Fast mixing of weakly interacting fermionic systems at any temperature},
  author={Tong, Yu and Zhan, Yongtao},
  journal={PRX Quantum},
  volume={6},
  number={3},
  pages={030301},
  year={2025},
  publisher={APS}
}

@inproceedings{bergamaschi2025fast,
  title={Fast Mixing of Quantum Spin Chains at All Temperatures},
  author={Bergamaschi, Thiago and Chen, Chi-Fang},
  booktitle={Proceedings of the 58th Annual ACM Symposium on Theory of Computing},
  pages={835--844},
  year={2026},
  doi={10.1145/3798129.3800798}
}

@article{PhysRevLett.93.246403,
  title = {Dynamical Mean Field Theory with the Density Matrix Renormalization Group},
  author = {Garc\'{\i}a, Daniel J. and Hallberg, Karen and Rozenberg, Marcelo J.},
  journal = {Physical Review Letters},
  volume = {93},
  issue = {24},
  pages = {246403},
  numpages = {4},
  year = {2004},
  month = {Dec},
  publisher = {American Physical Society},
  doi = {10.1103/PhysRevLett.93.246403},
  url = {https://link.aps.org/doi/10.1103/PhysRevLett.93.246403}
}

@article{ding2025efficient,
  title={Efficient Quantum Gibbs Samplers with Kubo--Martin--Schwinger Detailed Balance Condition: Z. Ding, B. Li, L. Lin},
  author={Ding, Zhiyan and Li, Bowen and Lin, Lin},
  journal={Communications in Mathematical Physics},
  volume={406},
  number={3},
  pages={67},
  year={2025},
  publisher={Springer}
}

@article{wilson1975renormalization,
  title={The renormalization group: Critical phenomena and the {Kondo} problem},
  author={Wilson, Kenneth G},
  journal={Reviews of Modern Physics},
  volume={47},
  number={4},
  pages={773--840},
  year={1975},
  publisher={APS},
  doi={10.1103/RevModPhys.47.773}
}

@article{sun2020finite,
  title={Finite-temperature density matrix embedding theory},
  author={Sun, Chong and Ray, Ushnish and Cui, Zhi-Hao and Stoudenmire, Miles and Ferrero, Michel and Chan, Garnet Kin-Lic},
  journal={Physical Review B},
  volume={101},
  number={7},
  pages={075131},
  year={2020},
  publisher={American Physical Society},
  doi={10.1103/PhysRevB.101.075131},
  url={https://doi.org/10.1103/PhysRevB.101.075131}
}

@article{maier2005systematic,
  title={Systematic Study of $d$-Wave Superconductivity in the {2D} Repulsive {Hubbard} Model},
  author={Maier, T. A. and Jarrell, M. and Schulthess, T. C. and Kent, P. R. C. and White, J. B.},
  journal={Physical Review Letters},
  volume={95},
  number={23},
  pages={237001},
  year={2005},
  doi={10.1103/PhysRevLett.95.237001},
  url={https://doi.org/10.1103/PhysRevLett.95.237001}
}
\end{document}